\documentclass[a4paper,12pt]{article}
\usepackage{jheppub}
\usepackage{amsmath,amssymb,mathrsfs,amsthm,tikz,shuffle,paralist}
\usepackage{dsfont}
\usepackage{color}
\usepackage{enumitem}
\definecolor{darkred}{RGB}{173,34,48}

\usetikzlibrary{shapes.misc}
\usetikzlibrary{cd}
\usetikzlibrary{snakes}
\usetikzlibrary{calc}
\usetikzlibrary{decorations}

\usepackage[all]{xypic}

\usepackage{subcaption}
\usepackage{caption}

\makeatletter
\DeclareRobustCommand*{\bfseries}{%
  \not@math@alphabet\bfseries\mathbf
  \fontseries\bfdefault\selectfont
  \boldmath
}

\usepackage[textsize=tiny]{todonotes}

\newtheorem*{theorem}{Theorem}

\newcommand{\otimesprime}{\otimes'}

\newcommand{\me}{\mathrm{e}}  
\newcommand{\mi}{i}
\newcommand{\dif}{\mathrm{d}} 

\DeclareMathOperator{\Li}{Li} 

\newcommand{\Ef}[3]{{\textrm{E}_4}(\begin{smallmatrix}#1\\#2%
\end{smallmatrix};#3)} 
\newcommand{\cEf}[3]{{\mathcal{E}_4}(\begin{smallmatrix}#1\\#2%
\end{smallmatrix};#3)} 
\newcommand{\gamt}[3]{{\widetilde{\Gamma}}(\begin{smallmatrix}#1\\#2%
\end{smallmatrix};#3)}

\usepackage[utf8]{inputenc}
\usepackage{graphicx}
\usepackage{dcolumn}
\usepackage{bm}

\usepackage{todonotes}

\newcommand{\db}{%
\begin{tikzpicture}[scale=0.1,label distance=-1mm,baseline={([yshift=-.5ex]current bounding box.center)}]
\clip (0,14) rectangle (3,16);
		\node (0) at (0, 15.5) {};
		\node (1) at (1.5, 16) {};
		\node (2) at (0.5, 16) {};
		\node (3) at (2.5, 16) {};
		\node (4) at (3.0, 15.5) {};
		\node (5) at (0.5, 14) {};
		\node (6) at (0, 14.5) {};
		\node (7) at (1.5, 14) {};
		\node (8) at (3, 14.5) {};
		\node (9) at (2.5, 14) {};
        \node (11) at (5, 15) {};
		\node (12) at (6, 15) {};
		\draw[thick] (0.center) to (4.center);
		\draw[thick] (6.center) to (8.center);
		\draw[thick] (2.center) to (5.center);
		\draw[thick] (1.center) to (7.center);
		\draw[thick] (3.center) to (9.center);
\end{tikzpicture}%
}

\newcommand{\softdb}{
    \begin{tikzpicture}[scale=0.1,label distance=-1mm]
        \clip (0,14) rectangle (3,16);
                \node (0) at (0, 15.5) {};
                \node (1) at (1.5, 15.5) {};
                \node (2) at (0.5, 16) {};
                \node (3) at (2.5, 16) {};
                \node (4) at (3.0, 15.5) {};
                \node (5) at (0.5, 14) {};
                \node (6) at (0, 14.5) {};
                \node (7) at (1.5, 14) {};
                \node (8) at (3, 14.5) {};
                \node (9) at (2.5, 14) {};
                \node (11) at (5, 15) {};
                \node (12) at (6, 15) {};
                \draw[thick] (0.center) to (4.center);
                \draw[thick] (6.center) to (8.center);
                \draw[thick] (2.center) to (5.center);
                \draw[thick] (1.center) to (7.center);
                \draw[thick] (3.center) to (9.center);
        \end{tikzpicture}%
    }

\newcommand{\sr}{%
\begin{tikzpicture}[scale=0.1,label distance=-1mm]
\clip (0,14.6) rectangle (3,16.4);
		\node (0) at (0, 15.5) {};
		\node (4) at (3.0, 15.5) {};
		\draw[thick] (0.center) to (4.center);
		\draw[thick] (1.5,15.5) circle (0.75);
\end{tikzpicture}%
}

\begin{document}

\title{Symbology for elliptic multiple polylogarithms and the symbol prime}

\author{Matthias Wilhelm,}%
\emailAdd{matthias.wilhelm@nbi.ku.dk}
\author{Chi Zhang}%
\emailAdd{chi.zhang@nbi.ku.dk}

 \affiliation{%
Niels Bohr International Academy, Niels Bohr Institute, Copenhagen University, Blegdamsvej 17, 2100 Copenhagen \O{}, Denmark}

\date{\today}%

\abstract{Elliptic multiple polylogarithms occur in Feynman integrals and in particular in scattering amplitudes. They can be characterized by their symbol, a tensor product in the so-called symbol letters. In contrast to the non-elliptic case, the elliptic letters themselves satisfy highly non-trivial identities, which we discuss in this paper.
Moreover, we introduce the symbol prime, an analog of the symbol for elliptic symbol letters, which makes these identities manifest. We demonstrate its use in two explicit examples at two-loop order: the unequal-mass sunrise integral in two dimensions and the ten-point double-box integral in four dimensions. Finally, we also report the result of the polylogarithmic nine-point double-box integral, which arises as the soft limit of the ten-point integral.}

\maketitle

\section{Introduction}

 Scattering amplitudes as well as further quantities in Quantum Field Theory contain a rich mathematical structure, whose understanding has frequently expanded our calculational reach -- benefiting both phenomenological tests of the Standard Model of Particle Physics at the LHC as well as more formal studies.
 
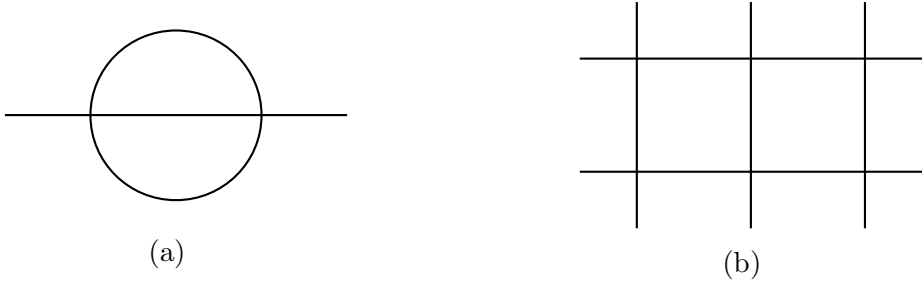
\begin{figure}
 \begin{subfigure}{.5\textwidth}
  \centering
  \begin{tikzpicture}[scale=1.5,label distance=-1mm]
\clip (0,14.6) rectangle (3,16.4);
		\node (0) at (0, 15.5) {};
		\node (4) at (3.0, 15.5) {};
		\draw[thick] (0.center) to (4.center);
		\draw[thick] (1.5,15.5) circle (0.75);
\end{tikzpicture}  
  \caption{\textcolor{white}{.}}
  \label{subfig:sunrise}
\end{subfigure}\begin{subfigure}{.5\textwidth}
  \centering
  \begin{tikzpicture}[scale=1.5,label distance=-1mm]
\clip (0,14) rectangle (3,16);
		\node (0) at (0, 15.5) {};
		\node (1) at (1.5, 16) {};
		\node (2) at (0.5, 16) {};
		\node (3) at (2.5, 16) {};
		\node (4) at (3.0, 15.5) {};
		\node (5) at (0.5, 14) {};
		\node (6) at (0, 14.5) {};
		\node (7) at (1.5, 14) {};
		\node (8) at (3, 14.5) {};
		\node (9) at (2.5, 14) {};
        \node (11) at (5, 15) {};
		\node (12) at (6, 15) {};
		\draw[thick] (0.center) to (4.center);
		\draw[thick] (6.center) to (8.center);
		\draw[thick] (2.center) to (5.center);
		\draw[thick] (1.center) to (7.center);
		\draw[thick] (3.center) to (9.center);
\end{tikzpicture}
  \caption{\textcolor{white}{.}}
  \label{subfig:doublebox}
\end{subfigure}
\caption{The sunrise integral in two dimensions with unequal internal masses (\subref{subfig:sunrise}) as well as the ten-point double-box integral in four dimensions with massless internal propagators (\subref{subfig:doublebox}).}
\label{fig: diagrams intro} 
\end{figure}

 At one-loop order, and in certain cases also at higher loop orders, the functions that occur in Feynman integrals and thus in Quantum Field Theory are multiple polylogarithms (MPLs) \cite{Chen:1977oja,G91b,Goncharov:1998kja,Remiddi:1999ew,Borwein:1999js,Moch:2001zr}, which are by now very well understood.
 Increasing the loop order, the next class of functions we encounter are elliptic multiple polylogarithms (eMPLs), on which there has been much recent progress \cite{Laporta:2004rb,MullerStach:2012az,brown2011multiple,Bloch:2013tra,Adams:2013nia,Adams:2014vja,Adams:2015gva,Adams:2015ydq,Adams:2016xah,Adams:2017ejb,Adams:2017tga,Bogner:2017vim,Broedel:2017kkb,Broedel:2017siw,Remiddi:2017har,Chen:2017soz,Bourjaily:2017bsb,Adams:2018yfj,Broedel:2018iwv,Broedel:2018qkq,Honemann:2018mrb,Bogner:2019lfa,Broedel:2019hyg,Duhr:2019rrs,Walden:2020odh,Weinzierl:2020fyx,Kristensson:2021ani,Frellesvig:2021hkr},
 in particular by studying the two-loop massive sunrise integral in two dimensions \cite{SABRY1962401,Broadhurst:1993mw,Laporta:2004rb,Muller-Stach:2011qkg,Adams:2013nia,Bloch:2013tra,Remiddi:2013joa,Bloch:2014qca,Adams:2014vja,Adams:2015gva,Adams:2015ydq,Bloch:2016izu,Remiddi:2016gno,Adams:2017ejb,Broedel:2017siw} (see figure \ref{subfig:sunrise}). More recently, also more complicated Feynman integrals are starting to be understood,
  in particular the 
 two-loop ten-point double-box integral with massless internal propagators in four dimensions \cite{Bourjaily:2017bsb,Kristensson:2021ani} (see figure \ref{subfig:doublebox}).
 Beyond eMPLs, also integrals over more complicated geometries than elliptic curves occur \cite{Brown:2009ta,Brown:2010bw,Bourjaily:2018yfy,Bourjaily:2018ycu,Festi:2018qip,Broedel:2019kmn,Besier:2019hqd,Bourjaily:2019hmc,Vergu:2020uur,mirrors_and_sunsets,
Broadhurst:1987ei,Adams:2018kez,Adams:2018bsn,Huang:2013kh,Klemm:2019dbm,Bonisch:2020qmm,Bonisch:2021yfw,Muller:2022gec,Chaubey:2022hlr}; an understanding of the corresponding spaces of functions is still in its infancy.
 For a recent review on functions in scattering amplitudes beyond MPLs, see \cite{Bourjaily:2022bwx}.
 
Our good understanding of MPLs is to a large extend due to the Hopf algebra structure underlying these functions \cite{Gonch2,Goncharov:2010jf,Brown:2011ik,Duhr:2011zq,Duhr:2012fh}, and in particular the symbol \cite{Goncharov:2010jf}.
The symbol map associates to each MPL $f$ a simple tensor product, $\mathcal{S}(f)=\sum \log(\phi_{i_1})\otimes\dots\otimes \log(\phi_{i_k})$.%
\footnote{Note that in contrast to much of the literature on MPLs, we are not suppressing the $\log$ in the notation.}
The entries in this tensor product, called symbol letters, are logarithms of rational or algebraic functions $\phi_i$ of the kinematic invariants.
Since a tensor product is easy to manipulate, and the identities of the symbol letters, $\log(a)+\log(b)=\log(ab)$, are well understood, the symbol provides a powerful way of finding identities between MPLs and for simplifying expressions. 
The symbol moreover manifests physical properties of the corresponding function. For example, the first entry of the symbol describes the discontinuities, which are heavily restricted in particular in massless theories resulting in so-called first-entry conditions \cite{Gaiotto:2011dt}.
Moreover, discontinuities in overlapping channels are forbidden by the so-called Steinmann conditions \cite{Steinmann,Steinmann2}, restricting which symbol letter in the second entry can follow a particular letter in the first entry.
The symbol has made possible enormous progress for quantities consisting of MPLs, both in relation to phenomenology and more formal studies, including in particular powerful bootstrap techniques \cite{Dixon:2011pw,Dixon:2011nj,Dixon:2013eka,Dixon:2014voa,Dixon:2014iba,Drummond:2014ffa,Dixon:2015iva,Caron-Huot:2016owq,Dixon:2016apl,Dixon:2016nkn,Drummond:2018caf,Caron-Huot:2019vjl,Li:2016ctv,Almelid:2017qju,Dixon:2020bbt,Dixon:2022rse,Abreu:2020jxa,Chicherin:2020umh,Dixon:2012yy,Chestnov:2020ifg,He:2021fwf,Drummond:2017ssj,Chicherin:2017dob,Caron-Huot:2018dsv,Henn:2018cdp,He:2021non,He:2021eec,Heller:2019gkq,Heller:2021gun,Duhr:2021fhk}.

While the symbol for eMPLs was defined in \cite{BrownNotes,Broedel:2018iwv}, it has so far not been put to much use, and is still much less understood than its analog for MPLs.
One reason is that the symbol letters $\Omega^{(i)}$ for eMPLs are themselves elliptic functions of the kinematic invariants. 
In particular, $\Omega^{(-1)}=-2\pi i \tau$ occurs as a letter, where $\tau=\omega_2/\omega_1$ is the ratio of the two periods of the elliptic curve. 
The letters $\Omega^{(0)}$ satisfy simple identities as the consequence of the group law on the elliptic curve.
In \cite{Kristensson:2021ani}, some identities for the elliptic letters $\Omega^{(i)}$ with $i=1,2$ were observed numerically in the study of the symbol of the ten-point elliptic double-box integral. 
Using these identities, it was found that the elliptic letters in the first two entries simplify to logarithms, manifesting the same first-entry condition as for polylogarithmic amplitudes as well as the Steinmann conditions.
Moreover, the last entries were found to be given by simple elliptic integrals $\Omega^{(0)}$, with $\Omega^{(2)}$ only occurring in the third entry preceding the modular parameter $\tau$ in the last entry.

In this paper, we show how the identities observed in \cite{Kristensson:2021ani} for $\Omega^{(1)}$ are a consequence of Abel's theorem \cite{abel1841}.
Moreover, we demonstrate that the identities observed in \cite{Kristensson:2021ani} for $\Omega^{(2)}$ are a consequence of the elliptic Bloch relation \cite{Zagier2000,bloch2011higher} for the elliptic dilogarithm, which generalizes the Bloch relation for the classic dilogarithm and which have also been studied in the context of finding identities between elliptic multiple polylogarithms \cite{Broedel:2019tlz,Bolbachan:2019dsu}.
While the identities for $\Omega^{(1)}$ can be reduced to three-term identities similar to $\log(a)+\log(b)=\log(ab)$ in the case of the logarithm, the elliptic Bloch relation, and thus the identities for $\Omega^{(2)}$, are five-term identities similar to the Bloch relation for the classical dilogarithm, which are made manifest only by the symbol.
Thus, we introduce a symbol prime $\mathcal{S}'$ for the symbol letters $\Omega^{(2)}$
(and similarly for $\Omega^{(n>2)}$)
in analogy to the symbol for MPLs and eMPLs, which makes the identities due to the elliptic Bloch relation manifest. 

In general, eMPLs transform under modular transformations of $\tau$ in a complicated way \cite{Duhr:2019rrs,Weinzierl:2020fyx}, and results given in terms of eMPLs are not manifestly double periodic.
However, in the examples we studied, we find that the symbol prime makes both double periodicity as well as a simple behavior under modular transformations manifest.
Finally, it makes also part of the integrability conditions manifest, which result from the requirement that partial derivatives commute.

To illustrate the use of the symbol for eMPLs, the application of the identities of elliptic letters as well as the symbol prime, we study two concrete examples.
The first example is the two-loop sunrise integral in two dimensions with all internal masses being unequal (see figure \ref{subfig:sunrise}). 
The second example is the ten-point two-loop double-box integral in four dimensions with massless internal propagators (see figure \ref{subfig:doublebox}).
In addition to the aforementioned properties and techniques, we also demonstrate how the (elliptic) symbol reduces to polylogarithmic symbol in kinematic limits where the elliptic curve degenerates.

The remainder of this paper is organized as follows. We review 
 elliptic multiple polylogarithms in section \ref{sec:2}. In section \ref{sec:3}, we derive identities for elliptic symbol letters -- based on Abel's addition theorem for $\Omega^{(1)}$'s and by introducing the symbol prime map for $\Omega^{(2)}$'s. We illustrate the use of these techniques for the unequal-mass sunrise integral in section \ref{sec: example 1} and for the ten-point double-box integral in section \ref{sec: example 2}.
 In particular, we provide analytic results for the non-elliptic nine-point double-box integral and its symbol, which result from taking the soft limit of the ten-point double-box integral.
 We conclude with a discussion and an  outlook on open questions in section \ref{sec:5}.
In appendix \ref{app:sunrise}, we review the calculation of the unequal-mass sunrise integral via Feynman parameters using our conventions and notation.  
The details of simplifying the symbols for the sunrise integral, as well as the expressions of the functions and 
symbols for the ten-point elliptic integral and its soft limit, are included as ancillary files (\texttt{sunrise\_symbol.nb}, \texttt{doublebox\_omega2} and \texttt{doublebox\_soft}).%
\footnote{In this article, 
we only provide the expression for the ten-point double box in the normalization by the period $-\omega_2$ since the corresponding expressions in the normalization by the period $\omega_1$ can be found in the ancillary files of \cite{Kristensson:2021ani}.}

\section{Review of Elliptic Multiple Polylogarithms}  \label{sec:2}

Let us first review several elementary facts about elliptic multiple polylogarithms; see \cite{Broedel:2017kkb,Broedel:2017siw,Broedel:2018iwv,Broedel:2018qkq} for further details. We follow the notations and conventions of \cite{Kristensson:2021ani}, which differ slightly from those of \cite{Broedel:2017kkb,Broedel:2017siw,Broedel:2018iwv,Broedel:2018qkq}.

\subsection{Elliptic multiple polylogarithms on the elliptic curve}
\label{subsec: 2.3}

In this paper, the elliptic curves $\mathcal{C}$ are described by monic quartic polynomials:
\begin{equation}
    y^{2} =P_{4}(x)= x^{4} + a_{3}x^{3} + a_{2}x^{2} + a_{1}x + a_{0} \:. \label{quarticEcurve}
\end{equation}
Such elliptic curves can be birationally mapped to Weierstrass form 
\begin{equation}
   Y^{2}=4X^{3}-g_{2}X-g_{3}                    \end{equation}
using the rational point at $(x,y)=(+\infty,+\infty)$, where $(X,Y)$ are related to $(x,y)$ by 
\begin{equation} \label{XYtoxy}
    \begin{aligned}
        X &= \frac{1}{12}\bigl( a_{2} + 3a_{3}x + 6x^{2}+6y\bigr) \:,  \\
        Y &= \frac{1}{4}\bigl( a_{1} + 2a_{2}x + 3a_{3}x^{2} + 4x^{3} +a_{3}y+4xy\bigr) \:.
    \end{aligned}    
\end{equation}

On the curve $\mathcal{C}$, we can introduce elliptic multiple polylogarithms $\mathrm{E}_{4}$, which are recursively defined as \cite{Broedel:2017kkb}%
\footnote{The subscript ``$4$'' indicates that the elliptic curve is given by a quartic polynomial. Analogous functions for a cubic polynomial were also defined in \cite{Broedel:2017kkb}.}
\begin{equation}
\label{Eiterateddefinition}
    \Ef{n_1 & \ldots & n_k}{c_1 & \ldots& c_k}{x}=
    \int_{0}^{x}\dif  x'\,\psi_{n_{1}}(c_{1},x')\Ef{n_{2} & \ldots & n_k}{c_{2} & \ldots& c_k}{x'}
\end{equation}
with $\mathrm{E}_{4}(;x)=1$, where
\begin{equation} \label{psibasis}
    \begin{split}
        &\psi_{0}(0,x)=\frac{1}{y} \,,\qquad  \:\:\psi_{-1}(\infty,x)=\frac{x}{y}\,, \\
        &\psi_{1}(c,x)=\frac{1}{x-c}\,,\quad 
        \psi_{-1}(c,x)=\frac{y_{c}}{y(x-c)}\,,
    \end{split} 
\end{equation}
with $y_{c}=y\vert_{x=c}$. The definitions of $\psi_{n}(c,x)$ for $n=\pm2,\pm3,\dots$ can be found in \cite{Broedel:2017kkb}; the kernels \eqref{psibasis} are sufficient for the purpose of this paper, though. 

The class of elliptic multiple polylogarithms $\Ef{n_1 & \ldots & n_k}{c_1 & \ldots& c_k}{x}$ contains in particular all non-elliptic (Goncharov) multiple polylogarithms, defined by 
\begin{equation}
 G(c_1,\dots,c_n;x)=\int_0^x\frac{\dif x'}{x'-c_1}G(c_2,\dots,c_n;x')
\end{equation}
with $G(;x)=1$,
since by definition $\Ef{1 & \ldots & 1}{c_1 & \ldots& c_k}{x}\equiv G(c_1,\dots,c_k;x)$.

In general, any integral of the form
\begin{equation}
    \int\frac{\dif x}{y} \:\mathcal{G}(x,y)\,, \label{1dIntofPolylog}
\end{equation}
where $\mathcal{G}$ is a polylogarithm whose letters are rational functions of $x$ and $y$, can be converted to $\mathrm{E}_{4}$ functions with only the four kinds of integration kernels defined in \eqref{psibasis}.%
\footnote{Since $\mathcal{G}$ is a polylogarithm, the integration kernels have only simple poles, in addition to being rational in $x$ and $y$. While all integration kernels $\psi_n$ have only simple poles, only $\psi_{-1,0,+1}$ are rational functions of $x,y$.}
 In particular, this is the case for the (unequal-mass) sunrise integral and the ten-point double-box integral, which we will study as examples in sections \ref{sec: example 1}--\ref{sec: example 2}.
 
 \subsection{From the elliptic curve to the torus}

The functions $\mathrm{E}_{4}$ are defined on the elliptic curve and directly arise from the Feynman-parameter representation of Feynman integrals. However, the purity of some elliptic Feynman integrals, such as integrals of the form \eqref{1dIntofPolylog}, is not visible when expressed  in terms of $\mathrm{E}_{4}$ functions since taking the total derivative of a $\mathrm{E}_{4}$ function does not necessarily decrease its length \cite{Broedel:2018iwv}.%
\footnote{This can be seen concretely by how the integration kernels $\psi_{-1}(\infty,x)\dif x$ and $\psi_{-1}(c,x) \dif x$ are related to the kernels of pure functions given below in \eqref{psitog1}.
}
On the other hand, iterated integrals defined on the torus, such as the $\tilde{\Gamma}$ functions we will review below, are manifestly pure and hence allow a symbol map defined via the total derivative. 

To connect both sides, we first need a bijection between the elliptic curve $\mathcal{C}$ and the torus $\mathbb{C}/\Lambda$, where $\Lambda$ is the lattice generated by the periods $\omega_{1}$ and $\omega_{2}$ of the elliptic curve. For an elliptic curve of the form \eqref{quarticEcurve}, one can find such a map through the birationally 
equivalent curve in the Weierstrass normal form: first solve $(x,y)$ in terms of $(X,Y)$ from  \eqref{XYtoxy}, then replace $X$ and $Y$ with the Weierstrass elliptic function $\wp(z)$ and its derivative $\wp'(z)$, respectively. 
This gives \begin{equation}
            z\mapsto (x,y)=(\kappa(z),\kappa'(z))\,,
           \end{equation}
where
 \begin{equation}
     \kappa(z)=\frac{6a_{1}-a_{2}a_{3}+12 a_{3}\wp(z)-24\wp'(z)}{3a_{3}^{2}-8a_{2}-48\wp(z)}.
 \end{equation} 
It is easy to see that $\kappa(0)=\infty$, and hence all lattice points are mapped to the infinity point in the $(x,y)$-space. Furthermore, each point $c$ in the $x$-space corresponds to two points $(c,\pm y_{c})$ on the elliptic curve $\mathcal{C}$ and hence to two images on the torus $\mathbb{C}/\Lambda$, which we denote by $z_{c}^{\pm}$;
 these two images satisfy 
 \begin{equation} \label{wpwmrelation}
     z_{c}^{+}+z_{c}^{-}= z^{-}_{\infty} + z^{+}_{\infty}\equiv z^{-}_{\infty} \:  \operatorname{mod} \Lambda \:,
 \end{equation}
 since the corresponding points $(X_{c}^{\pm},Y_{c}^{\pm})$ and
 \begin{equation}
    (X_{\infty}^{-},Y_{\infty}^{-})  = \bigl(\tfrac{1}{48}(3a_{3}^{2}-8a_{2}),\tfrac{1}{32}(-8a_{1}+4a_{2}a_{3}-a_{3}^{3}) \bigr)  
 \end{equation}
 are on the same line in $(X,Y)$-space.

\tikzset{cross/.style={cross out, thick, draw=black, fill=none, minimum size=2*(#1-\pgflinewidth), inner sep=0pt, outer sep=0pt}, cross/.default={2pt}}

\begin{figure}
   \centering
   \begin{tikzpicture}
           \draw[->, thick] (0,0) to (0,3);
           \draw[->, thick] (-3,1.5) to (1,1.5);
           \draw[<-,line width=0.23mm,red] (-0.8,0.0) to (-0.8,3.0);
           \draw[->,line width=0.23mm,blue] (-3,1.4) .. controls (-1,1.4) ..   (-1.0,0.0) ; 
           \node[cross,label=above:$r_{1}$] at (-1.5, 2.0) {};
           \node[cross,label=below:$r_{4}$] at (-1.5, 1.0) {};
           \node[cross,label=above:$r_{2}$] at (-0.5, 0.8) {};
           \node[cross,label=below:$r_{3}$] at (-0.5, 2.2) {};
           \node at (-0.5,0.1) {\textcolor{red}{$\gamma_{1}$}};
           \node at (-1.3,0.1) {\textcolor{blue}{$\gamma_{-}$}};
           \node at (0.4,3) {$\Im x$};
           \node at (1.3,1.5) {$\Re x$};
   \end{tikzpicture}
   \caption{The distribution of the four roots of $y^{2}(x)$ and the two integration contours $\gamma_1$ and $\gamma_-$ defining the period $\omega_1$ and $z_\infty^-$. The contour $\gamma_2$ which defines the period $\omega_2$ runs along the real axis.
   }
   \label{fig: contours}
   \end{figure}
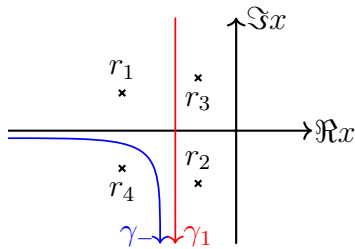

The inverse map from the torus to the elliptic curve is simply given by the Abel-Jacobi map. We assume that the four roots of $y^{2}(x)$ come in complex conjugate pairs as shown in figure \ref{fig: contours}.%
\footnote{For a discussion of other possible distributions of roots, see \cite{Broedel:2017kkb}.}
Then, the torus image $z_{c}^{+}$ for any real $c$ is given by
\begin{equation}
    z_{c}^{+}=\int_{-\infty}^{c} \frac{\dif x}{y}\,.
\end{equation} 
Hence, $z_{\infty}^{+}$ is one period of the torus, and we choose it to be $\omega_{2}$. The image $z_{c}^{-}$ can be obtained by \eqref{wpwmrelation} together with $z_{\infty}^{-}=\int_{\gamma_{-}} \dif x/y$, and the other period is $\omega_{1}=\int_{\gamma_{1}} \dif x/y$; see figure \ref{fig: contours} for the definitions of the integration contours $\gamma_{-}$ and $\gamma_{1}$.
Due to the distribution of roots, $\omega_{2}$ and $\mi\omega_{1}$ are \emph{positive} reals.

\subsection{Elliptic multiple polylogarithms on the torus}

Due to the equivalence between the elliptic curve and the torus,  
another way to define elliptic multiple polylogarithms is via iterated integrals on a torus. Such iterated integrals can be formulated in several ways. 
In this paper, we use the so-called $\tilde{\Gamma}$ functions \cite{Broedel:2017kkb,Broedel:2018iwv}, which are defined as\footnote{Another variant of such iterated integrals extensively used in one-loop string amplitudes  are the so-called $\Gamma$ functions, whose integration kernels $f^{(n)}$ are double periodic but \emph{not} meromorphic; see e.g.\ \cite{Broedel:2014vla}.} 
\begin{equation}
    \gamt{n_1 & \ldots & n_k}{w_1 & \ldots& w_k}{w|\tau}=
    \int_{0}^{w} \dif w'\,g^{(n_{1})}(w'{-}w_{1},\tau)\gamt{n_{2} & \ldots & n_k}{w_{2} & \ldots& w_k}{w'|\tau}
\end{equation}
with $\tilde{\Gamma}(;w|\tau)=1$; we will frequently suppress the dependence on $\tau$ for ease of notation. Such an iterated integral is said to have length $k$ and weight $\sum_kn_k$, and in contrast to the case of MPLs both quantities are not necessarily equal. 
The integration kernels $g^{(n)}(w,\tau)$ are generated by the \emph{Eisenstein-Kronecker series}
\begin{equation}
    \frac{\partial_{w}\theta_{1}(0|\tau)\theta_{1}(w+\alpha|\tau)}{\theta_{1}(w|\tau)\theta_{1}(\alpha|\tau)} = \sum_{n\geq 0}\alpha^{n-1}g^{(n)}(w,\tau)\:,
\end{equation}
where $\theta_{1}(w|\tau)$ is the odd Jacobi theta function. 
All the integration kernels $g^{(n)}$ except $g^{(0)}=1$ are quasi double periodic,
\begin{equation}
    g^{(n)}(w+1)=g^{(n)}(w)\:, \qquad  g^{(n)}(w+\tau)=\sum_{j=0}^{n}\frac{(-2\pi \mi)^{j}}{j!}g^{(n-j)}(w)\:,
\end{equation}
but meromorphic with only logarithmic poles at most \cite{Broedel:2017kkb,Broedel:2018iwv}. 

Note that the functions $\tilde\Gamma$ and the integration kernels $g^{(n)}$ are defined on the torus with one period rescaled to be $1$, and that the torus with periods $(\omega_{1},\omega_{2})$ has the two possible rescalings $[1:\tau=\omega_{2}/\omega_{1}]$ and $[1:\tau'=-\omega_{1}/\omega_{2}]$, which are related by the modular $S$-transformation $\tau\to -\tau^{-1}$. We denote the images of $c$ on $[1:\tau=\omega_{2}/\omega_{1}]$ as $w^{\pm}_{c}=z^{\pm}_{c}/\omega_1$ and the images on $[1:\tau'=-\omega_{1}/\omega_{2}]$ as $\xi_{c}^{\pm}=z^{\pm}_{c}/(-\omega_2)$. The two are related by $\xi_c^\pm=\tau' w_c^\pm$. In what follows, most of the results are written in terms of $w$-coordinates, but the corresponding results with $w$ replaced by $\xi$ also hold unless otherwise indicated.

The integration kernels $\psi_{n}$ can be identified as combinations of $g^{(j)}$'s by matching poles on both sides. On the torus $[1:\tau=\omega_{2}/\omega_{1}]$, one can easily find the following relations between $g^{(j)}$'s and $\psi_{n}$'s,
\begin{subequations} \label{psitog1}
    \begin{align}
        \psi_{1}(c,x)\dif x &=\Bigl(g^{(1)}(w-w_{c}^{+})+g^{(1)}(w-w_{c}^{-}) \nonumber \\ 
                            &\qquad \qquad \qquad \qquad -g^{(1)}(w-w_{\infty}^{+}) -g^{(1)}(w-w_{\infty}^{-})\Bigr)\dif w \,, 
                            \label{psitog1a}\\
        \psi_{-1}(c,x)\dif x &= 
        \Bigl( g^{(1)}(w-w_{c}^{+})-g^{(1)}(w-w_{c}^{-}) + g^{(1)}(w_{c}^{+}) -g^{(1)}(w_{c}^{-}) \Bigr) \dif w  \,,  \label{psitog1b}\\
        \psi_{-1}(\infty,x)\dif x &=   \Bigl(g^{(1)}(w-w_{\infty}^{-})-g^{(1)}(w) + g^{(1)}(w_{\infty}^{-})-\omega_{1}a_{3}/4 \Bigr)\dif w\,, \\
        \psi_{0}(0,x)\dif x&= \omega_{1}\dif w \,.
    \end{align}    
\end{subequations}
On the torus $[1:\tau']$, the corresponding relations can be obtained by replacing $w\to\xi$ and $\omega_1\to -\omega_2$ in \eqref{psitog1}.

Sometimes, it is more convenient to combine $\tilde{\Gamma}$ functions into the so-called $\mathcal{E}_{4}$ functions \cite{Broedel:2018qkq}, especially if the $\tilde{\Gamma}$ functions originally arose from an expression of $\mathrm{E}_{4}$ functions. The elliptic multiple polylogarithms $\mathcal{E}_{4}$ are defined in complete analogy to \eqref{Eiterateddefinition}:
\begin{equation}
\label{curlyEiterateddefinition}
    \cEf{n_1 & \ldots & n_k}{c_1 & \ldots& c_k}{x}=
    \int_{0}^{x}\dif  x'\,\Psi_{n_{1}}(c_{1},x')\cEf{n_{2} & \ldots & n_k}{c_{2} & \ldots& c_k}{x'}
\end{equation}
with $\mathcal{E}_{4}(;x)=1$,
\begin{align}
 \Psi_{\pm (n>0)}(c,x) \dif  x=\Bigl(&g^{(n)}(w-w_{c}^{+})\pm g^{(n)}(w-w_{c}^{-})  
\nonumber \\
    &
    - \delta_{\pm n,1}\bigl( g^{(1)}(w-w_{\infty}^{+})+g^{(1)}(w-w_{\infty}^{-} )\bigr) \Bigr) \dif w  
\label{Psikernels} 
\end{align}
and $\Psi_0(x)\dif x=\dif w$, as well as analogous  expressions in terms of $\xi$.
The weight of a function $\cEf{n_1 & \ldots & n_k}{c_1 & \ldots& c_k}{x}$ is defined as $\sum_i |n_i|$.

\subsection{Symbol}

By construction,
 the total derivative of $\tilde{\Gamma}$ admits a recursive structure \cite{Broedel:2018iwv},
\begin{align} \nonumber\label{devoftG}
    &\quad \dif\tilde{\Gamma}(A_{1},\ldots,A_{k};w)  \\
    &= \sum_{p=1}^{k-1}(-1)^{n_{p+1}}\tilde{\Gamma}(A_{1},\ldots,A_{p-1},\vec{0},A_{p+2},\ldots,A_{k};w) 
    \times\omega^{(n_{p}+n_{p+1})}(w_{p+1,p})
      \\
    &\quad+ \sum_{p=1}^{k}\sum_{r=0}^{n_{p}+1} \Biggl[ 
        \binom{n_{p-1}{+}r{-}1}{n_{p-1}{-}1}\tilde{\Gamma}(A_{1},\ldots, A_{p-1}^{[r]},A_{p+1},\ldots,A_{k};w) 
         \times\omega^{(n_{p}-r)}(w_{p-1,p}) \nonumber \\
    &\qquad \qquad  - \binom{n_{p+1}{+}r{-}1}{n_{p+1}{-}1} \tilde{\Gamma}(A_{1},\ldots, A_{p-1},A_{p+1}^{[r]},\ldots,A_{k};w) 
   \times \omega^{(n_{p}-r)}(w_{p+1,p})
    \Biggr],\nonumber
\end{align}
where $\vec{0}\equiv\bigl(\begin{smallmatrix}
    0 \\
    0       
    \end{smallmatrix}\bigr)$, $w_{i,j}\equiv w_{i}-w_{j}$, $(w_0,w_{k+1})\equiv(w,0)$, $(n_0,n_{k+1})\equiv(0,0)$ as well as 
\begin{equation}
    A_{i}^{[r]}\equiv\bigl(\begin{smallmatrix}
    n_{i}+r \\
    w_{i}       
    \end{smallmatrix}\bigr)\:, \qquad A_{i}^{[0]}\equiv A_{i}\:.
\end{equation}
The forms $\omega^{(j)}(w)$ are exact, and we can thus write them as 
\begin{equation}
 \omega^{(j)}(w,\tau)=(2\pi i)^{j-1}\dif\Omega^{(j)}(w,\tau)\,,
\end{equation}
with
\begin{align} \label{wdef}
    \Omega^{(-1)}(w,\tau) &= -2\pi\mi\tau\:, \quad \Omega^{(0)}(w,\tau) =2\pi\mi w\:, \quad \Omega^{(1)}(w,\tau)=\log \frac{\theta_{1}(w|\tau)}{\eta(\tau)} \:, \nonumber \\
    \Omega^{(\text{odd } j>1)}(w,\tau)&=- \frac{2j\zeta_{j+1}\tau}{(2\pi \mi)^{j}} +
    \frac{1}{(j{-}1)!}\sum_{n=1}^{\infty}
    n^{j-1}\log\bigl((1-\me^{2\pi \mi (n\tau-w)})(1-\me^{2\pi \mi (n\tau+w)})\bigr), \nonumber \\ 
    \Omega^{(\text{even } j)}(w,\tau)&=-\frac{2\zeta_{j}w}{(2\pi i)^{j-1}} 
    + \frac{1}{(j{-}1)!}\sum_{n=1}^{\infty} n^{j-1}\log\frac{1-\me^{2\pi \mi (n\tau+w)}}{1-\me^{2\pi \mi (n\tau-w)}} ,
\end{align}
where $\eta(\tau)$ is the Dedekind eta function and $\zeta_{j}=\sum_{n\in\mathbb{Z}_{+}} n^{-j}$ are the Riemann zeta values.\footnote{Recall that $\zeta_{2n}=\frac{(-1)^{n+1}B_{2n}(2\pi)^{2n}}{2(2n)!}$ with $B_{2n}$ being the $(2n)^{\rm th}$ Bernoulli number, such that the first terms in \eqref{wdef} can equivalently be written in terms of Bernoulli numbers.}
The functions $\Omega^{(j)}$ satisfy   
\begin{equation}
    g^{(j)}(w,\tau)=(2\pi\mi)^{j-1}\partial_{w}\Omega^{(j)}(w,\tau)=\frac{(2\pi\mi)^{j-1}}{j-1}\partial_{\tau}\Omega^{(j-1)}(w,\tau).
\end{equation}
The sum representation \eqref{wdef} can be derived using the sum representation of the $g^{(n)}$ functions given in \cite{Broedel:2018iwv}.
In particular, 
\begin{equation}
 (2\pi i)^{1-n}\gamt{n}{0}{w}=\Omega^{(n)}(w)-\Omega^{(n)}(0)\,,
\end{equation}
where $\Omega^{(n)}(0)$ vanishes for even $n$ and is the primitive of the Eisenstein series for odd $n$; see \cite{Broedel:2018iwv}.
We will see below that the functions $\Omega^{(j)}$ play the role of elliptic symbol letters.
As can be seen from \eqref{wdef}, $\Omega^{(1)}$ has a logarithmic singularity at all lattice points, while 
$\Omega^{(j>1)}$ has a logarithmic singularity at all lattice points except for the origin \cite{Broedel:2018iwv}.

For a function $\widetilde{\Gamma}_{k}^{(n)}$ of weight $n$ and length $k$, we can define 
$ \widetilde{\underline{\Gamma}}_{k}^{(n)}=(2\pi i)^{k-n}\widetilde{\Gamma}_{k}^{(n)}.$
Schematically, the differential of $\widetilde{\underline{\Gamma}}_{k}^{(n)}$ then takes the form 
\begin{equation}
     \dif \widetilde{\underline{\Gamma}}_{k}^{(n)}=\sum_i   \widetilde{\underline{\Gamma}}^{(n-j_{i})}_{k-1} \dif \Omega^{(j_i)}(w_i) \,,
\end{equation}
Thus, it is natural to define the symbol of $\tilde{\underline{\Gamma}}_{k}^{(n)}$ as 
\begin{equation}
\label{eq: elliptic symbol}
     \mathcal{S}(\widetilde{\underline{\Gamma}}_{k}^{(n)})=\sum_i \mathcal{S}(\tilde{\underline{\Gamma}}_{k-1}^{(n-j_{i})})\otimes \Omega^{(j_i)}(w_i) \,.
\end{equation}

Note that in contrast to \cite{Broedel:2018iwv} we have included additional factors of $(2\pi\mi)$ in the definition of the elliptic letters $\Omega^{(n)}$ and consider the symbol of $ \widetilde{\underline{\Gamma}}_{k}^{(n)}$ rather than $\widetilde{\Gamma}_{k}^{(n)}$.\footnote{In \cite{Broedel:2018iwv}, there is also a projection operator $\pi_{k}$ in the definition of the symbol for $\tilde{\Gamma}$ functions due to the fact that some eMPLs of weight $0$ evaluate to rational numbers, such as $\gamt{0}{0}{1}=1$. Here we exclude it by introducing these $2\pi \mi$ factors.} This is such that the elliptic letters and symbols degenerate to logarithms and polylogarithmic symbols without additional factors of $(2\pi i)$ in the limit where the elliptic curve degenerates, see sections \ref{sec: example 1}--\ref{sec: example 2}.

\subsection{Shuffle regularization}

Let us close this section by remarking on shuffle regularization. One can easily see that $\gamt{1}{0}{z}=\Omega^{(1)}(z)-\Omega^{(1)}(0)$ is divergent since $\Omega^{(1)}(0)$ is singular according to \eqref{wdef}. The shuffle regularization used in  \cite{brown2011multiple,Broedel:2018iwv} takes $\Omega^{(1)}(0)\equiv2\log\eta(\tau)$.
However, that regularization leads to an issue if we start with an integral of the form \eqref{1dIntofPolylog} since it is inconsistent with the usual shuffle regularization for polylogarithms, $G(0;x)\equiv\log x$. To see this, we apply \eqref{psitog1} to $G(0;1)=\log 1=0$ and find
\begin{equation}
    0 \stackrel{?}{=} \gamt{1}{0}{w_{1}^{+}-w_{0}^{+}} + \gamt{1}{w_{0}^{-}-w_{0}^{+}}{w_{1}^{+}-w_{0}^{+}} - \gamt{1}{w_{\infty}^{+}-w_{0}^{+}}{w_{1}^{+}-w_{0}^{+}} - \gamt{1}{w_{\infty}^{-}-w_{0}^{+}}{w_{1}^{+}-w_{0}^{+}}  \:,
\end{equation}
which is in general not true if we use $\Omega^{(1)}(0)\equiv2\log\eta(\tau)$.
To reconcile both sides, we expand $\gamt{1}{w'}{w}=\Omega^{(1)}(w-w')-\Omega^{(1)}(-w')$ to arrive at the 
following regularization for elliptic multiple polylogarithms:
\begin{align}
    \Omega^{(1)}(0)&\equiv\Omega^{(1)}(w_{0}^{+}-w_{\infty}^{-})+\Omega^{(1)}(w_{0}^{+}-w_{\infty}^{+}) -\Omega^{(1)}(w_{0}^{+}-w_{0}^{-}) \nonumber \\
    &\quad +\Omega^{(1)}(w_{1}^{+}-w_{0}^{-}) +\Omega^{(1)}(w_{1}^{+}-w_{0}^{+}) -\Omega^{(1)}(w_{1}^{+}-w_{\infty}^{-}) - \Omega^{(1)}(w_{1}^{+}-w_{\infty}^{+})  \nonumber \\
    &=2\log \eta(\tau)+\log\frac{2\pi\mi}{\omega_{1}}-\log y_{0} = \frac{1}{12}\log \Delta -\log y_{0}\:, \label{shuffreg2}
\end{align} 
where $\Delta= g_{2}^{3}-27g_{3}^{2}$ is the discriminant of the elliptic curve. The second equality will be explained in the next section and the third equality shows that this regularization is actually independent of the way we rescale the torus.

\section{Identities of Elliptic Symbol Letters and the Symbol Prime} 
\label{sec:3}

We have briefly reviewed several elementary facts about elliptic multiple polylogarithms, and we saw that the symbol letters of elliptic multiple polylogarithms are the functions $\Omega^{(n)}(w,\tau)$. These functions stand in the way of analyzing the elliptic symbols. For one thing, the relations among $\Omega^{(n)}$'s are much more complicated than the manipulation rules $\log a+\log b=\log ab$ for the symbol letters of multiple polylogarithms. For another, they depend on the kinematics in a rather indirect way -- their arguments $w$ and $\tau$ are (ratios) of elliptic integrals involving kinematics.

In this section, we investigate the identities of the elliptic letter $\Omega^{(n)}(w,\tau)$. 
The most trivial identities these letters satisfy are the following: 
\begin{align}
    \text{Parity :}&\quad \Omega^{(n)}(-w)=(-1)^{n+1} \Omega^{(n)}(w) \:, 
    \label{parity}\\
    \text{Quasi periodicity :} &\quad \Omega^{(n)}(w+\tau) = \sum_{j=0}^{n+1}\frac{(-1)^{j}}{j!}\Omega^{(n-j)}(w) \:. \label{quasiperiodicity}
\end{align}
They immediately follow from \eqref{wdef}.
Our investigation of more non-trivial identities will be focussed on the cases $n=0,1,2$ since -- for the two examples considered in this paper, namely the sunrise integral and the double-box integral -- the identities among $\Omega^{(n\leq 2)}$ are sufficient to simplify the symbols after using \eqref{parity} and \eqref{quasiperiodicity}. We comment on a generalization to identities among $\Omega^{(n>2)}$ at the end of this section.

Let us start with the slightly trivial identity
\begin{align} \label{identity1}
    \quad \log\frac{c-b}{c-a} &= \sum_{\sigma\in \pm} \Bigl(\Omega^{(1)}(w_{c}^{\sigma}-w_{b}^{+})-\Omega^{(1)}(w_{c}^{\sigma}-w_{a}^{+}) 
     \nonumber 
     \\
     &\qquad \qquad \qquad 
    -\Omega^{(1)}(w_{\infty}^{\sigma}-w_{b}^{+})+\Omega^{(1)}(w_{\infty}^{\sigma}-w_{a}^{+})\Bigr)\,,
\end{align}
which is a simple consequence of applying \eqref{psitog1a} to $\int_{a}^{b}\psi_{1}(c,x)\dif x = \log\frac{c-b}{c-a} $. 
The identity \eqref{identity1} has two important special cases. One is obtained by taking $a\to \infty$, giving
\begin{align} \label{identity2}
    \sum_{\sigma\in\pm} \Omega^{(1)}(w_{c}^{\sigma}-w_{b}^{+}) 
    &=\Omega^{(1)}(w_{\infty}^{-}-w_{\infty}^{+})+\sum_{\sigma\in\pm}\left[\Omega^{(1)}(w_{c}^{\sigma}-w_{\infty}^{+})+\Omega^{(1)}(w_{b}^{\sigma}-w_{\infty}^{+}) \right] \nonumber \\ 
    &\quad - \log\biggl(\frac{2\pi\mi}{\omega_{1}}\biggr)  -2\log\eta(\tau)+\log(c-b)\,;
\end{align}
the other one is obtained by further taking $b\to c$, which yields
\begin{align} \label{identity3}
    \Omega^{(1)}(w_{c}^{-}{-}w_{c}^{+}) &= 2\biggl(\Omega^{(1)}(w_{c}^{-}-w_{\infty}^{+})+\Omega^{(1)}(w_{c}^{+}-w_{\infty}^{+})
    -2\log\eta(\tau) -\log\frac{2\pi \mi}{\omega_{1}}  \biggr) \nonumber \\ 
    &\quad  - \Omega^{(1)}(w_{\infty}^{-}-w_{\infty}^{+}) - \log y_{c} \:.
\end{align}
Now one can easily see that the second equality in \eqref{shuffreg2} is the consequence of applying \eqref{identity2} and \eqref{identity3}. The two special cases are particularly useful since the letters $\Omega^{(1)}$ on their right-hand sides always involve $w^{\pm}_{\infty}$ and hence can serve as a basis.

\subsection{Abel's addition theorem} \label{sec:3.1}

Surprisingly, a very classical and powerful theorem, Abel's addition theorem \cite{abel1841}, 
yields other identities for $\Omega^{(0)}$ and $\Omega^{(1)}$.\footnote{See e.g.\ \cite{GriffithsHarris} for a textbook treatment of Abel's addition theorem and \cite{Tarasov:2017yyd,Tarasov:slides} for previous applications to Feynman integrals.}

Let us first spell out this theorem: Let $\mathcal{C}$ and $\mathcal{C}'$ be curves given by two polynomial equations
\begin{align}
    \mathcal{C}:& \quad  F(x,y)=0 \:, \\
    \mathcal{C}':&\quad  Q(x,y)=0\:,
\end{align}
where $\mathcal{C}$ is viewed as a \emph{fixed} curve and $\mathcal{C}'$ as a \emph{variable} curve with coefficients collectively denoted as $\{b_{i}\}$. 
Suppose that these two curves intersect at $n$ points $(x_{1},y_{1})$, ..., $(x_{n},y_{n})$. Let $R(x,y)$ be a rational function defined on $\mathcal{C}$. Then the following holds.
\begin{theorem}[Abel]
    The integral 
    \begin{equation}
        I(\{b_{i}\}) =\sum_{i=1}^{n} \int_{x_{\ast}}^{x_{i}}R(x,y)\:\dif x\:,
    \end{equation}
    where $x_{\ast}$ is an arbitrary base point, contains at most rational functions and logarithms of $\{b_{i}\}$. 
\end{theorem}
\noindent This theorem can be proven by showing that $\partial_{b_{\nu}}I$ is always a rational function of $\{b_{i}\}$ for all $b_{\nu}$.

If a symbol letter $\phi(u)$ can be expressed as $\int^{u} R(x,y)\dif x$, one can try to find an addition formula for $\phi(u)$ through Abel's addition theorem.
Of all applications of this theorem, we are most interested in the cases that $\mathcal{C}'$ has only \emph{two} degrees of freedom and intersects $\mathcal{C}$ at \emph{three} points. 
For this case, Abel's addition theorem gives 
\begin{equation}
    \phi(u)+\phi(v)=\phi\bigl(T(u,v)\bigr)+\cdots,
\end{equation}
where $T(u,v)$ is an algebraic function of $u$ and $v$ and `$\cdots$' denotes simpler objects, like logarithms. 

An example is the addition formula for logarithms, $\log(x_1)+\log(x_2)=\log(x_1x_2)$,\footnote{This example can e.g.\ be found in \cite{Tarasov:slides}.} which is given by choosing 
\begin{align}
    \mathcal{C}:&\quad y= \frac{1}{x} \:, \\
    \mathcal{C}': &\quad y=x^{2}+b_{1}x+b_{2} \:.
\end{align}
These two curves intersect at the three points  $x_1,x_2,x_3$ that solve 
\begin{equation}
 x^3+b_1x^2+b_2x-1=0,
\end{equation}
and hence satisfy $x_1 x_2 x_3=1$.
Now consider
 \begin{equation}
 I=\sum_{i=1}^3\int_1^{x_i}\frac{\dif x}{x},
  \end{equation}
  which satisfies
\begin{equation}
 \partial_{b_j}I=\sum_{i=1}^3\frac{1}{x_i}\frac{\partial x_i}{\partial b_j}=\frac{1}{x_1 x_2 x_3}\frac{\partial}{\partial b_j}x_1x_2x_3=0 ,
\end{equation}
since $x_1 x_2 x_3=1$. Thus, $I$ is a  constant. To fix this constant, we can pick $b_1=-3,b_2=3$, such that $x_1=x_2=x_3=1$, yielding $I=0$.
Again using $x_1 x_2 x_3=1$, we thus have 
\begin{equation}
 0=I=\log(x_1)+\log(x_2)+\log(x_3)=\log(x_1)+\log(x_2)-\log(x_1 x_2),
\end{equation}
as claimed.

For the case we are most interested in, the fixed curve $\mathcal{C}$ is given by  \eqref{quarticEcurve}, and we find that a convenient choice for $\mathcal{C}'$ is 
\begin{equation}
    y=-x^{2}+b_{1}x+b_{2} \:.
\end{equation} 
One can easily check that these two curves intersects at three point at most.
Suppose that two intersection points are $(x_{1},y_{1}=\sqrt{P_{4}(x_{1})})$ and $(x_{2},y_{2}=\sqrt{P_{4}(x_{2})})$, then
\begin{align}
    b_{1}&=\frac{y_{1}-y_{2}}{x_{1}-x_{2}}+x_{1}+x_{2} \:, &&
    b_{2}= \frac{x_{1}y_{2}-x_{2}y_{1}}{x_{1}-x_{2}}-x_{1}x_{2} \:, && \text{for }x_{1}\neq x_{2} \:, \\
    b_{1} &= \frac{P_{4}'(x_{1})}{2 y_{1}}+2x_{1} \:, && b_{2}=y_{1}+x_{1}^{2}-b_{1}x_{1} \:, && \text{for }x_{1}= x_{2} \:,\intertext {and}    
    x_{3} &= \frac{b_{1}^{2}-2b_{2}-a_{2}}{2b_{1}+a_{3}} - x_{1}-x_{2} \:, && 
    y_{3} = -\sqrt{P_{4}(x_{3})} \:.&&
\end{align}
Since $z_{c}^{+}=\int_{-\infty}^{c} \dif x/y$, Abel's addition theorem tells us 
\begin{subequations} \label{zIdentity}
    \begin{align}
        z_{x_{1}}^{+}+ z^{+}_{x_{2}} &\equiv z^{+}_{x_{3}} \operatorname{mod} \Lambda \:, && \text{for }b_{1}\neq -a_{3}/2 \:,  \\ 
        z_{x_{1}}^{+}+ z^{+}_{x_{2}} &\equiv 0 \operatorname{mod} \Lambda \:, && \text{for }b_{1}= -a_{3}/2  \:,
        \end{align} 
\end{subequations}
which is the well-known group law on the elliptic curve.
Furthermore, if we take $b_{2}=(a_{3}^{2}-4 a_{2})/8$ aside $b_{1}=-a_{3}/2$, then $\mathcal{C}$ and $\mathcal{C}'$ only intersect at one point,
\begin{equation}
    \chi=\frac{a_{3}^{4}-8a_{2}a_{3}^{2}+16a_{2}^{2}-64a_{0}}{8\bigl(a_{3}^{3}-4a_{2}a_{3}+8a_{1}\bigr)} \:.
\end{equation}
Together with a little divisor theory, 
this gives\footnote{For any meromorphic function $F$ on a torus, by using $\oint  \dif \log F(z)=0$ and $\oint z \:\dif \log F(z)=0$, one can conclude that the number and the sum of its poles are the same as of its zeros, where poles and zeros of order $n$ are counted $n$ times. Now consider the function 
\[ 
    F=-\kappa'(z)-\kappa(z)^{2}-a_{3}\kappa(z)/2+(a_{3}^{2}-4a_{2})/8\:,
\]
which has poles at lattice points but vanishes only at $z_{\chi}^{-}$ and $z_{\infty}^{-}$, two intersection points of the curve $ y=-x^{2}-a_{3}x/2+(a_{3}^{2}-4a_{2})/8$ and the elliptic curve. We then obtain \eqref{minfid} by using \eqref{wpwmrelation}.}
\begin{equation}
    2z_{\infty}^{-} \equiv \omega_{1}+z_{\chi}^{+} \operatorname{mod} \omega_{2} \:. \label{minfid}
\end{equation}

Similarly, for the integral $\int \psi_{-1}(c,x)\dif x$, the same procedure gives 
\begin{align} \label{eq1}
    \int_{\ast}^{x_{1}}\frac{y_{c}\,\dif x}{y(x-c)}+\int_{\ast}^{x_{2}}\frac{y_{c}\,\dif x}{y(x-c)}-\int_{\ast}^{x_{3}}\frac{y_{c}\,\dif x}{y(x-c)} &= \log\frac{c^{2}-b_{1} c-b_{2}+y_{c}}{c^{2}-b_{1}c-b_{2}-y_{c}} + \text{const.} 
\end{align}
If $z_{x_{1}}^\pm+z_{x_{2}}^\pm=z_{x_{3}}^\pm$, applying \eqref{psitog1b} to \eqref{eq1} gives%
\begin{align}
    &\quad \sum_{i=1}^{2}\Omega^{(1)}(w_{c}^{+}-w_{x_{i}}^{+})-\Omega^{(1)}(w_{c}^{+}-w_{x_{i}}^{-}) \label{Abel1}\\
    &=\Omega^{(1)}(w_{c}^{+}{-}w_{x_{3}}^{+})-\Omega^{(1)}(w_{c}^{+}{-}w_{x_{3}}^{-}) 
   +\Omega^{(1)}(w_{c}^{+})-\Omega^{(1)}(w_{c}^{-})+\log\frac{c^{2}-b_{1}c-b_{2}+y_{c}}{c^{2}-b_{1}c-b_{2}-y_{c}}. \nonumber 
\end{align}
If $z_{x_{1}}^\pm+z_{x_{2}}^\pm\equiv z_{x_{3}}^\pm \mod \Lambda$, a corresponding identity can be found from \eqref{Abel1} using the quasi double periodicity of $\Omega^{(1)}$ \eqref{quasiperiodicity}.

Three boundary cases of \eqref{Abel1} require special care: \\
(i) taking $c\to\infty$ gives
\begin{align}
    \sum_{i=1}^{2}\Omega^{(1)}(w_{x_{i}}^{+})-\Omega^{(1)}(w_{x_{i}}^{-})
    &=\Omega^{(1)}(w_{x_{3}}^{+})-\Omega^{(1)}(w_{x_{3}}^{-})  \nonumber \\
    &\quad-\Omega^{(1)}(\omega_{\infty}^{-}-\omega_{\infty}^{+})-\log\frac{2b_{1}+a_{3}}{4}+\frac{1}{12}\log\Delta \:, \label{Abel2} 
\end{align}
(ii) taking $x_{3}\to \infty$ gives 
\begin{align}
    & \quad \sum_{i=1}^{2}\Omega^{(1)}(w_{c}^{+}-w_{x_{i}}^{+})-\Omega^{(1)}(w_{c}^{+}-w_{x_{i}}^{-}) \label{Abel3} \\
    &=\Omega^{(1)}(w_{c}^{+}{-}w_{\infty}^{+})-\Omega^{(1)}(w_{c}^{+}{-}w_{\infty}^{-})  
    +\Omega^{(1)}(w_{c}^{+})-\Omega^{(1)}(w_{c}^{-})+\log\frac{c^{2}+a_{3}c/2-b_{2}+y_{c}}{c^{2}+a_{3}c/2-b_{2}-y_{c}} \:.  \nonumber 
\end{align} 
(iii) taking $c\to\infty$ and $x_{3}\to \infty$ gives 
\begin{align}
    \sum_{i=1}^{2}\left[\Omega^{(1)}(w_{\infty}^{+}-w_{x_{i}}^{+})-\Omega^{(1)}(w_{\infty}^{+}-w_{x_{i}}^{-})
   \right] &=-2\Omega^{(1)}(w_{\infty}^{+}-w_{\infty}^{-})+\Omega^{(0)}(w_{\infty}^{-}-w_{\infty}^{+})  \nonumber \\
    &\quad+\frac{1}{6}\log\Delta-\log\frac{4a_{2}-a_{3}^{2}+8b_{2}}{16} \:. \label{Abel4}
\end{align} 

Eqs.\ \eqref{zIdentity}, \eqref{identity1}--\eqref{identity3} as well as \eqref{Abel1}--\eqref{Abel4} explain the subset of the identities numerically found in \cite{Kristensson:2021ani} which only involve $\Omega^{(0)}$'s and $\Omega^{(1)}$'s.

Note that the identities we presented in this subsection can be equivalently formulated in terms of divisor theory, see e.g.\ \cite{Broedel:2019tlz}.

\subsection{Elliptic Bloch relation and the symbol prime}
\label{subsec: symbol prime}

In \cite{Kristensson:2021ani}, also five identities involving $\Omega^{(2)}$'s were observed which are much lengthier than the other identities; each of these five identities contains at least 100 terms in the form that they were found. It turns out all these identities are consequences of the so-called elliptic Bloch relation \cite{bloch2011higher,Zagier2000},  
an elliptic generalization of the five-term identity for dilogarithms,
\begin{equation}
\label{eq: five term identity}
    D(x)+D(y)+D\biggl(\frac{1-x}{1-xy}\biggr)+D(1-xy)+D\biggl(\frac{1-y}{1-xy}\biggr)=0 \:,
\end{equation}
where $D(z)=\Im(\Li_{2}(z))+\arg(1-z)\log |z|$ is the Bloch-Wigner function.%
\footnote{To show this concretely, one  would need to do the divisor-theory analog of finding a curve that intersects the elliptic curve at precisely the points given by the more than 100 terms in the identities.
An algorithm for doing this is given in \cite{Bolbachan:2019dsu}.}

In practice, it is difficult to simplify even expressions containing dilogarithms by using the above five-term identity directly. Instead, we introduce the symbol map \cite{Goncharov:2010jf} for polylogarithms as an assistance; we associate to each polylogarithm a tensor product whose entries satisfy simpler identities. We then exploit that the symbol of a combination of polylogarithms vanishes if that combination of polylogarithms vanishes.

A similar strategy can be used for the elliptic letters $\Omega^{(2)}(w)=(2\pi \mi)^{-1}\gamt{2}{0}{w}$, although they already serve as entries of the symbol for elliptic multiple polylogarithms. Inspired by the proof of the elliptic Bloch relation for $\gamt{2}{0}{w}$ in \cite{Broedel:2019tlz}, we associate to $\Omega^{(2)}(w)$ a rank-two tensor through the \emph{symbol prime} map,
\begin{equation} \label{symbolp1}
    \mathcal{S}'\bigl(\Omega^{(2)}(w)\bigr) = \Omega^{(0)}(w)\otimesprime \Omega^{(1)}(w) \:,
\end{equation}  
where we have added a prime on ``$\otimes$'' to distinguish it from the tensor product in the symbol.
This map has a property similar to that of the symbol map:
\begin{equation}
\label{eq: symbol prime property}
    \sum_{j}c_j \Omega^{(2)}(w_{j})=0  \quad 
    \text{``$\Rightarrow$''}
    \quad
    \sum_{j}c_j\mathcal{S}'(\Omega^{(2)}(w_{j}))\equiv
     \sum_{j}c_j\Omega^{(0)}(w_{j})\otimesprime \Omega^{(1)}(w_{j})=0
\end{equation}
for some rational coefficients $c_j$.
To show this, consider the sum $\sum_{j}c_j\gamt{1&0}{0&0}{w_{j}}$.
According to \eqref{devoftG}, 
\begin{align}
    \label{dev_tG2}
\mathcal{S}(2\pi i \gamt{1&0}{0&0}{w})= \Omega^{(0)}(w) \otimes \Omega^{(1)}(w)- \Omega^{(2)}(w) \otimes (2\pi i\tau)\:,
\end{align}
where we used that $\gamt{0}{0}{w}=w$, $\gamt{2}{0}{w}=2\pi \mi\Omega^{(2)}(w)$ and $\Omega^{(-1)}=-2\pi i \tau$.
If the arguments $w_j$ and coefficients $c_j$ are such that $\sum_{j}c_j\Omega^{(2)}(w_{j})=0$ due to an elliptic Bloch relation, $\sum_{j}c_j\gamt{1&0}{0&0}{w_{j}}=0$ according to an analogous elliptic Bloch relation \cite{Broedel:2019tlz}, which in turn implies that the second term on the right-hand side of \eqref{dev_tG2}, $\sum_{j}c_j\mathcal{S}'(\Omega^{(2)}(w_{j}))$, drops out in the sum.
In this sense, the symbol prime makes the elliptic Bloch relations manifest.

Note that we have assumed that $\sum_{j}c_j\Omega^{(2)}(w_{j})=0$ vanishes \emph{due to an elliptic Bloch relation} here in order to show that $\sum_{j}c_j\mathcal{S}'(\Omega^{(2)}(w_{j}))=0$. We have indicated this in \eqref{eq: symbol prime property} as ``$\Rightarrow$''.
However, we currently have no way of proving that all identities $\sum_{j}c_j\Omega^{(2)}(w_{j})=0$ are due to an elliptic Bloch relation \cite{bloch2011higher,Zagier2000}.
This is similar to the case of dilogarithms, where \eqref{eq: five term identity} is only conjectured but not proven to generate all functional identities among dilogarithms.%

The symbol map itself has a kernel, and the same is true for the symbol prime.
If $\sum_j \Omega^{(2)}(w_{j})$ is in the kernel of the symbol prime, i.e.\ $\mathcal{S}'(\sum_j \Omega^{(2)}(w_{j}))=0$, the first term on the right-hand side of \eqref{dev_tG2} drops out in the sum. This implies that $\sum_{j}\gamt{1&0}{0&0}{w_{j}}$ and thus $\sum_j \Omega^{(2)}(w_{j})$ is a function of $\tau$ \emph{only}. However, not all functions only of $\tau$ are in the kernel of the symbol prime; for example, $\Omega^{(2)}(\tau/n)$ with some positive integer $n$ only depends on $\tau$ but has a non-vanishing symbol prime.%
\footnote{However, the occurrence of such a letter means that the point on the elliptic curve that corresponds to $\tau/n$ on the torus, namely $\kappa(\omega_1\tau/n)$, should occur in the calculation of the $\mathrm{E}_4$ functions and hence is algebraic in kinematics. This is not the case for the examples of the unequal-mass sunrise integral and the ten-point double-box integral studied in sections \ref{sec: example 1}--\ref{sec: example 2}, but it is the case for the \emph{equal}-mass sunrise integral.}

One can find the action of the symbol prime map on the letters $\Omega^{(n<2)}$ by expressing them in terms of $\Omega^{(2)}$ using the quasi periodicity \eqref{quasiperiodicity} of $\Omega^{(n)}$:
\begin{align}
    \Omega^{(1)}(w) &= \tfrac{1}{6}\Omega^{(2)}(w+2\tau)-\Omega^{(2)}(w+\tau)+\tfrac{1}{2}\Omega^{(2)}(w)
    +\tfrac{1}{3}\Omega^{(2)}(w-\tau) \:, \\
    \Omega^{(0)}(w)&=\Omega^{(2)}(w+\tau)+\Omega^{(2)}(w-\tau)-2\Omega^{(2)}(w) \:, \\
    \Omega^{(-1)}&=-\Omega^{(2)}(w+2\tau)+3\Omega^{(2)}(w+\tau)-3\Omega^{(2)}(w)+\Omega^{(2)}(w-\tau)
    \:.
\end{align}
This yields 
\begin{subequations} \label{symbolp2}
    \begin{align} 
        \mathcal{S}'\bigl(\Omega^{(1)}(w)\bigr)&= \Omega^{(0)}(w)\otimesprime \Omega^{(0)}(w)
        +\Omega^{(-1)}\otimesprime \Omega^{(1)}(w) \:, \\
        \mathcal{S}'\bigl(\Omega^{(0)}(w)\bigr) &= \Omega^{(0)}(w)\otimesprime \Omega^{(-1)}
         + 2 \Omega^{(-1)} \otimesprime \Omega^{(0)}(w) \:, \\
         \mathcal{S}'\bigl(\Omega^{(-1)}\bigr)  &= 3\Omega^{(-1)}\otimesprime \Omega^{(-1)} \:,
    \end{align}    
\end{subequations}
where we have moreover used quasi periodicity to simplify the entries of the symbol prime.
In particular, by expressing a logarithm in terms of $\Omega^{(1)}$'s and $\Omega^{(0)}$'s either through \eqref{identity1} or \eqref{Abel1}, one finds
\begin{equation} \label{symbolp3}
    \mathcal{S}'\bigl(\log c\bigr) = \Omega^{(-1)} \otimesprime \log c\:.
\end{equation}
Thus, for a combination of $\Omega^{(n\leq 2)}$'s and logarithms, one can compute its symbol prime. 
It involves only $\Omega^{(n\leq1)}$ and can thus be simplified using the techniques discussed in subsection \ref{sec:3.1}. If the symbol prime is not zero, one may search for a simpler combination of $\Omega^{(n\leq 2)}$'s and logarithms with the same symbol prime according to \eqref{symbolp1}, \eqref{symbolp2} and \eqref{symbolp3}.%
\footnote{In particular, if the first entry of the symbol prime is only $\tau$, then the function is the sum of logarithms and a function of $\tau$.} The difference of these two combinations has to be a function of $\tau$ only, and a simple expression for this function can be obtained by sending the independent $w$-variables in the difference to any values, say $0$. In this way, we have proven the five identities involving $\Omega^{(2)}$'s found in \cite{Kristensson:2021ani}.

The current definitions for $\Omega^{(n\leq 2)}$ are sufficient for the two examples treated in this paper. 
For $n>2$, one might similarly define the symbol prime for $\Omega^{(n)}$'s as 
\begin{equation}
    \mathcal{S}^{(n-1)}\bigl(\Omega^{(n)}(w)\bigr) = \frac{1}{n-1}\Omega^{(0)}(w)\otimes^{(n-1)} \Omega^{(n-1)}(w),
\end{equation}
due to the fact that,
\begin{equation}
\mathcal{S}\bigl((2\pi i)^{2-n} \gamt{n-1&0}{0&0}{w}\bigr)=\Omega^{(0)}(w) \otimes \Omega^{(n-1)}(w)- (n-1)\bigl( \Omega^{(n)}(w)-\Omega^{(n)}(0)\bigr)\otimes (2\pi i\tau)\:,
\end{equation}
where $\Omega^{(n)}(0)$ is either zero or a function only depending on $\tau$ for even or odd $n$, respectively.
With the knowledge of the identities among $\Omega^{(n-1)}$, we can then find identities among $\Omega^{(n)}$ recursively. 
We leave the exploration of the symbol prime for $\Omega^{(n>2)}$ to future work.

\section{Example I: Unequal-Mass Sunrise Integral}
\label{sec: example 1}

Two particularly interesting cases of elliptic Feynman integrals are the unequal-mass sunrise integral in two dimensions and the double-box integral in four dimensions. We will investigate these two integrals through the tools developed so far. The main focus will be on the sunrise integral treated in this section, since this integral is simple enough such that the main results can be written within a couple of lines. After applying the symbol prime map, we will see that several properties, such as double periodic invariance, modular invariance (covariance) and part of  integrability are manifest.

We calculate the unequal-mass sunrise integral in terms of elliptic multiple polylogarithms $\mathcal{E}_{4}$ in appendix \ref{app:sunrise}.
This integral was originally calculated in terms of iterated integrals on the moduli space $\overline{\mathcal{M}}_{1,3}$  in \cite{Bogner:2019lfa}. 
We closely follow the Feynman-parameter approach of \cite{Broedel:2017siw} for the equal-mass case.

The resulting expression when rescaling the torus by the period $\omega_1$ is 
\begin{equation}
\label{eq: sunrise general}
    I_{\sr} = \frac{\omega_{1}}{2\pi\mi m_{1}^{2}} (2\pi \mi T^{(1)}_{\sr})
    \,,
\end{equation}
where the periods were defined in figure \ref{fig: contours} and $T_{\sr}^{(1)}$ is a pure combination of elliptic multiple polylogarithms of weight one and length two,
\begin{align}
    T_{\sr}^{(1)} &= \cEf{0&-1}{0&-1}{\infty|\tau}-\cEf{0&-1}{0&0}{\infty|\tau} + \cEf{0&-1}{0&r}{\infty|\tau} -\cEf{0&-1}{0&\infty}{\infty|\tau} \nonumber  \\
    &\quad +4\pi \mi\cEf{0&0}{0&0}{\infty|\tau}  -\cEf{0}{0}{\infty|\tau} \log\frac{t^{2}_{2}}{t_{3}^{2}} \:, \label{eq:sunrise normalization 1}
    \end{align}
where we introduced $t_{i}^{2}=m_{i}^{2}/p^{2}$ and $r=-t_{3}^{2}/t^{2}_{1}$. 
Note that we have included seemingly redundant factors of $(2\pi i)$ in the numerator and denominator of \eqref{eq: sunrise general} that ensure that the prefactor degenerates to an algebraic function in the limit where the elliptic curve degenerates, and the term in parentheses degenerates to a pure logarithm of transcendental weight two; see subsection \ref{sec:4.1}. 

However, we can also rescale the torus by the period $-\omega_2$, finding
\begin{equation}
\label{eq: sunrise general 2}
    I_{\sr}     = \frac{-\omega_{2}}{2\pi\mi m_{1}^{2}} (2\pi \mi T^{(2)}_{\sr})\,,
\end{equation}
with
\begin{align}
    T_{\sr}^{(2)} &= \cEf{0&-1}{0&-1}{\infty|\tau'}-\cEf{0&-1}{0&0}{\infty|\tau'} + \cEf{0&-1}{0&r}{\infty|\tau'} -\cEf{0&-1}{0&\infty}{\infty|\tau'} \nonumber \\ 
    &\quad -\cEf{0}{0}{\infty|\tau'} \log\frac{t^{2}_{2}}{t_{3}^{2}} \:,\label{eq:sunrise normalization 2}
\end{align}
(Recall from subsection \ref{subsec: 2.3} that $\tau'=-\omega_1/\omega_2$.)

According to \eqref{eq: sunrise general} and \eqref{eq: sunrise general 2}, the values of $T_{\sr}^{(1)}$ and $T_{\sr}^{(2)}$ are related by $T_{\sr}^{(1)}= -\tau T_{\sr}^{(2)}$, but this relation is not obvious from their expressions in terms of eMPLs. 
In general, eMPLs transform non-trivially under the modular $S$-transformation $\tau\to \tau'=-1/\tau$; for example,
\begin{equation}
\cEf{-1}{c}{x|\tau}=    \cEf{-1}{c}{x|\tau'}-\frac{2\pi \mi(\xi_{c}^{+}-\xi^{-}_{c})}{\tau'}  \cEf{0}{0}{x|\tau'} \:.
\end{equation}
See \cite{Duhr:2019rrs} for the cases of iterated integrals of modular forms. The same is true for the symbol, as we will see soon. 
However, we will see that the application of the symbol prime map makes the behavior under the modular $S$-transformation manifest.

\subsection{Symbol of the sunrise integral} \label{sec:4.1}

The symbol of $ T_{\sr}^{(1,2)}$ can be calculated by first rewriting $\mathcal{E}_4$'s in terms of $\tilde\Gamma$'s via \eqref{Psikernels} and then applying \eqref{devoftG}--\eqref{eq: elliptic symbol}.
For example,
\begin{equation}
 \cEf{0&-1}{0&c}{x}= \gamt{0&1}{0&w_{c}^{+}-w_{0}^{+}}{w_{x}^{+}-w_{0}^{+}}-  \gamt{0&1}{0&w_{c}^{-}-w_{0}^{+}}{w_{x}^{+}-w_{0}^{+}}
\end{equation}
and 
\begin{align}
 \mathcal{S}\bigl(2\pi i\gamt{0&1}{0&w_{1}}{w_{2}}\bigr)&= \Bigl(\Omega^{(2)}(-w_{1})-\Omega^{(2)}(w_{2}-w_{1})\Bigr)\otimes \Omega^{(-1)}
 -\Omega^{(0)}(w_{2})\otimes \Omega^{(1)}(w_{1}) \nonumber\\ 
 &\quad + \Bigl(\Omega^{(1)}(w_{2}-w_{1})-\Omega^{(1)}(-w_{1})\Bigr)\otimes \Omega^{(0)}(w_{2}-w_{1}) \,.
\end{align}

The simplification of the symbols in this case 
is slightly non-trivial: it involves some non-trivial relations of $\Omega^{(1)}$'s, $\Omega^{(0)}$'s and logarithms as described in 
section \ref{sec:3.1}; for instance, 
\begin{align}
    \log \frac{t_{1}}{t_{3}} &= \Omega^{(1)}(w_{-1}^{+}-w_{0}^{+})-\Omega^{(1)}(w_{-1}^{+}-w_{\infty}^{+}) 
    +\Omega^{(1)}(w_{-1}^{+}-w_{0}^{-})-\Omega^{(1)}(w_{-1}^{+}-w_{\infty}^{-}) \:,  \\
    \log \frac{t_{2}}{t_{3}} &= \Omega^{(1)}(w_{0}^{+}-w_{\infty}^{+}) - \Omega^{(1)}(w_{-1}^{+}-w_{\infty}^{+}) 
    +\Omega^{(1)}(w_{-1}^{+}-w_{0}^{-})-\Omega^{(1)}(w_{\infty}^{-}-w_{\infty}^{+}) \:.
\end{align}
(Recall from subsection \ref{subsec: 2.3} that $w_{c}^{+}=\omega_{1}^{-1}\int_{-\infty}^{c} \dif x/y$). 
All the relations involving $\Omega^{(2)}$ in this case are comparably trivial; they are the consequences of \eqref{quasiperiodicity}. 
We present the full simplification in the attached file \texttt{sunrise\_symbol.nb}.

The final result is
\begin{align}
    \mathcal{S}\bigl(2\pi\mi T_{\sr}^{(1)} \bigr) &= \log \frac{t_{2}^{2}}{t_{1}^{2}} \otimes \Omega^{(0)}(w_{0}^{+})
    + \log \frac{t_{1}^{2}}{t_{3}^{2}}\otimes \Omega^{(0)}(w_{-1}^{+}) \nonumber \\
    &\quad+\biggl[\frac{1}{2\pi\mi} \bigl(2\cEf{-2}{-1}{\infty}-\cEf{-2}{0}{\infty} - \cEf{-2}{\infty}{\infty}\bigr)+\log\frac{t_{3}^{2}}{t_{2}^{2}} \biggr]\otimes (2\pi\mi\tau)\:. \label{symbolw1} 
\end{align}
where we have moreover used 
\begin{align}
\label{eq:Ecal-gammat-relation}
 \frac{\cEf{-n}{c}{\infty}}{(2\pi\mi)^{n-1}}=\Omega^{(n)}(w_{\infty}^{+}-w_{c}^{+})-\Omega^{(n)}(w_{0}^{+}-w_{c}^{+})
 -\Omega^{(n)}(w_{\infty}^{+}-w_{c}^{-})+\Omega^{(n)}(w_{0}^{+}-w_{c}^{-}) \:.
\end{align}
Similarly, 
 \begin{align}
    \mathcal{S}\bigl(2\pi\mi T_{\sr}^{(2)} \bigr) &= \log \frac{t_{2}^{2}}{t_{1}^{2}} \otimes \Omega^{(0)}(\xi_{0}^{+})
    + \log \frac{t_{1}^{2}}{t_{3}^{2}}\otimes \Omega^{(0)}(\xi_{-1}^{+}) \nonumber \\
    &\quad+ \biggl[\frac{1}{2\pi \mi}\bigl(2\cEf{-2}{-1}{\infty}-\cEf{-2}{0}{\infty} - \cEf{-2}{\infty}{\infty}\bigr)\nonumber \\
    &\qquad \qquad +\log\frac{t_{1}}{t_{3}} +
    \Omega^{(1)}(\xi_{\infty}^{+}-\xi_{-1}^{+})-\Omega^{(1)}(\xi_{-1}^{+}-\xi_{0}^{+})\biggr] \otimes (2\pi\mi\tau^{\prime}) \:, \label{symbolw2}
\end{align}
where we used \eqref{eq:Ecal-gammat-relation} in terms of $\xi$-coordinates. 

At this point, the symbols of the sunrise integral partially show some desired properties; for example, the first entries of the first two terms in \eqref{symbolw1} and \eqref{symbolw2} indicate the physical first-entry conditions known from the massless case, and their last entries are related by simple $S$-transformations $w\to\xi$.
However, the first two terms on their own are neither double periodic nor integrable.

The first entries of the last terms in \eqref{symbolw1} and \eqref{symbolw2}, i.e., $\partial_\tau T_{\sr}^{(1)}$ and $\partial_{\tau'} T_{\sr}^{(2)}$,
are rather complicated and the main obstacles to understanding the entire symbols, since it is hard to see how they render the whole symbol double periodic and integrable. In this respect, it is instructive to consider the symbol primes of these entries:
\begin{align}
    \mathcal{S'}\bigl(\partial_\tau T_{\sr}^{(1)} \bigr) &=   \Omega^{(0)}(w_{0}^{+})\otimesprime \log\frac{t_{2}^2}{t_{1}^{2}}+\Omega^{(0)}(w_{-1}^{+})\otimesprime \log\frac{t_{1}^2}{t_{3}^{2}} \:, \label{spofsr1}\\
    \mathcal{S'}\bigl(\partial_{\tau'} T_{\sr}^{(2)} \bigr) &=\Omega^{(0)}(\xi_{0}^{+})\otimesprime \log\frac{t_{2}^2}{t_{1}^{2}}+\Omega^{(0)}(\xi_{-1}^{+})\otimesprime \log\frac{t_{1}^2}{t_{3}^{2}} \:.\label{spofsr2}
\end{align}
They have the following advantageous properties:
\begin{enumerate}
    \item It is obvious that $\mathcal{S'}\bigl(\partial_\tau T_{\sr}^{(1)} \bigr)$ differs from $\mathcal{S'}\bigl(\partial_{\tau'} T_{\sr}^{(2)} \bigr) $ only by a modular $S$-transformation $w\to\xi$, i.e.\ the symbol prime makes modular covariance manifest.
    \item If we shift $w_{-1}^{+}$ or $w_{0}^{+}$ by $\tau$, then $\partial_\tau T_{\sr}^{(1)}$ changes by $\log\frac{t_{3}^2}{t_{1}^{2}} $ or  $\log\frac{t_{1}^2}{t_{2}^{2}}$, respectively, (and similarly for $\partial_{\tau'} T_{\sr}^{(2)}$) since
        \begin{align}
            \mathcal{S'}\bigl(\partial_\tau T_{\sr}^{(1)}|_{w_{-1}^{+}\to w_{-1}^{+}+\tau}-\partial_\tau T_{\sr}^{(1)} \bigr) &=   (2\pi\mi\tau)\otimesprime \log\frac{t_{1}^2}{t_{3}^{2}}=\mathcal{S'}\biggl(-\log\frac{t_{1}^2}{t_{3}^{2}}\biggr) \:, \\
            \mathcal{S'}\bigl(\partial_\tau T_{\sr}^{(1)}|_{w_{0}^{+}\to w_{0}^{+}+\tau}-\partial_\tau T_{\sr}^{(1)} \bigr) &=   (2\pi\mi\tau)\otimesprime \log\frac{t_{2}^2}{t_{1}^{2}}=\mathcal{S'}\biggl(-\log\frac{t_{2}^2}{t_{1}^{2}}\biggr) \:.
        \end{align}
        The first two terms in the symbol change by corresponding terms with opposite sign that cancel these.  Thus, $\mathcal{S}(2\pi i T_{\sr}^{(1,2)})$ are \emph{double periodic}.
        \item Moreover, the symbol prime also makes integrability with respect to $\tau$ manifest. This is slightly trivial in the case of the length-two sunrise integral, and will thus be discussed in full generality for the case of the double-box integral in section \ref{sec:5.1}.
\end{enumerate}

Finally, note that the equal-mass case can be obtained smoothly by taking $t_1=t_2=t_3$; we will briefly comment on this case in section \ref{sec:5}.

\subsection{Degeneration at \texorpdfstring{$p^{2}=0$}{p**2=0} and pseudo-thresholds}

Next, let us see how the symbol of the unequal-mass sunrise integral behaves in kinematic limits where the elliptic curve degenerates.

The kinematic configurations where the elliptic curve degenerates can be easily read off from the discriminant
\begin{equation} \label{Discriminant}
    \Delta_{\sr}=\frac{t_{2}^{4}t_{3}^{4}}{t_{1}^{20}}\Bigl((t_{1}+t_{2}+t_{3})^{2}-1\Bigr)\Bigl((t_{1}+t_{2}-t_{3})^{2}-1\Bigr)\Bigl((t_{1}-t_{2}+t_{3})^{2}-1\Bigr)\Bigl((-t_{1}+t_{2}+t_{3})^{2}-1\Bigr) \:,
\end{equation}
where $t_i^2=m_i^2/p^2$ as before. 
In particular, the sunrise integral remains finite at $p^{2}=0$, at the pseudo-thresholds $p^{2}=\{(m_{1}+m_{2}-m_{3})^{2},(m_{1}+m_{3}-m_{2})^{2},(m_{2}+m_{3}-m_{1})^{2}\}$ and at the threshold $p^{2}=(m_{1}+m_{2}+m_{3})^{2}$, while it diverges for $m_i=0$.
The values at $p^{2}=0$ and the pseudo-thresholds were given in terms of MPLs in \cite{Bloch:2013tra}.  
In what follows, we will show how the symbols $\mathcal{S}(2\pi iT_{\sr}^{(1,2)})$ reproduce the corresponding symbols in these two limits.%
\footnote{The threshold can be treated in a similar way.}

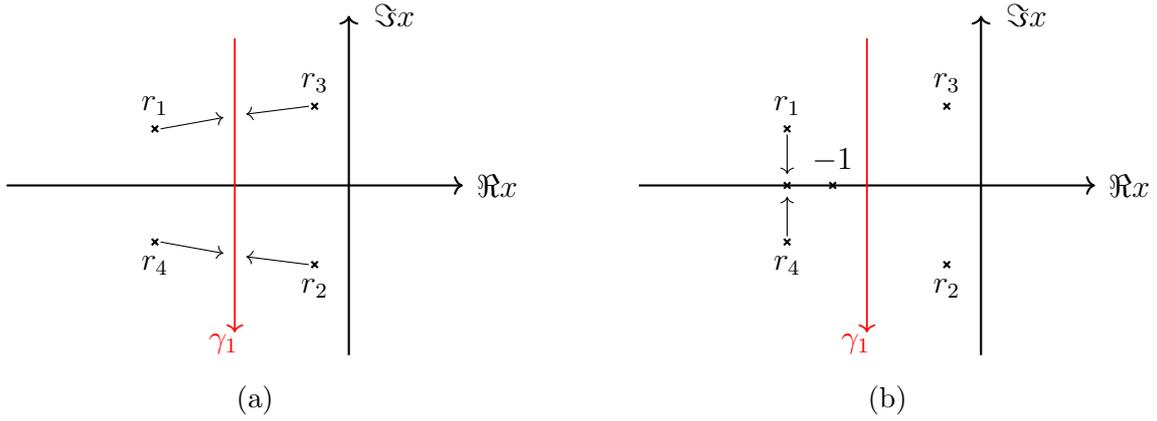
\begin{figure} 
      \begin{subfigure}[b]{0.45\textwidth}
         \centering
        \begin{tikzpicture}[scale=1.5]
         
                    \draw[->, thick] (0,0) to (0,3);
                    \draw[->, thick] (-3,1.5) to (1,1.5);
                     \draw[<-,line width=0.23mm,red] (-1.0,0.2) to (-1.0,2.8);
                    \node[cross,label=above:$r_{1}$] at (-1.7, 2.0) {};
                    \node[cross,label=below:$r_{4}$] at (-1.7, 1.0) {};
                    \node[cross,label=below:$r_{2}$] at (-0.3, 0.8) {};
                    \node[cross,label=above:$r_{3}$] at (-0.3, 2.2) {};
                     \draw[->] (-1.65,2.0) to (-1.1,2.1);
                     \draw[->] (-1.65,1.0) to (-1.1,0.9);
                     \draw[->] (-0.35,0.8) to (-0.9,0.87);
                     \draw[->] (-0.35,2.2) to (-0.9,2.13);
                    \node at (-1.1,0.1) {\textcolor{red}{$\gamma_{1}$}};
                    \node at (0.4,3) {$\Im x$};
                    \node at (1.3,1.5) {$\Re x$};
            \end{tikzpicture}
         \caption{\phantom{.}}
         \label{Fig:nullmomentumlimit}
     \end{subfigure}
     \hfill
      \begin{subfigure}[b]{0.45\textwidth}
         \centering
            \begin{tikzpicture}[scale=1.5]
             
                        \draw[->, thick] (0,0) to (0,3);
                        \draw[->, thick] (-3,1.5) to (1,1.5);
                       \draw[<-,line width=0.23mm,red] (-1.0,0.2) to (-1.0,2.8);
                        \node[cross,label=above:$r_{1}$] at (-1.7, 2.0) {};
                        \node[cross,label=below:$r_{4}$] at (-1.7, 1.0) {};
                        \node[cross,label=below:$r_{2}$] at (-0.3, 0.8) {};
                        \node[cross,label=above:$r_{3}$] at (-0.3, 2.2) {};
                         \draw[->] (-1.7,1.95) to (-1.7,1.6);
                         \draw[->] (-1.7,1.05) to (-1.7,1.4);
             \node at (-1.1,0.1) {\textcolor{red}{$\gamma_{1}$}};
                        \node at (0.4,3) {$\Im x$};
                        \node at (1.3,1.5) {$\Re x$};
                        \node[cross] at (-1.7, 1.5) {};
                        \node[cross,label=above:$-1$] at (-1.3, 1.5) {};
                \end{tikzpicture}
         \caption{\phantom{.}}
         \label{Fig:pseudothreshold}
     \end{subfigure}
    \caption{The roots of $y^{2}(x)$ for the sunrise integral coincide in the null-momentum limit (\subref{Fig:nullmomentumlimit}) and pseudo-thresholds (\subref{Fig:pseudothreshold}). In the latter case, the position of the coinciding roots relative to $-1$ is shown for the case $t_3>t_1$.}
    \label{fig: sr_degeneration}
\end{figure}

\paragraph{Null-momentum limit} As $p^{2}\to 0$, $t_{1}$, $t_{2}$ and $t_{3}$ approach infinity while their ratios remain finite. The elliptic curve degenerates in a way that $r_{1} \to r_{3}$ and $r_{2}\to r_{4}$; cf.\ figure \ref{Fig:nullmomentumlimit}.
In this case, $\omega_{1}\to \infty$ since the roots pinch the corresponding integration contour $\gamma_1$, while  
    \begin{equation}
      \omega_{2} \to \int_{-\infty}^{\infty} \frac{\dif x}{x^{2}+\bigl(1+(t_{3}/t_{1})^{2}-(t_{2}/t_{1})^{2}\bigr)x+(t_{3}/t_{1})^{2}}
        = \frac{2\pi \mi}{\sqrt{\Bigl(1-\frac{t_{2}^{2}}{t_{1}^{2}}-\frac{t_{3}^{2}}{t_{1}^{2}}\Bigr)^{2}-4\frac{t_{2}^{2}t_{3}^{2}}{t_{1}^{4}}}} \:.
    \end{equation}
    Then $-(2\pi \mi)^{-1}m_{1}^{-2}\omega_{2}$ reproduces the same normalization factor as in \cite{Bloch:2013tra} (up to a sign).
        Thus, we should expect that $\mathcal{S}(2\pi iT_{\sr}^{(2)})$ reduces to the corresponding symbol.
        In this limit,  $q=\exp(2\pi \mi \tau')$ vanishes, and hence all $\cEf{-2}{c}{x}$ in \eqref{symbolw1} vanish; cf.\ \eqref{eq:Ecal-gammat-relation} and \eqref{wdef}. Furthermore, 
    \begin{align}
        &\Omega^{(0)}(\xi_{c}^{+} )\to \log \frac{2c+1-u+v+\sqrt{(1-u-v)^{2}-4uv}}{2c+1-u+v-\sqrt{(1-u-v)^{2}-4uv}} \:, \\
        &\log\frac{t_{1}}{t_{3}} +
    \Omega^{(1)}(\xi_{\infty}^{+}-\xi_{-1}^{+})-\Omega^{(1)}(\xi_{-1}^{+}-\xi_{0}^{+}) 
    \to 0\,,
    \end{align}
    where we have introduced $u = (t_{2}/t_{1})^{2}=z\bar{z}$ and $v=(t_{3}/t_{1})^{2}=(1-z)(1-\bar{z})$. 
    Then,
\begin{align}
    \mathcal{S}\bigl(2\pi\mi T_{\sr}^{(2)} \bigr) \to \log u \otimes \log\frac{1-\bar{z}}{1-z}-\log v \otimes \log \frac{\bar{z}}{z} \:.
\end{align}
which is the symbol of $-4i$ times the Bloch-Wigner dilogarithm $D(z)$, in perfect agreement with \cite{Bloch:2013tra}.

    \paragraph{Pseudo-thresholds} Without loss of generality, we consider the pseudo-threshold $p^{2}=(m_{1}+m_{2}-m_{3})^{2}$. In terms of $t_{i}$, this pseudo-threshold is equal to the condition $(t_{1}+t_{2}-t_{3})^{2}=1$. We only consider the solution $t_{3}=t_{1}+t_{2}-1$ since the treatment for the other solution is similar. At $t_{3}=t_{1}+t_{2}-1$, the roots $r_{1}$ and $r_{4}$ pinch the real axis; cf.\ figure \ref{Fig:pseudothreshold}. Thus $\omega_2$ diverges and we should consider the rescaling of the torus by $\omega_1$, $T_{\sr}^{(1)}$. We can then close the contour $\gamma_{1}$ with a large semi-circle in the left half-plane and evaluate the integral via residues:
    \begin{align}
        \int_{\gamma_{1}}\frac{t_{1}\dif x}{\Bigl(x+\frac{t_{3}}{t_{1}}\Bigr)\sqrt{\Bigl(t_{1}^{2}x^{2}+\bigl(t_{1}^{2}+t_{3}^{2}-(t_{2}+1)^{2}\bigr)x+t_{3}^{2}\Bigr)}} =-2\pi \mi\sqrt{\frac{t_{1}^{3}}{4t_{2}t_{3}}} \:.
    \end{align}
    Again $(2\pi\mi)^{-1}m_{1}^{-2}\omega_{1}$ reproduces the same normalization factor as in \cite{Bloch:2013tra} at $p^{2}=(m_{1}+m_{2}-m_{3})^{2}$. Thus, we should expect that $\mathcal{S}(2\pi i T_{\sr}^{(1)})$ reduces to the corresponding symbol. 
    At the pseudo-threshold, $q=\exp(2\pi \mi \tau)$ vanishes. Furthermore, we assume $t_{3}>t_{1}$;\footnote{The case $t_{1}<t_{3}$ can be obtained by an analytic continuation.} then all $\cEf{-2}{c}{x}$ appearing in \eqref{symbolw1} also vanish in this limit. 
    Each of the three last entries in \eqref{symbolw1} is divergent in this limit since the roots pinch the integration contour, but all these divergences cancel in the following combinations 
    \begin{align}
        \Omega^{(0)}(w_{\infty}^{+}-w_{-1}^{+}) &=2\pi i\int_{-1}^{\infty} \frac{\dif x}{y}\to \log\frac{1+\sqrt{\frac{t_{3}}{{t_{1}t_{2}}}}}{1-\sqrt{\frac{t_{3}}{{t_{1}t_{2}}}}} \:, \\
        \Omega^{(0)}(w_{\infty}^{+}-w_{0}^{+}) & =2\pi i\int_{0}^{\infty} \frac{\dif x}{y}\to\log\frac{1+\sqrt{\frac{t_{2}}{{t_{1}t_{3}}}}}{1-\sqrt{\frac{t_{2}}{{t_{1}t_{3}}}}} \:, \\ 
        \Omega^{(0)}(w_{0}^{+}-w_{-1}^{+}) &=2\pi i\int_{-1}^{0} \frac{\dif x}{y}\to \log\frac{1+\sqrt{\frac{t_{1}}{{t_{2}t_{3}}}}}{1-\sqrt{\frac{t_{1}}{{t_{2}t_{3}}}}} \:;
    \end{align}
cf.\ figure \ref{Fig:pseudothreshold}.
Correspondingly, the symbol is reduced to
\begin{align}
    \mathcal{S}\bigl(2\pi\mi T_{\sr}^{(1)} \bigr)  &\to \log \frac{t_{1}}{t_{2}t_{3}} \otimes \log\frac{1-\sqrt{\frac{t_{1}}{{t_{2}t_{3}}}}}{1+\sqrt{\frac{t_{1}}{{t_{2}t_{3}}}}} 
    +(t_1\leftrightarrow t_2)-(t_1\leftrightarrow t_3)
\:,
\end{align}
where each term is the symbol of $\Li_{2}(z)-\Li_{2}(-z)+\log(z)\log\bigl((1-z)/(1+z)\bigr)$ -- in perfect agreement with \cite{Bloch:2013tra}.

\section{Example II: Double-Box Integral}   \label{sec: example 2}

As a second example of Feynman integrals that evaluate to  elliptic multiple polylogarithms, we consider the two-loop ten-point double-box integral:
\begin{align} \label{dbIntegrand}
    I_{\db}=
\begin{aligned}
\begin{tikzpicture}[scale=0.85,label distance=-1mm]
		\node[label=left:$8$] (0) at (0, 15.5) {};
		\node[label=above:$10$] (1) at (1.5, 16) {};
		\node[label=above:$9$] (2) at (0.5, 16) {};
		\node[label=above:$1$] (3) at (2.5, 16) {};
		\node[label=right:$2$] (4) at (3.0, 15.5) {};
		\node[label=below:$6$] (5) at (0.5, 14) {};
		\node[label=left:$7$] (6) at (0, 14.5) {};
		\node[label=below:$5$] (7) at (1.5, 14) {};
		\node[label=right:$3$] (8) at (3, 14.5) {};
		\node[label=below:$4$] (9) at (2.5, 14) {};
        \node(11) at (5, 15) {};
		\node (12) at (6, 15) {};
		\draw[thick] (0.center) to (4.center);
		\draw[thick] (6.center) to (8.center);
		\draw[thick] (2.center) to (5.center);
		\draw[thick] (1.center) to (7.center);
		\draw[thick] (3.center) to (9.center);
        \node[label=left:\textcolor{blue!50}{$x_8$}] (10) at (0.2, 15) {};
        \node[label=above:\textcolor{blue!50}{$x_{10}$}] (13) at (1.0, 15.8) {};
        \node[label=above:\textcolor{blue!50}{$x_1$}] (14) at (2.0, 15.8) {};
        \node[label=right:\textcolor{blue!50}{$x_3$}] (15) at (2.8, 15) {};
        \node[label=below:\textcolor{blue!50}{$x_5$}] (16) at (2.0, 14.3) {};
        \node[label=below:\textcolor{blue!50}{$x_6$}] (17) at (1.0, 14.3) {};
		 \draw[very thick,blue!50] (10.center) to (15.center);
		 \draw[very thick,blue!50] (13.center) to (17.center);
		 \draw[very thick,blue!50] (14.center) to (16.center);
\end{tikzpicture} 
\end{aligned}
\hspace{-0.14\textwidth}
&=\int \dif^4 x_{l} \dif^4x_{k}\frac{(x_1-x_5)^2}{(x_1-x_l)^2(x_3-x_l)^2(x_5-x_l)^2}\nonumber 
\\[-1.5\baselineskip]
&\quad \times \frac{(x_3-x_8)^2(x_6-x_{10})^2}{(x_l-x_k)^2(x_6-x_k)^2(x_8-x_k)^2(x_{10}-x_k)^2} \:,
\end{align}
where we have introduced the dual coordinates $x_{i}$ defined as $x_{i}-x_{i+1}=p_{i}$ and the notation $x_{i,j}=x_{i}-x_{j}$. 
This integral is a particular component of the two-loop ten-point N$^{3}$MHV amplitude in planar maximally  supersymmetric Yang-Mills theory ($\mathcal{N}=4$ sYM theory) \cite{CaronHuot:2012ab}, and was recently integrated in terms of elliptic multiple polylogarithms in \cite{Kristensson:2021ani}. It depends on seven dual conformal cross-ratios
\begin{gather}  \label{uvdef}
    u_{1}=\frac{x_{1,3}^{2}x_{5,8}^{2}}{x_{1,5}^{2}x_{3,8}^{2}}\:, \qquad 
    u_{2}=\frac{x_{3,6}^{2}x_{8,10}^{2}}{x_{3,8}^{2}x_{6,10}^{2}} \:, \qquad 
    v_{1}=\frac{x_{1,8}^{2}x_{3,5}^{2}}{x_{1,5}^{2}x_{3,8}^{2}} \:, \qquad 
    v_{2}=\frac{x_{3,10}^{2}x_{6,8}^{2}}{x_{3,8}^{2}x_{6,10}^{2}} \:, \qquad  \nonumber \\
    u_{3}=\frac{x_{1,3}^{2}x_{5,10}^{2}}{x_{1,5}^{2}x_{3,10}^{2}} \:, \qquad 
    u_{4}=\frac{x_{1,6}^{2}x_{3,5}^{2}}{x_{1,5}^{2}x_{3,6}^{2}} \:, \qquad 
    u_{5}= \frac{x_{1,5}^{2}x_{6,10}^{2}}{x_{1,6}^{2}x_{5,10}^{2}}
\end{gather}
and contains the elliptic curve defined by
\begin{align}
    y^{2}&=x^{4}+a_{3}x^{3}+a_{2}x^{2}+a_{1}x+a_{0}  \nonumber  \\
&=\biggl(\frac{v_{1}}{u_{4}}\bigl((1{-}u_{4})(x{+}1{-}v_{2}){-}u_{1}{+}u_{3}v_{2}\bigr){+}h_{1}{+}h_{2}\biggr)^{2}{-}4h_{1}h_{2}, \label{ellcurve}
\end{align}
where
\begin{equation}
 \begin{aligned}
    h_{1}&=\frac{u_{2}u_{4}}{v_{1}}\bigl(x^{2}+(1{-}u_{1}{+}v_{1})x+v_{1}\bigr),  \\
    h_{2}&=\Bigl(x{+}\frac{v_{1}}{u_{4}}\Bigr)\Bigl((1{+}x{-}u_{1})\Bigl(\frac{u_{2}u_{4}}{v_{1}}-1\Bigr)+(1{-}u_{3})v_{2}\Bigr). 
\end{aligned}
\end{equation}

As shown in \cite{Kristensson:2021ani}, the elliptic double-box integral can be evaluated in terms of $\mathcal{E}_{4}$ functions 
whose arguments make up the set
\begin{align} \label{arguset}
   \{c_i\}= \biggl\{&0,-1,\infty,-u_{2},-v_{1},-\frac{v_{1}}{u_{4}},-1+\frac{u_{1}}{u_{3}},-u_{2}u_{4}u_{5}, 
   {-}u_{2}(1{-}u_{4}){-}v_{1},  \nonumber \\
   &\frac{u_{2}(u_{3}{+}u_{4}{-}1){-}v_{1}}{1-u_{3}},
   \frac{u_{2}u_{3}u_{4}u_{5}{-}v_{1}}{1-u_{3}}, 
   \frac{u_{2}u_{3}u_{4}u_{5}{-}v_{1}}{u_{4}(1-u_{3}u_{5})},\frac{u_{2}(u_{3}u_{4}u_{5}v_{2}{-}u_{1})}{u_{3}v_{2}-u_{1}},
   \nonumber \\
   &\frac{v_{1}(u_{3}u_{4}u_{5}v_{2}{-}u_{1})}{u_{4}(u_{1}{-}u_{3}u_{5}v_{2})}, 
   \frac{u_{4}u_{5}(u_{2}(u_{4}{-}1){-}v_{1}){+}v_{1}}{u_{4}(u_{5}-1)},\frac{u_{1}u_{2}(u_{4}{-}1){-}v_{1}(u_{1}{-}u_{3}v_{2})}{u_{1}-u_{3}v_{2}}, \nonumber \\
   &z_{1}{-}1,\bar{z}_{1}{-}1,z_{1,3,6,8}{-}1,\bar{z}_{1,3,6,8}{-}1, {-}z_{3,5,8,10},{-}\bar{z}_{3,5,8,10},{-}z_{2},{-}\bar{z}_{2},\nonumber \\
   & \frac{u_{2}u_{3}u_{4}u_{5}{-}v_{1}+ r_{+}}{1-u_{3}} ,\frac{u_{2}u_{3}u_{4}u_{5}{-}v_{1}+ r_{-}}{1-u_{3}} \biggr\}\,,
\end{align} 
where $z_{a,b,c,d}\bar{z}_{a,b,c,d}=x_{a,b;c,d}$, $(1-z_{a,b,c,d})(1-\bar{z}_{a,b,c,d})=x_{d,a;b,c}$, with the abbreviation
\[
x_{a,b;c,d}=\frac{x_{a,b}^{2}x_{c,d}^{2}}{x_{a,c}^{2}x_{b,d}^{2}}\:.     
\]
Note that $u_{1}=x_{1,3;5,8}, v_{1}=x_{3,5;8,1}$ and $ u_{2}=x_{3,6;8,10}, v_{1}=x_{6,8;10,3}$, and we thus have abbreviated $z_1\equiv z_{1,3,5,8}$ and $z_2\equiv z_{3,6,8,10}$.
The expressions for $r_{\pm}$ are slightly lengthy,
\begin{equation}
    r_{\pm} = \frac{\mathbf{G}^{-1}_{45}\operatorname{det}\mathbf{G}\pm\sqrt{\operatorname{det}\mathbf{G}^{(45)}}\sqrt{-\operatorname{det}\mathbf{G}}}{2(1-u_{5})x_{1,5}^{2}x_{3,10}^{2}x_{1,6}^{2}x_{3,8}^{2}x_{5,10}^{2}} \:,
\end{equation}
where $\mathbf{G}$ denotes the Gram matrix $(x_{a,b}^{2})$ with $a$ and $b$ running over the index set of dual coordinates $\{1,3,5,6,8,10\}$, $\mathbf{G}^{-1}_{ij}$ denotes the elements of the inverse of the Gram matrix $(\mathbf{G}^{-1})_{ij}$, and $\mathbf{G}^{(ij)}$ denotes the matrix obtained from $\mathbf{G}$ by deleting the $i$'th and $j$'th rows and columns. 
By using \eqref{psitog1}, the double-box integral was also expressed in terms of $\tilde{\Gamma}$ functions on the torus $[1:\tau=\omega_{2}/\omega_{1}]$ \cite{Kristensson:2021ani}.

The last two arguments in \eqref{arguset} can be written in a slightly more compact form if we introduce the \emph{momentum twistor} variables~\cite{Hodges:2009hk},
\begin{equation}
    Z_{i}^{a}=(\lambda_{i}^{\alpha},x_{i}^{\alpha\dot{\alpha}}\lambda_{i \alpha}) \:,\qquad \alpha,\dot{\alpha}=1,2\,,
\end{equation}
as well as the $\mathrm{SL}(4)$-invariant $\langle ijkl\rangle=\epsilon_{abcd}Z_{i}^{a}Z_{j}^{b}Z_{k}^{c}Z_{l}^{d}$, where $\lambda_{i}$ is the usual spinor-helicity variable $p_{i}^{\mu}\sigma_{\mu}^{\alpha\dot{\alpha}}=\lambda^{\alpha}_{i}\tilde{\lambda}_{i}^{\dot{\alpha}}$. 
In terms of these variables, the last two arguments in \eqref{arguset} can be written as 
\begin{equation}
 \frac{\langle 9,10,1,(7,8)\cap (2,3,5)\rangle}{\langle 1,5,9,10\rangle\langle 2,3,7,8\rangle}, 
   \frac{\langle 4,5,6, (2,3)\cap(7,8,10) \rangle}{ \langle 2,3,7,8\rangle \langle 4,5,6,10\rangle}\,,
\end{equation}
where $  (ab)\cap(ijk)=Z_{a}\langle bijk\rangle+Z_{b}\langle ijka\rangle $ is the intersection point of the line $(ab)$ and the plane $(ijk)$~\cite{ArkaniHamed:2010kv}.

In the present paper, we compute  $I_{\db}$ in terms of 
$\tilde{\Gamma}$ functions on the torus rescaled as $[1:\tau'=-\omega_{1}/\omega_{2}] $, as well as the corresponding symbol. 
The computation is straightforward; however, both expressions are too lengthy to be recorded in the main text and are given in the ancillary file \texttt{doublebox\_omega2}.
In what follows, we will give a schematic expression of the symbol as well as the symbol prime and comment on several interesting aspects, such as the integrability condition and the soft limit $p_{10}\to0$.

\subsection{Symbol of the double-box integral} \label{sec:5.1}

Let us first review some results and notations from \cite{Kristensson:2021ani}. After normalization by $\omega_{1}$, $T^{(1)}_{\db}=I_{\db}/\omega_{1}$ is a pure combination of $\tilde{\Gamma}$ functions of length four and weight three. The corresponding symbol $\mathcal{S}(2\pi i T_{\db}^{(1)})$ satisfies the same physical first-entry conditions as in the MPL cases~\cite{Gaiotto:2011dt}; namely, the first entries are given by $\log u_i$ and $\log v_i$. The first two entries are $\Li_{2}(1-x_{a,b;c,d})$, $\log x_{a,b;c,d} \log x_{a',b';c',d'}$ 
or four-mass-box functions,
\begin{equation}
 -\Li_{2}(z_{a,b,c,d})+\Li_{2}(\bar{z}_{a,b,c,d})-
\frac{1}{2}\log (x_{a,b;c,d})\log\biggl(\frac{1-z_{a,b,c,d}}{1-\bar{z}_{a,b,c,d}}\biggr) \:.
\end{equation}
The last entries of this symbol consist of seven letters of elliptic type $\Omega^{(0)}$. According to these last entries, the symbol can be organized as\footnote{Here we use a slightly different notation for $\mathcal{S}(2\pi i T_{\db}^{(1)})$ than the one in \cite{Kristensson:2021ani} for bookkeeping.} 
\begin{align}
    \mathcal{S}(2\pi i T_{\db}^{(1)})&= \mathcal{S}(I_{\text{hex}})\otimes \Omega^{(0)}(w_{c_{25}}^{+}) +
    \mathcal{S}(F_{-})\otimes \Omega^{(0)}(w_{\infty}^{-}) + \mathcal{S}(F_{\tau})\otimes (2\pi i\tau) \nonumber \\
    &\quad + \mathcal{S}(F_{z_{1}-1})\otimes \Omega^{(0)}(w_{z_{1}-1}^{+})+
 \mathcal{S}(F_{\bar{z}_{1}-1})\otimes \Omega^{(0)}(w_{\bar{z}_{1}-1}^{+}) \nonumber  \\
  &\quad + \mathcal{S}(F_{-z_{2}})\otimes \Omega^{(0)}(w_{-z_{2}}^{+})+
 \mathcal{S}(F_{-\bar{z}_{2}})\otimes \Omega^{(0)}(w_{-\bar{z}_{2}}^{+}) \:, \label{symbol_db_1}
\end{align}
where $I_{\text{hex}}$ is the 6D hexagon integral (normalized to be
pure), $c_{25}=\frac{\langle 9,10,1,(7,8)\cap (2,3,5)\rangle}{\langle 1,5,9,10\rangle\langle 2,3,7,8\rangle}$ is the $25$'th element of the set \eqref{arguset} and we used $\Omega^{(0)}(w_{\infty}^{+})=2\pi i\tau$. In particular, the symbols of all weight-three functions except $F_{\tau}$, 
that are $I_{\text{hex}}$, $F_{-}$, etc., are \emph{polylogarithmic}; their symbol entries are logarithms of kinematics. Furthermore, $I_{\text{hex}}$ and $F_{-}$ are invariant up to a sign under the two reflections
\begin{align}
    & R_{1}:\quad p_{i}\to p_{15-i} \:,\\
    & R_{2}:\quad p_{i}\to p_{10-i} \:,
\end{align}
where $p_{i}\equiv p_{i+10}$, while $F_{z_{1}-1}$, $F_{\bar{z}_{1}-1}$, $F_{-z_{2}}$ and $F_{-\bar{z}_{2}}$ form an orbit (up to a sign) under both reflections. %
Besides the seven elliptic last entries, elliptic letters only appear at the third entry of $\mathcal{S}(F_{\tau})$ and come in only 13 linear independent combinations of $\Omega^{(2,1,0)}$'s.
For the full symbol alphabet and a form that is manifestly invariant under the two reflections, see \cite{Kristensson:2021ani}.

In the other normalization, $T^{(2)}_{\db}=-I_{\db}/\omega_{2}$, the structure of the symbol is almost the same as \eqref{symbol_db_1},
\begin{align}
    \mathcal{S}(2\pi i T_{\db}^{(2)})&= \mathcal{S}(I_{\text{hex}})\otimes \Omega^{(0)}(\xi_{c_{25}}^{+}) +
    \mathcal{S}(F_{-})\otimes \Omega^{(0)}(\xi_{\infty}^{-}) + \mathcal{S}(F_{\tau'})\otimes (2\pi i\tau') \nonumber \\
    &\quad + \mathcal{S}(F_{z_{1}-1})\otimes \Omega^{(0)}(\xi_{z_{1}-1}^{+})+
 \mathcal{S}(F_{\bar{z}_{1}-1})\otimes \Omega^{(0)}(\xi_{\bar{z}_{1}-1}^{+})  \label{symbol_db_2} \\
  &\quad + \mathcal{S}(F_{-z_{2}})\otimes \Omega^{(0)}(\xi_{-z_{2}}^{+})+
 \mathcal{S}(F_{-\bar{z}_{2}})\otimes \Omega^{(0)}(\xi_{-\bar{z}_{2}}^{+}) \:. \nonumber
\end{align}
Besides the last entries, which are simple modular $S$-transformations of the last entries of $ \mathcal{S}(2\pi i T_{\db}^{(1)})$, the only difference arises from the third entry of $F_{\tau'}$ which consist of \emph{only} $\Omega^{(2)}$'s. Meanwhile, the $\Omega^{(2)}$'s appearing in the third entry of $\mathcal{S}(F_\tau')$ also come in 13 linear independent combinations. 

It is not surprising that all weight-three functions appearing in $\mathcal{S}(2\pi i T_{\db}^{(1)})$ except~$F_{\tau}$ are 
the same as in $\mathcal{S}(2\pi i T_{\db}^{(2)})$. Actually, for any elliptic multiple polylogarithm $I$ given by an integral of the form \eqref{1dIntofPolylog}, its two normalizations, $T^{(1)}=I/\omega_{1}$ and $T^{(2)}=-I/\omega_{2}$, are related by $T^{(2)}=\tau' T^{(1)}$. 
With the last entry of $T^{(1)}$ being of elliptic type $\Omega^{(0)}$, i.e.,
\begin{equation}
    \dif T^{(1)} = \sum_{i} F_{i} \dif w_{i} + F_{\tau}\dif \tau ,
\end{equation}
where $F_{i}$ and $F_{\tau}$ are elliptic multiple polylogarithms of lower length, a straightforward computation gives
\begin{equation}
    \dif T^{(2)} = \sum_{i}F_{i}\dif \xi_{i} + \underbrace {\biggl(T^{(1)}-\sum_{i}F_{i}w_{i}-\tau F_\tau\biggr)}_{F_{\tau'}}\dif \tau' \:.
    \label{eq: S transformation Ftau}
\end{equation}
While it is not manifest from \eqref{eq: S transformation Ftau}, this particular expression for $F_{\tau'}$ has length three.

Once we evaluate the symbol prime for the third entry of $F_{\tau}$ and $F_{\tau}'$, a simple connection emerge. Both $ \mathcal{S}(2\pi i T_{\db}^{(1)})$ and $ \mathcal{S}(2\pi i T_{\db}^{(2)})$ have similar structures as in the case of the sunrise integral: 
\begin{align}
    \mathcal{S}(2\pi i  T_{\db}^{(1)})&= \sum_{ij} \mathcal{S}(f_{i})\otimes \Bigl( \log a_{ij} \otimes \Omega^{(0)}(w_{j}) + \bm{\Omega}_{i}\otimes (2 \pi i\tau)   \Bigr), \label{symbol_prime_db}
    \intertext{with}
    \mathcal{S}'(\bm{\Omega}_{i})&=\Omega^{(0)}(w_{j})\otimesprime \log a_{ij}\,,
\end{align}
and a similar expression for $\mathcal{S}(2\pi i  T_{\db}^{(2)})$ with $\xi_{j}$ and $\tau'$ in place of $w_{j}$ and $\tau$. Here $w_{j}$ are the six last entries except $\tau$, $a_{ij}$ are some algebraic functions of kinematics, $f_{i}$ are of the form $\Li_{2}(1-x_{ab;cd})$, $\log x_{ab;cd} \log x_{a'b';c'd'}$ or four-mass-box functions, and $\bm{\Omega}_{i}$ are combinations of $\Omega^{(2,1,0)}$ that occur in the last entry of $F_{\tau}$ associated with $\mathcal{S}(f_{i})$.

Let us close this subsection by remarking some advantages of the form \eqref{symbol_prime_db}. Firstly, the symbol prime is manifestly double-periodic due to the same argument used in section \ref{sec:4.1}: under the translation $w_{j}\to w_{j}+\tau$,
\begin{align}
    & \underbrace{[\Omega^{(-1)}\otimesprime\log a_{ij}]}_{\mathcal{S}'(\log a_{ij})} \otimes \Omega^{(0)}(w_{j}) + \underbrace{[\Omega^{(0)}(w_{j})\otimesprime \log a_{ij}]}_{\mathcal{S}'(\bm{\Omega}_{i})} \otimes (2 \pi i\tau) \nonumber\\
    &\to [\Omega^{(-1)}\otimesprime\log a_{ij}] \otimes \Omega^{(0)}(w_{j}+\tau) + [\Omega^{(0)}(w_{j}+\tau)\otimesprime \log a_{ij}] \otimes (2 \pi i\tau) \\
    & ={}\,[\Omega^{(-1)}\otimesprime\log a_{ij}] \otimes \Omega^{(0)}(w_{j}) + [\Omega^{(0)}(w_{j})\otimesprime \log a_{ij}] \otimes (2 \pi i\tau) \:,\nonumber
\end{align}
where we have used \eqref{symbolp3} and $\Omega^{(-1)}=-\Omega^{(0)}(\tau)=-2\pi i\tau$.
Secondly, this form makes a part of the integrability conditions manifest and sheds light on the (13 linearly independent) combinations of $\Omega^{(2,1,0)}$ that occur in the last entry of $F_{\tau}$. 
The integrability condition requires 
\begin{equation}
\partial_{w_j}\bm{\Omega}_{i} + \partial_{\tau} \log a_{ij} =0\,,
\end{equation}
which is a consequence of 
\begin{equation}
    \mathcal{S}'(\bm{\Omega}_{i})= \sum_{j}\Omega^{(0)}(w_{j})\otimesprime \log a_{ij}\:,
\end{equation}
since 
\begin{equation}
    \bm{\Omega}_{i}\otimes \tau - \sum_{j}\Omega^{(0)}(w_{j})\otimes \log a_{ij} =
    \bm{\Omega}_{i}\otimes \tau +\sum_{j} \log a_{ij} \otimes \Omega^{(0)}(w_{j}) - \mathcal{S}(\log a_{ij} \Omega^{(0)}(w_{j}))
\end{equation}
is an integrable symbol by the definition of the symbol prime.
We will discuss the remaining integrability conditions in upcoming work \cite{integrability_in_progress}.

\subsection{Soft limit}

In this subsection, we will consider the soft limit $p_{10}\to 0$, in which the elliptic double-box integral remains finite and becomes polylogarithmic:
\begin{equation}
\begin{aligned}
\begin{tikzpicture}[scale=0.85,label distance=-1mm]
		\node[label=left:$8$] (0) at (0, 15.5) {};
		\node[label=above:$10$] (1) at (1.5, 16) {};
		\node[label=above:$9$] (2) at (0.5, 16) {};
		\node[label=above:$1$] (3) at (2.5, 16) {};
		\node[label=right:$2$] (4) at (3.0, 15.5) {};
		\node[label=below:$6$] (5) at (0.5, 14) {};
		\node[label=left:$7$] (6) at (0, 14.5) {};
		\node[label=below:$5$] (7) at (1.5, 14) {};
		\node[label=right:$3$] (8) at (3, 14.5) {};
		\node[label=below:$4$] (9) at (2.5, 14) {};
        \node(11) at (5, 15) {};
		\node (12) at (6, 15) {};
		\draw[thick] (0.center) to (4.center);
		\draw[thick] (6.center) to (8.center);
		\draw[thick] (2.center) to (5.center);
		\draw[thick] (1.center) to (7.center);
		\draw[thick] (3.center) to (9.center);
        \node[label=left:\textcolor{blue!50}{$x_8$}] (10) at (0.2, 15) {};
        \node[label=above:\textcolor{blue!50}{$x_{10}$}] (13) at (1.0, 15.8) {};
        \node[label=above:\textcolor{blue!50}{$x_1$}] (14) at (2.0, 15.8) {};
        \node[label=right:\textcolor{blue!50}{$x_3$}] (15) at (2.8, 15) {};
        \node[label=below:\textcolor{blue!50}{$x_5$}] (16) at (2.0, 14.3) {};
        \node[label=below:\textcolor{blue!50}{$x_6$}] (17) at (1.0, 14.3) {};
		 \draw[very thick,blue!50] (10.center) to (15.center);
		 \draw[very thick,blue!50] (13.center) to (17.center);
		 \draw[very thick,blue!50] (14.center) to (16.center);
\end{tikzpicture} 
\end{aligned}
\hspace{-0.1\linewidth}
\xrightarrow{p_{10}\to 0}\qquad
\begin{aligned}
\begin{tikzpicture}[scale=0.85,label distance=-1mm]
		\node[label=left:$8$] (0) at (0, 15.5) {};
		\node[] (1) at (1.5, 15.5) {};
		\node[label=above:$9$] (2) at (0.5, 16) {};
		\node[label=above:$1$] (3) at (2.5, 16) {};
		\node[label=right:$2$] (4) at (3.0, 15.5) {};
		\node[label=below:$6$] (5) at (0.5, 14) {};
		\node[label=left:$7$] (6) at (0, 14.5) {};
		\node[label=below:$5$] (7) at (1.5, 14) {};
		\node[label=right:$3$] (8) at (3, 14.5) {};
		\node[label=below:$4$] (9) at (2.5, 14) {};
        \node(11) at (5, 15) {};
		\node (12) at (6, 15) {};
		\draw[thick] (0.center) to (4.center);
		\draw[thick] (6.center) to (8.center);
		\draw[thick] (2.center) to (5.center);
		\draw[thick] (1.center) to (7.center);
		\draw[thick] (3.center) to (9.center);
        \node[label=left:\textcolor{blue!50}{$x_8$}] (10) at (0.2, 15) {};
        \node[label=above:\textcolor{blue!50}{$x_{1}$}] (13) at (1.5, 16) {};
        \node[label=right:\textcolor{blue!50}{$x_3$}] (15) at (2.8, 15) {};
        \node[label=below:\textcolor{blue!50}{$x_5$}] (16) at (2.0, 14.3) {};
        \node[label=below:\textcolor{blue!50}{$x_6$}] (17) at (1.0, 14.3) {};
		 \draw[very thick,blue!50] (10.center) to (15.center);
		 \draw[very thick,blue!50] (13.center) to (1,15) to (17.center);
		 \draw[very thick,blue!50] (13.center) to (2,15) to (16.center);
\end{tikzpicture} 
\end{aligned}
\hspace{-0.1\linewidth}
\end{equation}

In terms of momentum twistors, this limit amounts to first setting 
\begin{equation}
    Z_{10} \to Z_{9}+\alpha Z_{1}+\epsilon (Z_{8}+\beta Z_{2})  \label{soft_limit_mt}
\end{equation}
with finite $\alpha$ and $\beta$, and then taking the limit $\epsilon\to 0$~\cite{Drummond:2010mb}. In this limit, 
the seven cross-ratios become
\begin{gather}  \label{uvdefsoft}
    u_{1}=\frac{x_{1,3}^{2}x_{5,8}^{2}}{x_{1,5}^{2}x_{3,8}^{2}}\:, \qquad 
    u_{2}=\frac{x_{3,6}^{2}x_{8,1}^{2}}{x_{3,8}^{2}x_{6,1}^{2}} \:, \qquad 
    v_{1}=\frac{x_{1,8}^{2}x_{3,5}^{2}}{x_{1,5}^{2}x_{3,8}^{2}} \:, \qquad 
    v_{2}=\frac{x_{3,1}^{2}x_{6,8}^{2}}{x_{3,8}^{2}x_{6,1}^{2}} \:, \qquad  \nonumber \\
    u_{3}=1 \:, \qquad 
    u_{4}=\frac{x_{1,6}^{2}x_{3,5}^{2}}{x_{1,5}^{2}x_{3,6}^{2}}=\frac{v_{1}}{u_{2}} \:, \qquad 
    u_{5}=1 \:.
\end{gather}
Correspondingly, the elliptic curve \eqref{ellcurve} degenerates to 
\renewcommand{\rho}{r}
\begin{equation}
    y^{2}=\bigl(x^{2}+(1-u_{1}+u_{2})x+u_{2}-u_{1}u_{2}+v_{1}v_{2}\bigr)^{2}=\bigl((x-\rho)(x-\bar{\rho})\bigr)^{2} \:, \label{ellcurvesoft}
\end{equation}
where 
\begin{align}
    \rho       &=-\frac{1}{2}\bigl(1-u_{1}+u_{2}+\sqrt{(1-u_{1}-u_{2})^{2}-4v_{1}v_{2}}\bigr)  \:,\\
    \bar{\rho} &=-\frac{1}{2}\bigl(1-u_{1}+u_{2}-\sqrt{(1-u_{1}-u_{2})^{2}-4v_{1}v_{2}}\bigr)  \:.
\end{align}

The double-box integral in this soft limit can be easily integrated to multiple polylogarithms starting from the one-fold 
integral representation in \cite{Kristensson:2021ani} by using, for instance, 
{\sc HyperInt} \cite{Panzer:2014caa} or {\sc PolyLogTools} \cite{Duhr:2019tlz}.
We record both the function and the symbol result in the ancillary file \texttt{doublebox\_soft}. 
In the following, we demonstrate how the same result for the symbol is obtained as a limit of the elliptic symbol of the ten-point double-box integral \eqref{symbol_db_1}.

We work in the region given by positive momentum-twistor kinematics~\cite{Arkani-Hamed:2013jha}, where the four roots of $y^{2}(x)$ in \eqref{ellcurve} come in complex conjugate pairs as shown in figure \ref{fig: contours}, but both $\rho$ and $\bar{\rho}$ are negative real numbers, i.e., 
$r_{1},r_{4}\to \rho$, while $r_{2},r_{3}\to\bar{\rho}$ in the soft limit. Therefore, $\omega_{2}\to\infty$ in this limit, 
and 
\begin{equation}
  \omega_{1}\to  \int_{\gamma_{1}} -\frac{\dif x}{(x-\rho)(x-\bar{\rho})} = \frac{2\pi i}{\rho-\bar{\rho}} 
\end{equation}
where we made the following choice for the degenerated curve \eqref{ellcurvesoft} in the different regions:
\begin{align}
  y = \begin{cases}
    -(x-\rho)(x-\bar{\rho}) \:, & \text{for} \quad \rho<\operatorname{Re} x <\bar{\rho} \\
     (x-\rho)(x-\bar{\rho}) \:, & \text{otherwise} \,.
   \end{cases}
\end{align}
The reason for this choice is that we want to keep $y$ non-negative on the real $x$-axis for the degenerated curve, following the convention we chose for the original curve in \eqref{ellcurve}.
Again, $(2\pi \mi)^{-1} \omega_{1}$ gives the correct normalization factor for this nine-point double-box integral, and we should expect that $\mathcal{S}(2\pi i T_{\db}^{(1)})$ will reproduce the correct symbol in this soft limit.

\begin{figure}
   \centering
   \begin{tikzpicture}[scale=2]

           \draw[->, thick] (0,0) to (0,3);
           \draw[->, thick] (-3,1.5) to (1,1.5);
           \draw[->,line width=0.23mm,blue] (-3,1.4) .. controls (-1,1.4) ..   (-1.0,0.0) ; 
           \node[cross,label=above:$r_{1}$] at (-1.7, 2.0) {};
           \node[cross,label=below:$r_{4}$] at (-1.7, 1.0) {};
           \node[cross,label=below:$r_{2}$] at (-0.3, 0.8) {};
           \node[cross,label=above:$r_{3}$] at (-0.3, 2.2) {};
            \draw[->] (-1.7,1.95) to (-1.7,1.7);
            \draw[->] (-1.7,1.05) to (-1.7,1.3);
            \draw[->] (-0.3,0.85) to (-0.3,1.45);
            \draw[->] (-0.3,2.15) to (-0.3,1.55);
           \node at (-1.3,0.1) {\textcolor{blue}{$\gamma_{-}$}};
           \node at (0.4,3) {$\Im x$};
           \node at (1.3,1.5) {$\Re x$};
           \node[cross,label=above:${-z_{2}}$] at (-2.5, 1.5) {};
           \node[cross,label=above:${\bar{z}_{1}-1}$] at (-1.3, 1.5) {};
           \node[cross,label=above:${-\bar{z}_{2}}$] at (-0.7, 1.5) {};
           \node[cross,label=above:${z_{1}-1}$] at (0.5, 1.5) {};
   \end{tikzpicture}
   \caption{In the soft limit $p_{10}\to0$, the roots of $y^{2}(x)$ pairwise pinch the integration contours for $w^+_{\bar{z}_{1}-1}, w^+_{-\bar{z}_{2}}$ and $w^+_{z_{1}-1}$, which run along the real axis.
   By subtracting $w^-_{\infty}$ and $w^+_{\infty}$, respectively, we obtain integration contours that can be deformed such they are not pinched, thus resulting in finite integrals in the soft limit.  
   }
   \label{fig: degeneration}
   \end{figure}
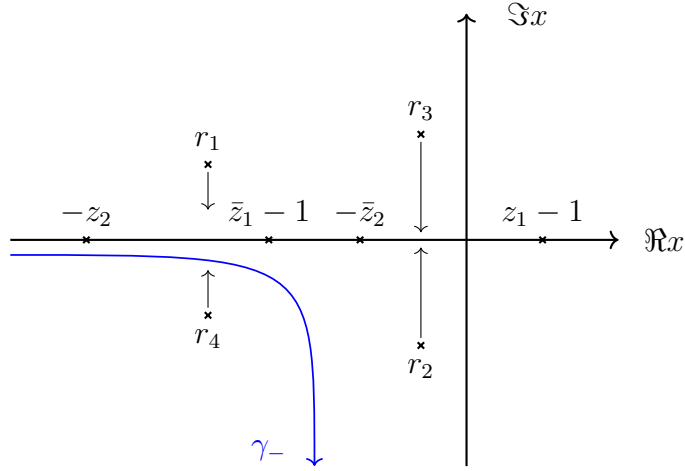

The immediate problem we encounter in taking the soft limit for $\mathcal{S}(2\pi i T_{\db}^{(1)})$ is that 
not all of the seven last entries in \eqref{symbol_db_1} have smooth, definite and finite limits.
While the soft limit of $c_{25}$ depends on the arbitrary parameters $\alpha$ and $\beta$ in \eqref{soft_limit_mt}, 
the term $\mathcal{S}(I_{\text{hex}})\otimes \Omega^{(0)}(w_{c_{25}}^{+})$ does not introduce a problem since $\mathcal{S}(I_{\text{hex}})$ vanishes in the soft limit.
The integration contours for $w_{\infty}^{-}$, $w_{\bar{z}_{1}-1}^{+}$ as well as $w_{-\bar{z}_{2}}^{+}$ go through the pole at $x=\rho$, and the integration contours for $w_{z_{1}-1}^{+}$ as well as $w_{\infty}^{+}=\tau$ go through both poles at $x=\rho$ and $x=\bar{\rho}$; see figure \ref{fig: degeneration}. 
 To cancel the resulting singularities, we reorganize $\mathcal{S}(2\pi i T_{\db}^{(1)})$ as 
\begin{align}
    \mathcal{S}(2\pi i T_{\db}^{(1)}) &= \mathcal{S}(F_{-z_{2}})\otimes \Omega^{(0)}(w_{-z_{2}}^{+})+
    \mathcal{S}(F_{z_{1}-1})\otimes \Omega^{(0)}(w_{z_{1}-1}^{+}-w_{\infty}^{+})  \nonumber \\
    &\quad + \mathcal{S}(F_{\bar{z}_{1}-1})\otimes \Omega^{(0)}(w_{\bar{z}_{1}-1}^{+}-w_{\infty}^{-}) 
    + \mathcal{S}(F_{-\bar{z}_{2}})\otimes \Omega^{(0)}(w_{-\bar{z}_{2}}^{+}-w_{\infty}^{-}) \nonumber \\
    &\quad 
    +\mathcal{S}(F_{\tau}+F_{z_{1}-1}) \otimes (2\pi i\tau) +\mathcal{S}(F_{-}+F_{\bar{z}_{1}-1}+F_{-\bar{z}_{2}})\otimes \Omega^{(0)}(w_{\infty}^{-}) \nonumber \\
    &\quad+\mathcal{S}(I_{\text{hex}}) \otimes \Omega^{(0)}(w_{c_{25}}^{+}) \:,
\end{align}
cf.\ figure \ref{fig: degeneration}.
One can easily check that not only the last term but the  last three terms do not contribute in the soft limit since the three weight-three symbols making up their first three entries vanish in the soft limit. The first four terms yield the correct polylogarithmic symbol in the soft limit with last entries
\begin{equation}
\begin{aligned}
    \Omega^{(0)}(w_{-z_{2}}^{+}) &\to (\rho-\bar{\rho}) \int_{-\infty}^{-z_{2}}\frac{\dif x}{(x-\rho)(x-\bar{\rho})}\:= 
    \log \frac{\rho+z_{2}}{\bar{\rho}+z_{2}} \:, \\
    \Omega^{(0)}(w_{z_{1}-1}^{+}-w_{\infty}^{+})& \to (\rho-\bar{\rho}) \int_{+\infty}^{z_{1}-1}\frac{\dif x}{(x-\rho)(x-\bar{\rho})} = \log\frac{1+\rho-z_{1}}{1+\bar{\rho}-z_{1}} \:, \\
    \Omega^{(0)}(w_{\bar{z}_{1}-1}^{+}-w_{\infty}^{-}) & \to (\rho-\bar{\rho}) \int_{-i\infty}^{\bar{z}_{1}-1} 
    \frac{-\dif x}{(x-\rho)(x-\bar{\rho})}= \log\frac{1+\bar{\rho}-\bar{z}_{1}}{1+\rho-\bar{z}_{1}} \:, \\
    \Omega^{(0)}(w_{-\bar{z}_{2}}^{+}-w_{\infty}^{-})&\to (\rho-\bar{\rho}) \int_{-i\infty}^{-\bar{z}_{2}}\frac{-\dif x}{(x-\rho)(x-\bar{\rho})} \: = \log \frac{\bar{\rho}+\bar{z}_{2}}{\rho+\bar{z}_{2}} \:.
\end{aligned} 
\end{equation}
Note that in the soft limit $z_{1}\equiv z_{1,3,5,8}$ and $z_{2}\equiv z_{3,6,8,1}$ and the reflection symmetry $R_{1}$ is broken while $R_{2}$ survives; 
thus, the symbol for this nine-point double-box integrals can be expressed as
\begin{align}
    \mathcal{S}\Bigl((\rho-\bar{\rho})I_{\softdb}\Bigr) &= \mathcal{S}(F_{-z_{2}}\vert_{p_{10}\to 0})\otimes \log \frac{\rho+z_{2}}{\bar{\rho}+z_{2}} +\mathcal{S}(F_{-\bar{z}_{2}}\vert_{p_{10}\to 0})\otimes \log \frac{\bar{\rho}+\bar{z}_{2}}{\rho+\bar{z}_{2}} \nonumber \\
    & \quad 
    + (\text{images under }R_2
    ) \:,
\end{align}
where $R_2$ acts on the last entries via  $\rho\leftrightarrow -(1{+}\bar{\rho})$, $z_{1}\leftrightarrow z_{2}$ and $\bar{z}_{1}\leftrightarrow \bar{z}_{2}$.
Furthermore, $F_{-z_{2}}\vert _{p_{10}\to 0}$ and $F_{-\bar{z}_{2}}\vert _{p_{10}\to 0}$ are related by exchanging $z_{2}$ and $\bar{z}_{2}$, same as the corresponding last entries.
The reason is that $z$ and $\bar{z}$ occur symmetrically in their definition $\{z\bar{z}=u, (1-z)(1-\bar{z})=v\}$ , and thus have to occur symmetrically in the symbol as well.

The symbol alphabet of the nine-point double-box integral consists of 10 rational letters and 11 algebraic letters:
\begin{enumerate}
    \item Rational letters:
\begin{equation}
    \begin{gathered}
        u_{1}\,,\:\: u_{2} \,,\:\:  v_{1} \,,\:\: v_{2} \,,\:\: u_{1}-v_{2} \,,\:\: v_{1}-u_{2} \,,\:\: 
        u_{1}u_{2}-v_{1}v_{2} \,,\:\: \Delta_{1} \,,\:\:  \Delta_{2}\:, \\
        \frac{\langle 5(91)(23)(78)\rangle \langle \bar{5}(91)(23)(78)\rangle\langle 1239\rangle \langle 1789\rangle}{ \langle 1459\rangle^{2}\langle 1569\rangle^{2}\langle 2378\rangle^{3}} \:,
    \end{gathered} 
    \end{equation}
    where we introduced the following notations:\footnote{Here we use $(\bar{a})\equiv Z_{a-1}{\wedge}Z_{a}{\wedge} Z_{a+1}$ to denote the dual plane of $Z_{a}$. Then a vanishing $\langle \bar{a} (i\,i{+}1)(j\,j{+}1)(k\,k{+}1) \rangle$ means that the three intersection points $(i\,i{+1}){\cap} (\bar{a})$, $(j\,j{+1}){\cap} (\bar{a})$ and $(k\,k{+1}){\cap} (\bar{a})$ are on the same line, which is the dual picture of the vanishing of $\langle a (i\,i{+}1)(j\,j{+}1)(k\,k{+}1) \rangle$. We are grateful to Cristian Vergu for pointing this out.}
    \begin{align}
        \langle a(bc)(de)(fg)\rangle &=\langle abde\rangle \langle acfg \rangle-\langle acde\rangle \langle abfg \rangle , \\
    \langle \bar{a} (i\,i{+}1)(j\,j{+}1)(k\,k{+}1) \rangle &= \langle (i\,i{+}1)\cap (\bar{a}) \,j\,j{+}1\,(k,k{+}1)\cap (\bar{a})\rangle 
    \end{align}
and 
\begin{align}
    \Delta_{i}=(1-u_{i}-v_{i})^{2}-4u_{i}v_{i}=(z_{i}-\bar{z}_{i})^{2} \:, \qquad i=1,2\,.
\end{align}
    \item Algebraic letters:
    \begin{itemize}
        \item $\displaystyle \frac{z_{1}}{\bar{z}_{1}}$, $\displaystyle \frac{1-z_{1}}{1-\bar{z}_{1}}$, $\displaystyle \frac{1+\bar{\rho}-z_{1}}{1+\rho-z_{1}}$, $\displaystyle \frac{1+\bar{\rho}-\bar{z}_{1}}{1+\rho-\bar{z}_{1}}$, 
        $\displaystyle \frac{\frac{\langle 5(23)(46)(78)\rangle \langle 1239\rangle }{\langle 5(19)(23)(78)\rangle \langle 2378\rangle}-z_{1}}{\frac{\langle 5(23)(46)(78)\rangle \langle 1239\rangle }{\langle 5(19)(23)(78)\rangle \langle 2378\rangle}-\bar{z}_{1}}$, \\
         and five others generated by the reflection $R_2$. 
        \item $\displaystyle \frac{(z_{1}-1+\bar{z}_{2})(\bar{z}_{1}-1+z_{2})}{(z_{1}-1+z_{2})(\bar{z}_{1}-1+\bar{z}_{2})}$.
    \end{itemize}
\end{enumerate}
We find that there are three different square roots in this alphabet; two of them are of four-mass-box type and the other, that is the square root in $\rho$ and $\bar{\rho}$, arises from the leading singularity of the whole Feynman diagram. Furthermore, the new type of square root \emph{only} appears in the last entries.
The symbol alphabet is organized such that the symbol is manifestly invariant (up to a sign) under the reflection $R_{2}$, as well as under each of the three transformations $z_{1} \leftrightarrow \bar{z}_{1}$, 
$z_{2} \leftrightarrow \bar{z}_{2}$, and $\rho \leftrightarrow \bar{\rho}$.
For an analysis of these letters through Schubert problems, see \cite{Yang:2022gko}.

\section{Discussion and Outlook} \label{sec:5}

In this paper, we have investigated various techniques for manipulating and simplifying the symbol of Feynman integrals that evaluate to elliptic multiple polylogarithms. In particular, we studied identities between the elliptic symbol letters $\Omega^{(i)}$.

In contrast to ordinary multiple polylogarithms, the length of an elliptic multiple polylogarithm is not necessarily equal to its weight. A symbol letter $\Omega^{(i)}$, whose length is by definition one, can have weight $i\neq 1$.
Identities for $\Omega^{(0)}$ follow from the well-known group law on the elliptic curve.
Moreover, we found that various identities for $\Omega^{(1)}$ can be derived from Abel's theorem, which generalize the identity $\log(a)+\log(b)=\log(ab)$ in the polylogarithmic case.
The higher-weight letters $\Omega^{(2)}$ satisfy significantly more intricate identities, closer to those of $\Li_{2}(a)$ than those of $\log(a)$, which are harder to exploit in a direct fashion.
We have thus introduced the \emph{symbol prime} $\mathcal{S}'$ for elliptic symbol letters $\Omega^{(2)}$, which plays the same role the symbol $\mathcal{S}$ plays for  $\Li_{2}(a)$. We also introduced a symbol prime for $\Omega^{(i>2)}$ but leave its exploration for future work.

We studied two concrete examples at two-loop order, namely the sunrise integral in two dimensions and the ten-point double-box integral in four dimensions. In particular, we provided proofs for the identities between symbol letters numerically found in \cite{Kristensson:2021ani}.

In addition to identities between symbol letters, we also studied how the symbol behaves under kinematic limits in which the elliptic curve degenerates. We recovered the known symbols of the sunrise integral in the null-momentum limit $p^{2}\to 0$ and the pseudo threshold $p^2\to(m_{1}+m_{2}-m_{3})^{2}$ \cite{Bloch:2013tra}, as well as the nine-point limit of the double box, which has not previously appeared in the literature.

The numeric values of elliptic Feynman integrals are of course independent of whether we rescale the torus by the period $\omega_1$ or $-\omega_2$; the corresponding modular parameters $\tau=\omega_{2}/\omega_{1}$ and $\tau'=-1/\tau$ are related by a modular $S$-transformation. In particular, for an elliptic integral of the form $\int \mathcal{G}(x,y)\, \dif x/y$ as in \eqref{1dIntofPolylog}, its two normalizations $T^{(1)}$ and $T^{(2)}$, which are obtained by dividing by $\omega_{1}$ and $-\omega_{2}$ respectively, are simply related by $T^{(2)} = \tau' T^{(1)}$. However, this property is not manifest when expressed in terms of elliptic multiple polylogarithms or their symbols. 
Instead, we found that the application of the symbol prime to the two examples in this paper yields symbols of the form
\begin{align}
    & \sum_{ij} \mathcal{S}(f_{i})\otimes \Bigl( \log a_{ij} \otimes \Omega^{(0)}(w_{j}) + \bm{\Omega}_{i}\otimes (2 \pi i\tau)   \Bigr), 
    \intertext{with}
    &\qquad\mathcal{S}'(\bm{\Omega}_{i})=\Omega^{(0)}(w_{j})\otimesprime \log a_{ij}\,.
\end{align}
Not only the modular covariance is manifest in this form, but also the double-periodic invariance and the integrability conditions involving $\tau$. However, one could \emph{not} expect that the application of the symbol prime to the one-fold integral of general polylogarithms, of the form $\int \mathcal{G}(x,y) \,\dif x/y$, yields such a structure.
As a simple counter-example, consider the symbol of the integral $\int_{0}^{c} \log(x+a)\,\dif x/y$, with arbitrary values of $a$ and $c$ as well as the elliptic curve given by \eqref{ellcurve}, which does \emph{not} follow the above structure after applying the symbol prime map.
It would be very interesting to investigate why the two elliptic Feynman integrals we considered in this paper turn out to exhibit such a structure after applying the symbol prime map, and to study whether this property extends to further Feynman integrals.

Already the polylogarithmic symbol has a kernel, which is given by $i\pi$, multiple zeta values (MZVs) and their products with MPLs.
Similarly, also the symbol prime has a kernel. 
As discussed in section \ref{subsec: symbol prime}, all functions in the kernel necessarily depend only on the modular parameter $\tau$, but not all functions that depend only on $\tau$ are in the kernel. 
To see that the kernel can be non-trivial, 
consider the symbol of the \emph{equal-mass} sunrise integral, which is an iterated integral of modular forms and \emph{only} depends on $\tau$:
\begin{align}
    \mathcal{S}\bigl(2\pi\mi T_{\sr}^{(1)} \bigr) &= \biggl[\frac{1}{2\pi\mi} \bigl(2\cEf{-2}{-1}{\infty}-\cEf{-2}{0}{\infty} - \cEf{-2}{\infty}{\infty}\bigr) \biggr]\otimes (2\pi\mi\tau)\:. \label{eq:symbol equal mass sunrise} 
\end{align}
where the $\mathcal{E}_4$ are specific combinations of $\Omega^{(2)}$'s given in \eqref{eq:Ecal-gammat-relation}.
The application of the symbol prime map to the first entry in \eqref{eq:symbol equal mass sunrise} yields $0$.
We leave a 
comprehensive treatment of the kernel of the symbol prime map to a future study.

Another interesting problem is to lift simplified symbols to simplified functions for elliptic multiple polylogarithms. 
As a primary example, let us consider how to lift the simplified symbol \eqref{symbolw1} to a simplified function for the sunrise integral. By rewriting the logarithms in \eqref{symbolw1} in terms of $\Omega^{(1)}$'s, we find a (slightly) simpler expression  for $T^{(1)}_{\sr}$,
\begin{equation}
\begin{aligned}
    T^{(1)}_{\sr} &= 2\cEf{0&-1}{0&-1}{\infty|\tau}-\cEf{0&-1}{0&0}{\infty|\tau} -\cEf{0&-1}{0&\infty}{\infty|\tau} \\ 
      &\quad  - \biggl(2\log\frac{t_{2}}{t_{3}}+\cEf{-1}{-1}{\infty}\biggr)\cEf{0}{0}{\infty|\tau}  \:,
\end{aligned}
\end{equation}
(and a similar expression for $T^{(2)}_{\sr}$), in which the $\mathcal{E}_4$ functions only contain $c=0,-1,\infty$ but not the fourth argument $r$ that occurs in  \eqref{eq:sunrise normalization 1}. 
However, the general prescription of uplifting more complicated elliptic symbols to functions is still underexplored.%
\footnote{In particular, there are symbols that can be written in terms of only logarithms as letters that can nevertheless not be lifted to polylogarithmic functions, only to elliptic ones \cite{Duhr:2020gdd}.
}

Although the simplified symbols of the elliptic Feynman integrals manifest some desired properties, such as double periodicity and modular covariance, after applying the symbol prime map, the singularity structures are not completely manifest. For example, the sunrise integral becomes singular at $m_{i}=0$ as well as at the threshold $p^{2}=(m_{1}+m_{2}+m_{3})^{2}$, as can be seen through a Landau analysis~\cite{OKUN1960261}. One can see the branch cut at $m_{i}=0$ explicitly from eq.\ \eqref{symbolw1} or \eqref{symbolw2}; however, the branch cut at the threshold is not manifest from the symbol. 
In general, the logarithmic letter $\log(a)$ has a logarithmic singularity if $a=0$ or $a=\infty$.
In contrast, the elliptic letter $\Omega^{(1)}(w)$ has a logarithmic singularity at all lattice points, while $\Omega^{(j\geq2)}(w)$ has a logarithmic singularity at all lattice points except for the origin.
However, $w$ is a function of the kinematics; typically, $w=w_c^+=1/\omega_1 \int_{-\infty}^c\dif x/y$, where $c$ is an algebraic function of the kinematics. If the configuration of roots in $y$ does not change as we vary $c$, $w_c^+=0$ if $c=-\infty$ and $w_c^+=\tau$ if $c=+\infty$. However, the configuration of roots may also vary as we vary $c$;  
we leave a comprehensive analysis to a future study.

It would be interesting to apply the techniques used in this paper to bootstrap the symbol of scattering amplitudes or Feynman integrals that can be expressed in terms of elliptic multiple polylogarithms, such as the twelve-point elliptic double box. 
On top of the integrability condition for the final entry $\tau$, which is made manifest by the symbol prime, this requires understanding the integrability condition for the other last entries  \cite{integrability_in_progress}.
Moreover, it requires an educated guess for the alphabet of symbol letters that occur in them.
For six- and seven-point amplitudes in $\mathcal{N}=4$ sYM theory, the symbol alphabet was shown to be given by cluster algebras \cite{Golden:2013xva,Golden:2014pua,Drummond:2017ssj,Drummond:2018dfd,Drummond:2018caf}, and similar techniques have recently been extended to Feynman integrals and amplitudes~\cite{Zhang:2019vnm,He:2020vob,He:2020lcu,Li:2021bwg} containing symbol letters that are given by logarithms of algebraic functions of the kinematics \cite{Chicherin:2020umh,He:2021esx,He:2021non,Drummond:2019cxm,Arkani-Hamed:2019rds,Henke:2019hve,Herderschee:2021dez,Henke:2021ity,Ren:2021ztg}.
It would be interesting to extend these techniques also to the elliptic case.

\acknowledgments

We thank Johannes Brödel, Simon Caron-Huot, Claude Duhr, Zhenjie Li, Robin Marzucca, Andrew McLeod, Cristian Vergu, Matthias Volk, Matt von Hippel and Stefan Weinzierl for interesting discussions as well as Andrew McLeod, Mark Spradlin and  Stefan Weinzierl for comments on the manuscript.
MW thanks the organizers of the conference ``Elliptics '21'', where part of this work was presented.
This work was supported by the research grant  00025445 from Villum Fonden and the ERC starting grant 757978.

\appendix

\section{Calculation of the Unequal-Mass Sunrise Integral}
\label{app:sunrise}

In this appendix, we calculate the unequal-mass sunrise integral in two dimensions in terms of eMPLs. 
This integral was originally calculated in terms of iterated integrals on the moduli space $\overline{\mathcal{M}}_{1,3}$  in \cite{Bogner:2019lfa}. 
We use a Feynman-parameter approach, closely following the calculation of \cite{Broedel:2017siw} in the equal-mass case.

The Feynman parameter representation for the unequal-mass sunrise in two dimensions is (see e.g.\ \cite{Broedel:2017siw})
\begin{equation}
    I_{\sr}= \int_0^\infty \frac{\dif x_{1}\,\dif x_{2}\quad \delta(x_{3}-1)}{-p^{2}\,x_{1}x_{2}x_{3}+(m_{1}^{2}x_{1}+m_{2}^{2}x_{2}+m_{3}^{2}x_{3})(x_{1}x_{2}+x_{1}x_{3}+x_{2}x_{3})} \:.
\end{equation}
Furthermore, we introduce $t_{i}^{2}=m_{i}^{2}/p^{2}$ and integrate out $x_{2}$. This gives
\begin{equation}
    I_{\sr}
    =  \frac{1}{m_{1}^{2}}\int \frac{\dif x}{y}\log \bigl(R_{\sr}(x,y)\bigr),\label{1dintforsr}
\end{equation}
where we have denoted $x_{1}$ by $x$ and $R_{\sr}(x,y)$ is a rational function of $x$ and $y$,
\begin{equation}
    R_{\sr}(x,y)= \frac{t_{3}^{2}+x(t^{2}_{1}+t^{2}_{2}+t^{2}_{3}-1)+t^{2}_{1}x^{2}+t^{2}_{1}y}{t_{3}^{2}+x(t^{2}_{1}+t^{2}_{2}+t^{2}_{3}-1)+t^{2}_{1}x^{2}-t^{2}_{1}y}
\end{equation}
with the elliptic curve defined by
\begin{align}
    (t_{1}^{2}y)^{2}&=\Bigl(t_{1}^{2}x^{2}+(t_{1}^{2}+t_{2}^{2}+t_{3}^{2}-1)x+t^{2}_{3}\Bigr)^{2}-4t^{2}_{2}(1+x)(t_{1}^{2}x+t^{2}_{3})x \\
     &=\Bigl(t_{1}^{2}x^{2}+\bigl(t_{1}^{2}+t_{3}^{2}-(t_{2}-1)^{2}\bigr)x+t_{3}^{2}\Bigr)
     \Bigl(t_{1}^{2}x^{2}+\bigl(t_{1}^{2}+t_{3}^{2}-(t_{2}+1)^{2}\bigr)x+t_{3}^{2}\Bigr) \:. \nonumber 
\end{align}
It is straightforward to rewrite the logarithm in  \eqref{1dintforsr} in terms of $\mathrm{E}_{4}$ functions by  expanding $\partial_{x}\log R_{\sr}(x,y)$ on the $\psi$-basis \eqref{psibasis} and integrating up again:%
\footnote{To fix the integration constant, we can consider the difference of the left and the right hand side in the limit $x\to0$.}
\begin{align}
  \log R_{\sr}(x,y)
    &= -\Ef{-1}{0}{x}+\Ef{-1}{\infty}{x}+\Ef{-1}{-1}{x}+\Ef{-1}{r}{x} \nonumber \\
    &\quad +\frac{1-t^{2}_{2}}{t^{2}_{1}}\Ef{0}{0}{x}
    -\log\frac{t^{2}_{2}}{t_{3}^{2}} \:,
\end{align}
where $r=-t_{3}^{2}/t^{2}_{1}$.

In the conventions introduced in section \ref{sec:2}, we have two rescaled tori. By using \eqref{psitog1} and \eqref{Psikernels}, we find 
\begin{align} \label{lognormalizebyw1}
    \log R_{\sr}(x,y) &= \cEf{-1}{-1}{x|\tau}-\cEf{-1}{0}{x|\tau} + \cEf{-1}{r}{x|\tau}  \nonumber \\
     &\quad -\cEf{-1}{\infty}{x|\tau}
    +4\pi \mi\cEf{0}{0}{x|\tau}  -\log\frac{t^{2}_{2}}{t_{3}^{2}} 
\end{align}  
on the torus $[1:\tau=\omega_{2}/\omega_{1}]$ with coordinates $w$, while, by using the analog of \eqref{psitog1} and \eqref{Psikernels},
\begin{align} \label{lognormalizebyw2}
    \log R_{\sr}(x,y) &= \cEf{-1}{-1}{x|\tau'}-\cEf{-1}{0}{x|\tau'} + \cEf{-1}{r}{x|\tau'}  \nonumber \\
    &\quad -\cEf{-1}{\infty}{x|\tau'}  -\log\frac{t^{2}_{2}}{t_{3}^{2}} 
\end{align}
on the torus $[1:\tau'=-\omega_{1}/\omega_{2}]$ with coordinates $\xi$. The equality of \eqref{lognormalizebyw1} and \eqref{lognormalizebyw2} is given by the $S$-transformation of $g^{(1)}$,
\begin{equation}
    g^{(1)}(w|\tau)= \tau' g^{(1)}(\xi|\tau')+2\pi \mi \xi \:,
\end{equation}
and the following identities on the (rescaled) tori,
\begin{align}
    z_{-1}^{+}-z_{0}^{+}+z_{r}^{+} =0\:, \quad w_{\infty}^{+}=\tau\:, \quad \xi_{\infty}^{+}=-1.
\end{align}
Furthermore, there are three independent variables $\{w_{-1}^{+}, w_{0}^{+},\tau \}$ and $\{\xi_{-1}^{+}, \xi_{0}^{+},\tau' \}$  on each torus, respectively, since \eqref{minfid} here gives
\begin{equation}
    z_{\infty}^{-}=z_{0}^{+}-\frac{1}{2}\omega_{1}\,.
\end{equation}

Now the integration in \eqref{1dintforsr} can be easily performed and gives the result \eqref{eq: sunrise general}--\eqref{eq:sunrise normalization 2} in the main text.

\bibliography{reference}

\providecommand{\noopsort}[1]{}\providecommand{\singleletter}[1]{#1}%

\providecommand{\href}[2]{#2}\begingroup\raggedright\begin{thebibliography}{100}

\bibitem{Chen:1977oja}
K.-T.~Chen, \emph{{Iterated path integrals}},
  \href{https://doi.org/10.1090/S0002-9904-1977-14320-6}{\emph{Bull. Am. Math.
  Soc.} {\bfseries 83} (1977) 831}.

\bibitem{G91b}
A.B.~Goncharov, \emph{{Geometry of Configurations, Polylogarithms, and Motivic
  Cohomology}},
  \href{https://doi.org/http://dx.doi.org/10.1006/aima.1995.1045}{\emph{Adv.
  Math.} {\bfseries 114} (1995) 197}.

\bibitem{Goncharov:1998kja}
A.B.~Goncharov, \emph{{Multiple polylogarithms, cyclotomy and modular
  complexes}}, \href{https://doi.org/10.4310/MRL.1998.v5.n4.a7}{\emph{Math.
  Res. Lett.} {\bfseries 5} (1998) 497}
  [\href{https://arxiv.org/abs/1105.2076}{{\ttfamily 1105.2076}}].

\bibitem{Remiddi:1999ew}
E.~Remiddi and J.A.M.~Vermaseren, \emph{{Harmonic polylogarithms}},
  \href{https://doi.org/10.1142/S0217751X00000367}{\emph{Int. J. Mod. Phys. A}
  {\bfseries 15} (2000) 725}
  [\href{https://arxiv.org/abs/hep-ph/9905237}{{\ttfamily hep-ph/9905237}}].

\bibitem{Borwein:1999js}
J.M.~Borwein, D.M.~Bradley, D.J.~Broadhurst and P.~Lisonek, \emph{{Special
  values of multiple polylogarithms}},
  \href{https://doi.org/10.1090/S0002-9947-00-02616-7}{\emph{Trans. Am. Math.
  Soc.} {\bfseries 353} (2001) 907}
  [\href{https://arxiv.org/abs/math/9910045}{{\ttfamily math/9910045}}].

\bibitem{Moch:2001zr}
S.~Moch, P.~Uwer and S.~Weinzierl, \emph{{Nested sums, expansion of
  transcendental functions and multiscale multiloop integrals}},
  \href{https://doi.org/10.1063/1.1471366}{\emph{J. Math. Phys.} {\bfseries 43}
  (2002) 3363} [\href{https://arxiv.org/abs/hep-ph/0110083}{{\ttfamily
  hep-ph/0110083}}].

\bibitem{Laporta:2004rb}
S.~Laporta and E.~Remiddi, \emph{{Analytic treatment of the two loop equal mass
  sunrise graph}},
  \href{https://doi.org/10.1016/j.nuclphysb.2004.10.044}{\emph{Nucl. Phys. B}
  {\bfseries 704} (2005) 349}
  [\href{https://arxiv.org/abs/hep-ph/0406160}{{\ttfamily hep-ph/0406160}}].

\bibitem{MullerStach:2012az}
S.~Muller-Stach, S.~Weinzierl and R.~Zayadeh, \emph{{From motives to
  differential equations for loop integrals}},
  \href{https://doi.org/10.22323/1.151.0005}{\emph{PoS} {\bfseries LL2012}
  (2012) 005} [\href{https://arxiv.org/abs/1209.3714}{{\ttfamily 1209.3714}}].

\bibitem{brown2011multiple}
F.~Brown and A.~Levin, \emph{{Multiple Elliptic Polylogarithms}},
  \href{https://arxiv.org/abs/1110.6917}{{\ttfamily 1110.6917}}.

\bibitem{Bloch:2013tra}
S.~Bloch and P.~Vanhove, \emph{{The elliptic dilogarithm for the sunset
  graph}}, \href{https://doi.org/10.1016/j.jnt.2014.09.032}{\emph{J. Number
  Theor.} {\bfseries 148} (2015) 328}
  [\href{https://arxiv.org/abs/1309.5865}{{\ttfamily 1309.5865}}].

\bibitem{Adams:2013nia}
L.~Adams, C.~Bogner and S.~Weinzierl, \emph{{The two-loop sunrise graph with
  arbitrary masses}}, \href{https://doi.org/10.1063/1.4804996}{\emph{J. Math.
  Phys.} {\bfseries 54} (2013) 052303}
  [\href{https://arxiv.org/abs/1302.7004}{{\ttfamily 1302.7004}}].

\bibitem{Adams:2014vja}
L.~Adams, C.~Bogner and S.~Weinzierl, \emph{{The two-loop sunrise graph in two
  space-time dimensions with arbitrary masses in terms of elliptic
  dilogarithms}}, \href{https://doi.org/10.1063/1.4896563}{\emph{J. Math.
  Phys.} {\bfseries 55} (2014) 102301}
  [\href{https://arxiv.org/abs/1405.5640}{{\ttfamily 1405.5640}}].

\bibitem{Adams:2015gva}
L.~Adams, C.~Bogner and S.~Weinzierl, \emph{{The two-loop sunrise integral
  around four space-time dimensions and generalisations of the Clausen and
  Glaisher functions towards the elliptic case}},
  \href{https://doi.org/10.1063/1.4926985}{\emph{J. Math. Phys.} {\bfseries 56}
  (2015) 072303} [\href{https://arxiv.org/abs/1504.03255}{{\ttfamily
  1504.03255}}].

\bibitem{Adams:2015ydq}
L.~Adams, C.~Bogner and S.~Weinzierl, \emph{{The iterated structure of the
  all-order result for the two-loop sunrise integral}},
  \href{https://doi.org/10.1063/1.4944722}{\emph{J. Math. Phys.} {\bfseries 57}
  (2016) 032304} [\href{https://arxiv.org/abs/1512.05630}{{\ttfamily
  1512.05630}}].

\bibitem{Adams:2016xah}
L.~Adams, C.~Bogner, A.~Schweitzer and S.~Weinzierl, \emph{{The kite integral
  to all orders in terms of elliptic polylogarithms}},
  \href{https://doi.org/10.1063/1.4969060}{\emph{J. Math. Phys.} {\bfseries 57}
  (2016) 122302} [\href{https://arxiv.org/abs/1607.01571}{{\ttfamily
  1607.01571}}].

\bibitem{Adams:2017ejb}
L.~Adams and S.~Weinzierl, \emph{{Feynman integrals and iterated integrals of
  modular forms}},
  \href{https://doi.org/10.4310/CNTP.2018.v12.n2.a1}{\emph{Commun. Num. Theor.
  Phys.} {\bfseries 12} (2018) 193}
  [\href{https://arxiv.org/abs/1704.08895}{{\ttfamily 1704.08895}}].

\bibitem{Adams:2017tga}
L.~Adams, E.~Chaubey and S.~Weinzierl, \emph{{Simplifying Differential
  Equations for Multiscale Feynman Integrals beyond Multiple Polylogarithms}},
  \href{https://doi.org/10.1103/PhysRevLett.118.141602}{\emph{Phys. Rev. Lett.}
  {\bfseries 118} (2017) 141602}
  [\href{https://arxiv.org/abs/1702.04279}{{\ttfamily 1702.04279}}].

\bibitem{Bogner:2017vim}
C.~Bogner, A.~Schweitzer and S.~Weinzierl, \emph{{Analytic continuation and
  numerical evaluation of the kite integral and the equal mass sunrise
  integral}},
  \href{https://doi.org/10.1016/j.nuclphysb.2017.07.008}{\emph{Nucl. Phys. B}
  {\bfseries 922} (2017) 528}
  [\href{https://arxiv.org/abs/1705.08952}{{\ttfamily 1705.08952}}].

\bibitem{Broedel:2017kkb}
J.~Broedel, C.~Duhr, F.~Dulat and L.~Tancredi, \emph{{Elliptic polylogarithms
  and iterated integrals on elliptic curves. Part I: general formalism}},
  \href{https://doi.org/10.1007/JHEP05(2018)093}{\emph{JHEP} {\bfseries 05}
  (2018) 093} [\href{https://arxiv.org/abs/1712.07089}{{\ttfamily
  1712.07089}}].

\bibitem{Broedel:2017siw}
J.~Broedel, C.~Duhr, F.~Dulat and L.~Tancredi, \emph{{Elliptic polylogarithms
  and iterated integrals on elliptic curves II: an application to the sunrise
  integral}}, \href{https://doi.org/10.1103/PhysRevD.97.116009}{\emph{Phys.
  Rev. D} {\bfseries 97} (2018) 116009}
  [\href{https://arxiv.org/abs/1712.07095}{{\ttfamily 1712.07095}}].

\bibitem{Remiddi:2017har}
E.~Remiddi and L.~Tancredi, \emph{{An Elliptic Generalization of Multiple
  Polylogarithms}},
  \href{https://doi.org/10.1016/j.nuclphysb.2017.10.007}{\emph{Nucl. Phys. B}
  {\bfseries 925} (2017) 212}
  [\href{https://arxiv.org/abs/1709.03622}{{\ttfamily 1709.03622}}].

\bibitem{Chen:2017soz}
L.-B.~Chen, J.~Jiang and C.-F.~Qiao, \emph{{Two-Loop integrals for CP-even
  heavy quarkonium production and decays: Elliptic Sectors}},
  \href{https://doi.org/10.1007/JHEP04(2018)080}{\emph{JHEP} {\bfseries 04}
  (2018) 080} [\href{https://arxiv.org/abs/1712.03516}{{\ttfamily
  1712.03516}}].

\bibitem{Bourjaily:2017bsb}
J.L.~Bourjaily, A.J.~McLeod, M.~Spradlin, M.~von Hippel and M.~Wilhelm,
  \emph{{Elliptic Double-Box Integrals: Massless Scattering Amplitudes beyond
  Polylogarithms}},
  \href{https://doi.org/10.1103/PhysRevLett.120.121603}{\emph{Phys. Rev. Lett.}
  {\bfseries 120} (2018) 121603}
  [\href{https://arxiv.org/abs/1712.02785}{{\ttfamily 1712.02785}}].

\bibitem{Adams:2018yfj}
L.~Adams and S.~Weinzierl, \emph{{The $\varepsilon$-form of the differential
  equations for Feynman integrals in the elliptic case}},
  \href{https://doi.org/10.1016/j.physletb.2018.04.002}{\emph{Phys. Lett. B}
  {\bfseries 781} (2018) 270}
  [\href{https://arxiv.org/abs/1802.05020}{{\ttfamily 1802.05020}}].

\bibitem{Broedel:2018iwv}
J.~Broedel, C.~Duhr, F.~Dulat, B.~Penante and L.~Tancredi, \emph{{Elliptic
  symbol calculus: from elliptic polylogarithms to iterated integrals of
  Eisenstein series}},
  \href{https://doi.org/10.1007/JHEP08(2018)014}{\emph{JHEP} {\bfseries 08}
  (2018) 014} [\href{https://arxiv.org/abs/1803.10256}{{\ttfamily
  1803.10256}}].

\bibitem{Broedel:2018qkq}
J.~Broedel, C.~Duhr, F.~Dulat, B.~Penante and L.~Tancredi, \emph{{Elliptic
  Feynman integrals and pure functions}},
  \href{https://doi.org/10.1007/JHEP01(2019)023}{\emph{JHEP} {\bfseries 01}
  (2019) 023} [\href{https://arxiv.org/abs/1809.10698}{{\ttfamily
  1809.10698}}].

\bibitem{Honemann:2018mrb}
I.~H\"onemann, K.~Tempest and S.~Weinzierl, \emph{{Electron self-energy in QED
  at two loops revisited}},
  \href{https://doi.org/10.1103/PhysRevD.98.113008}{\emph{Phys. Rev. D}
  {\bfseries 98} (2018) 113008}
  [\href{https://arxiv.org/abs/1811.09308}{{\ttfamily 1811.09308}}].

\bibitem{Bogner:2019lfa}
C.~Bogner, S.~M\"uller-Stach and S.~Weinzierl, \emph{{The unequal mass sunrise
  integral expressed through iterated integrals on $\overline{\mathcal
  M}_{1,3}$}},
  \href{https://doi.org/10.1016/j.nuclphysb.2020.114991}{\emph{Nucl. Phys. B}
  {\bfseries 954} (2020) 114991}
  [\href{https://arxiv.org/abs/1907.01251}{{\ttfamily 1907.01251}}].

\bibitem{Broedel:2019hyg}
J.~Broedel, C.~Duhr, F.~Dulat, B.~Penante and L.~Tancredi, \emph{{Elliptic
  polylogarithms and Feynman parameter integrals}},
  \href{https://doi.org/10.1007/JHEP05(2019)120}{\emph{JHEP} {\bfseries 05}
  (2019) 120} [\href{https://arxiv.org/abs/1902.09971}{{\ttfamily
  1902.09971}}].

\bibitem{Duhr:2019rrs}
C.~Duhr and L.~Tancredi, \emph{{Algorithms and tools for iterated Eisenstein
  integrals}}, \href{https://doi.org/10.1007/JHEP02(2020)105}{\emph{JHEP}
  {\bfseries 02} (2020) 105}
  [\href{https://arxiv.org/abs/1912.00077}{{\ttfamily 1912.00077}}].

\bibitem{Walden:2020odh}
M.~Walden and S.~Weinzierl, \emph{{Numerical evaluation of iterated integrals
  related to elliptic Feynman integrals}},
  \href{https://doi.org/10.1016/j.cpc.2021.108020}{\emph{Comput. Phys. Commun.}
  {\bfseries 265} (2021) 108020}
  [\href{https://arxiv.org/abs/2010.05271}{{\ttfamily 2010.05271}}].

\bibitem{Weinzierl:2020fyx}
S.~Weinzierl, \emph{{Modular transformations of elliptic Feynman integrals}},
  \href{https://doi.org/10.1016/j.nuclphysb.2021.115309}{\emph{Nucl. Phys. B}
  {\bfseries 964} (2021) 115309}
  [\href{https://arxiv.org/abs/2011.07311}{{\ttfamily 2011.07311}}].

\bibitem{Kristensson:2021ani}
A.~Kristensson, M.~Wilhelm and C.~Zhang, \emph{{Elliptic Double Box and
  Symbology Beyond Polylogarithms}},
  \href{https://doi.org/10.1103/PhysRevLett.127.251603}{\emph{Phys. Rev. Lett.}
  {\bfseries 127} (2021) 251603}
  [\href{https://arxiv.org/abs/2106.14902}{{\ttfamily 2106.14902}}].

\bibitem{Frellesvig:2021hkr}
H.~Frellesvig, \emph{{On epsilon factorized differential equations for elliptic
  Feynman integrals}},
  \href{https://doi.org/10.1007/JHEP03(2022)079}{\emph{JHEP} {\bfseries 03}
  (2022) 079} [\href{https://arxiv.org/abs/2110.07968}{{\ttfamily
  2110.07968}}].

\bibitem{SABRY1962401}
A.~Sabry, \emph{Fourth order spectral functions for the electron propagator},
  \href{https://doi.org/https://doi.org/10.1016/0029-5582(62)90535-7}{\emph{Nuclear
  Physics} {\bfseries 33} (1962) 401 }.

\bibitem{Broadhurst:1993mw}
D.J.~Broadhurst, J.~Fleischer and O.V.~Tarasov, \emph{{Two loop two point
  functions with masses: Asymptotic expansions and Taylor series, in any
  dimension}}, \href{https://doi.org/10.1007/BF01474625}{\emph{Z. Phys. C}
  {\bfseries 60} (1993) 287}
  [\href{https://arxiv.org/abs/hep-ph/9304303}{{\ttfamily hep-ph/9304303}}].

\bibitem{Muller-Stach:2011qkg}
S.~M\"uller-Stach, S.~Weinzierl and R.~Zayadeh, \emph{{A Second-Order
  Differential Equation for the Two-Loop Sunrise Graph with Arbitrary Masses}},
  \href{https://doi.org/10.4310/CNTP.2012.v6.n1.a5}{\emph{Commun. Num. Theor.
  Phys.} {\bfseries 6} (2012) 203}
  [\href{https://arxiv.org/abs/1112.4360}{{\ttfamily 1112.4360}}].

\bibitem{Remiddi:2013joa}
E.~Remiddi and L.~Tancredi, \emph{{Schouten identities for Feynman graph
  amplitudes; The Master Integrals for the two-loop massive sunrise graph}},
  \href{https://doi.org/10.1016/j.nuclphysb.2014.01.009}{\emph{Nucl. Phys. B}
  {\bfseries 880} (2014) 343}
  [\href{https://arxiv.org/abs/1311.3342}{{\ttfamily 1311.3342}}].

\bibitem{Bloch:2014qca}
S.~Bloch, M.~Kerr and P.~Vanhove, \emph{{A Feynman Integral via Higher Normal
  Functions}}, \href{https://doi.org/10.1112/S0010437X15007472}{\emph{Compos.
  Math.} {\bfseries 151} (2015) 2329}
  [\href{https://arxiv.org/abs/1406.2664}{{\ttfamily 1406.2664}}].

\bibitem{Bloch:2016izu}
S.~Bloch, M.~Kerr and P.~Vanhove, \emph{{Local Mirror Symmetry and the Sunset
  Feynman Integral}},
  \href{https://doi.org/10.4310/ATMP.2017.v21.n6.a1}{\emph{Adv. Theor. Math.
  Phys.} {\bfseries 21} (2017) 1373}
  [\href{https://arxiv.org/abs/1601.08181}{{\ttfamily 1601.08181}}].

\bibitem{Remiddi:2016gno}
E.~Remiddi and L.~Tancredi, \emph{{Differential equations and dispersion
  relations for Feynman amplitudes. The two-loop massive sunrise and the kite
  integral}},
  \href{https://doi.org/10.1016/j.nuclphysb.2016.04.013}{\emph{Nucl. Phys. B}
  {\bfseries 907} (2016) 400}
  [\href{https://arxiv.org/abs/1602.01481}{{\ttfamily 1602.01481}}].

\bibitem{Brown:2009ta}
F.C.S.~Brown, \emph{{On the periods of some Feynman integrals}},
  \href{https://arxiv.org/abs/0910.0114}{{\ttfamily 0910.0114}}.

\bibitem{Brown:2010bw}
F.~Brown and O.~Schnetz, \emph{{A K3 in $\phi^4$}},
  \href{https://doi.org/10.1215/00127094-1644201}{\emph{Duke Math. J.}
  {\bfseries 161} (2012) 1817}
  [\href{https://arxiv.org/abs/1006.4064}{{\ttfamily 1006.4064}}].

\bibitem{Bourjaily:2018yfy}
J.L.~Bourjaily, A.J.~McLeod, M.~von Hippel and M.~Wilhelm, \emph{{Bounded
  Collection of Feynman Integral Calabi-Yau Geometries}},
  \href{https://doi.org/10.1103/PhysRevLett.122.031601}{\emph{Phys. Rev. Lett.}
  {\bfseries 122} (2019) 031601}
  [\href{https://arxiv.org/abs/1810.07689}{{\ttfamily 1810.07689}}].

\bibitem{Bourjaily:2018ycu}
J.L.~Bourjaily, Y.-H.~He, A.J.~McLeod, M.~von Hippel and M.~Wilhelm,
  \emph{{Traintracks through Calabi-Yau Manifolds: Scattering Amplitudes beyond
  Elliptic Polylogarithms}},
  \href{https://doi.org/10.1103/PhysRevLett.121.071603}{\emph{Phys. Rev. Lett.}
  {\bfseries 121} (2018) 071603}
  [\href{https://arxiv.org/abs/1805.09326}{{\ttfamily 1805.09326}}].

\bibitem{Festi:2018qip}
D.~Festi and D.~van Straten, \emph{{Bhabha Scattering and a special pencil of
  K3 surfaces}},
  \href{https://doi.org/10.4310/CNTP.2019.v13.n2.a4}{\emph{Commun. Num. Theor.
  Phys.} {\bfseries 13} (2019) 463}
  [\href{https://arxiv.org/abs/1809.04970}{{\ttfamily 1809.04970}}].

\bibitem{Broedel:2019kmn}
J.~Broedel, C.~Duhr, F.~Dulat, R.~Marzucca, B.~Penante and L.~Tancredi,
  \emph{{An analytic solution for the equal-mass banana graph}},
  \href{https://doi.org/10.1007/JHEP09(2019)112}{\emph{JHEP} {\bfseries 09}
  (2019) 112} [\href{https://arxiv.org/abs/1907.03787}{{\ttfamily
  1907.03787}}].

\bibitem{Besier:2019hqd}
M.~Besier, D.~Festi, M.~Harrison and B.~Naskrecki, \emph{{Arithmetic and
  geometry of a K3 surface emerging from virtual corrections to
  Drell\textendash{}Yan scattering}},
  \href{https://doi.org/10.4310/CNTP.2020.v14.n4.a4}{\emph{Commun. Num. Theor.
  Phys.} {\bfseries 14} (2020) 863}
  [\href{https://arxiv.org/abs/1908.01079}{{\ttfamily 1908.01079}}].

\bibitem{Bourjaily:2019hmc}
J.L.~Bourjaily, A.J.~McLeod, C.~Vergu, M.~Volk, M.~Von~Hippel and M.~Wilhelm,
  \emph{{Embedding Feynman Integral (Calabi-Yau) Geometries in Weighted
  Projective Space}},
  \href{https://doi.org/10.1007/JHEP01(2020)078}{\emph{JHEP} {\bfseries 01}
  (2020) 078} [\href{https://arxiv.org/abs/1910.01534}{{\ttfamily
  1910.01534}}].

\bibitem{Vergu:2020uur}
C.~Vergu and M.~Volk, \emph{{Traintrack Calabi-Yaus from Twistor Geometry}},
  \href{https://doi.org/10.1007/JHEP07(2020)160}{\emph{JHEP} {\bfseries 07}
  (2020) 160} [\href{https://arxiv.org/abs/2005.08771}{{\ttfamily
  2005.08771}}].

\bibitem{mirrors_and_sunsets}
C.F.~Doran, A.Y.~Novoseltsev and P.~Vanhove, ``{Mirroring Towers: Calabi-Yau
  Geometry of the Multiloop Feynman Sunset Integrals}.'' To appear.

\bibitem{Broadhurst:1987ei}
D.J.~Broadhurst, \emph{{The Master Two Loop Diagram With Masses}},
  \href{https://doi.org/10.1007/BF01551921}{\emph{Z. Phys. C} {\bfseries 47}
  (1990) 115}.

\bibitem{Adams:2018kez}
L.~Adams, E.~Chaubey and S.~Weinzierl, \emph{{Analytic results for the planar
  double box integral relevant to top-pair production with a closed top loop}},
  \href{https://doi.org/10.1007/JHEP10(2018)206}{\emph{JHEP} {\bfseries 10}
  (2018) 206} [\href{https://arxiv.org/abs/1806.04981}{{\ttfamily
  1806.04981}}].

\bibitem{Adams:2018bsn}
L.~Adams, E.~Chaubey and S.~Weinzierl, \emph{{Planar Double Box Integral for
  Top Pair Production with a Closed Top Loop to all orders in the Dimensional
  Regularization Parameter}},
  \href{https://doi.org/10.1103/PhysRevLett.121.142001}{\emph{Phys. Rev. Lett.}
  {\bfseries 121} (2018) 142001}
  [\href{https://arxiv.org/abs/1804.11144}{{\ttfamily 1804.11144}}].

\bibitem{Huang:2013kh}
R.~Huang and Y.~Zhang, \emph{{On Genera of Curves from High-loop Generalized
  Unitarity Cuts}}, \href{https://doi.org/10.1007/JHEP04(2013)080}{\emph{JHEP}
  {\bfseries 04} (2013) 080} [\href{https://arxiv.org/abs/1302.1023}{{\ttfamily
  1302.1023}}].

\bibitem{Klemm:2019dbm}
A.~Klemm, C.~Nega and R.~Safari, \emph{{The $l$-loop Banana Amplitude from GKZ
  Systems and relative Calabi-Yau Periods}},
  \href{https://doi.org/10.1007/JHEP04(2020)088}{\emph{JHEP} {\bfseries 04}
  (2020) 088} [\href{https://arxiv.org/abs/1912.06201}{{\ttfamily
  1912.06201}}].

\bibitem{Bonisch:2020qmm}
K.~B\"onisch, F.~Fischbach, A.~Klemm, C.~Nega and R.~Safari, \emph{{Analytic
  structure of all loop banana integrals}},
  \href{https://doi.org/10.1007/JHEP05(2021)066}{\emph{JHEP} {\bfseries 05}
  (2021) 066} [\href{https://arxiv.org/abs/2008.10574}{{\ttfamily
  2008.10574}}].

\bibitem{Bonisch:2021yfw}
K.~B\"onisch, C.~Duhr, F.~Fischbach, A.~Klemm and C.~Nega, \emph{{Feynman
  Integrals in Dimensional Regularization and Extensions of Calabi-Yau
  Motives}},  \href{https://arxiv.org/abs/2108.05310}{{\ttfamily 2108.05310}}.

\bibitem{Muller:2022gec}
H.~M\"uller and S.~Weinzierl, \emph{{A Feynman integral depending on two
  elliptic curves}},  \href{https://arxiv.org/abs/2205.04818}{{\ttfamily
  2205.04818}}.

\bibitem{Chaubey:2022hlr}
E.~Chaubey, M.~Kaur and A.~Shivaji, \emph{{Master integrals for ${\mathcal
  O}(\alpha \alpha_s)$ corrections to $H \to ZZ^*$}},
  \href{https://arxiv.org/abs/2205.06339}{{\ttfamily 2205.06339}}.

\bibitem{Bourjaily:2022bwx}
J.L.~Bourjaily et~al., \emph{{Functions Beyond Multiple Polylogarithms for
  Precision Collider Physics}},  in \emph{{2022 Snowmass Summer Study}}, 3,
  2022 [\href{https://arxiv.org/abs/2203.07088}{{\ttfamily 2203.07088}}].

\bibitem{Gonch2}
A.B.~Goncharov, \emph{Galois symmetries of fundamental groupoids and
  noncommutative geometry},
  \href{https://doi.org/10.1215/S0012-7094-04-12822-2}{\emph{Duke Math. J.}
  {\bfseries 128} (2005) 209}
  [\href{https://arxiv.org/abs/math/0208144}{{\ttfamily math/0208144}}].

\bibitem{Goncharov:2010jf}
A.B.~Goncharov, M.~Spradlin, C.~Vergu and A.~Volovich, \emph{{Classical
  Polylogarithms for Amplitudes and Wilson Loops}},
  \href{https://doi.org/10.1103/PhysRevLett.105.151605}{\emph{Phys. Rev. Lett.}
  {\bfseries 105} (2010) 151605}
  [\href{https://arxiv.org/abs/1006.5703}{{\ttfamily 1006.5703}}].

\bibitem{Brown:2011ik}
F.~Brown, \emph{{On the decomposition of motivic multiple zeta values}},
  {\emph{{Adv. Studies in Pure Math.}} {\bfseries 63} (2012) 31}
  [\href{https://arxiv.org/abs/1102.1310}{{\ttfamily 1102.1310}}].

\bibitem{Duhr:2011zq}
C.~Duhr, H.~Gangl and J.R.~Rhodes, \emph{{From polygons and symbols to
  polylogarithmic functions}},
  \href{https://doi.org/10.1007/JHEP10(2012)075}{\emph{JHEP} {\bfseries 10}
  (2012) 075} [\href{https://arxiv.org/abs/1110.0458}{{\ttfamily 1110.0458}}].

\bibitem{Duhr:2012fh}
C.~Duhr, \emph{{Hopf algebras, coproducts and symbols: an application to Higgs
  boson amplitudes}},
  \href{https://doi.org/10.1007/JHEP08(2012)043}{\emph{JHEP} {\bfseries 08}
  (2012) 043} [\href{https://arxiv.org/abs/1203.0454}{{\ttfamily 1203.0454}}].

\bibitem{Gaiotto:2011dt}
D.~Gaiotto, J.~Maldacena, A.~Sever and P.~Vieira, \emph{{Pulling the straps of
  polygons}}, \href{https://doi.org/10.1007/JHEP12(2011)011}{\emph{JHEP}
  {\bfseries 12} (2011) 011} [\href{https://arxiv.org/abs/1102.0062}{{\ttfamily
  1102.0062}}].

\bibitem{Steinmann}
O.~Steinmann, \emph{{\"Uber den Zusammenhang zwischen den Wightmanfunktionen
  und der retardierten Kommutatoren}}, {\emph{Helv. Physica Acta} {\bfseries
  33} (1960) 257}.

\bibitem{Steinmann2}
O.~Steinmann, \emph{{Wightman-Funktionen und retardierten Kommutatoren. II}},
  {\emph{Helv. Physica Acta} {\bfseries 33} (1960) 347}.

\bibitem{Dixon:2011pw}
L.J.~Dixon, J.M.~Drummond and J.M.~Henn, \emph{{Bootstrapping the three-loop
  hexagon}}, \href{https://doi.org/10.1007/JHEP11(2011)023}{\emph{JHEP} (2011)
  023} [\href{https://arxiv.org/abs/1108.4461}{{\ttfamily 1108.4461}}].

\bibitem{Dixon:2011nj}
L.J.~Dixon, J.M.~Drummond and J.M.~Henn, \emph{{Analytic result for the
  two-loop six-point NMHV amplitude in $\mathcal{N}=4$ super Yang-Mills
  theory}}, \href{https://doi.org/10.1007/JHEP01(2012)024}{\emph{JHEP}
  {\bfseries 01} (2012) 024} [\href{https://arxiv.org/abs/1111.1704}{{\ttfamily
  1111.1704}}].

\bibitem{Dixon:2013eka}
L.J.~Dixon, J.M.~Drummond, M.~von Hippel and J.~Pennington, \emph{{Hexagon
  functions and the three-loop remainder function}},
  \href{https://doi.org/10.1007/JHEP12(2013)049}{\emph{JHEP} {\bfseries 12}
  (2013) 049} [\href{https://arxiv.org/abs/1308.2276}{{\ttfamily 1308.2276}}].

\bibitem{Dixon:2014voa}
L.J.~Dixon, J.M.~Drummond, C.~Duhr and J.~Pennington, \emph{{The four-loop
  remainder function and multi-Regge behavior at NNLLA in planar $\mathcal{N} =
  4$ super-Yang-Mills theory}},
  \href{https://doi.org/10.1007/JHEP06(2014)116}{\emph{JHEP} {\bfseries 06}
  (2014) 116} [\href{https://arxiv.org/abs/1402.3300}{{\ttfamily 1402.3300}}].

\bibitem{Dixon:2014iba}
L.J.~Dixon and M.~von Hippel, \emph{{Bootstrapping an NMHV amplitude through
  three loops}}, \href{https://doi.org/10.1007/JHEP10(2014)065}{\emph{JHEP}
  {\bfseries 10} (2014) 065} [\href{https://arxiv.org/abs/1408.1505}{{\ttfamily
  1408.1505}}].

\bibitem{Drummond:2014ffa}
J.M.~Drummond, G.~Papathanasiou and M.~Spradlin, \emph{{A Symbol of Uniqueness:
  The Cluster Bootstrap for the 3-Loop MHV Heptagon}},
  \href{https://doi.org/10.1007/JHEP03(2015)072}{\emph{JHEP} {\bfseries 03}
  (2015) 072} [\href{https://arxiv.org/abs/1412.3763}{{\ttfamily 1412.3763}}].

\bibitem{Dixon:2015iva}
L.J.~Dixon, M.~von Hippel and A.J.~McLeod, \emph{{The four-loop six-gluon NMHV
  ratio function}}, \href{https://doi.org/10.1007/JHEP01(2016)053}{\emph{JHEP}
  {\bfseries 01} (2016) 053}
  [\href{https://arxiv.org/abs/1509.08127}{{\ttfamily 1509.08127}}].

\bibitem{Caron-Huot:2016owq}
S.~Caron-Huot, L.J.~Dixon, A.~McLeod and M.~von Hippel, \emph{{Bootstrapping a
  Five-Loop Amplitude Using Steinmann Relations}},
  \href{https://doi.org/10.1103/PhysRevLett.117.241601}{\emph{Phys. Rev. Lett.}
  {\bfseries 117} (2016) 241601}
  [\href{https://arxiv.org/abs/1609.00669}{{\ttfamily 1609.00669}}].

\bibitem{Dixon:2016apl}
L.J.~Dixon, M.~von Hippel, A.J.~McLeod and J.~Trnka, \emph{{Multi-loop
  positivity of the planar $ \mathcal{N} $ = 4 SYM six-point amplitude}},
  \href{https://doi.org/10.1007/JHEP02(2017)112}{\emph{JHEP} {\bfseries 02}
  (2017) 112} [\href{https://arxiv.org/abs/1611.08325}{{\ttfamily
  1611.08325}}].

\bibitem{Dixon:2016nkn}
L.J.~Dixon, J.~Drummond, T.~Harrington, A.J.~McLeod, G.~Papathanasiou and
  M.~Spradlin, \emph{{Heptagons from the Steinmann Cluster Bootstrap}},
  \href{https://doi.org/10.1007/JHEP02(2017)137}{\emph{JHEP} {\bfseries 02}
  (2017) 137} [\href{https://arxiv.org/abs/1612.08976}{{\ttfamily
  1612.08976}}].

\bibitem{Drummond:2018caf}
J.~Drummond, J.~Foster, {\"{O}}.~G{\"{u}}rdo{\u{g}}an and G.~Papathanasiou,
  \emph{{Cluster adjacency and the four-loop NMHV heptagon}},
  \href{https://doi.org/10.1007/JHEP03(2019)087}{\emph{JHEP} {\bfseries 03}
  (2019) 087} [\href{https://arxiv.org/abs/1812.04640}{{\ttfamily
  1812.04640}}].

\bibitem{Caron-Huot:2019vjl}
S.~Caron-Huot, L.J.~Dixon, F.~Dulat, M.~von Hippel, A.J.~McLeod and
  G.~Papathanasiou, \emph{{Six-Gluon amplitudes in planar $ \mathcal{N} $ = 4
  super-Yang-Mills theory at six and seven loops}},
  \href{https://doi.org/10.1007/JHEP08(2019)016}{\emph{JHEP} {\bfseries 08}
  (2019) 016} [\href{https://arxiv.org/abs/1903.10890}{{\ttfamily
  1903.10890}}].

\bibitem{Li:2016ctv}
Y.~Li and H.X.~Zhu, \emph{{Bootstrapping Rapidity Anomalous Dimensions for
  Transverse-Momentum Resummation}},
  \href{https://doi.org/10.1103/PhysRevLett.118.022004}{\emph{Phys. Rev. Lett.}
  {\bfseries 118} (2017) 022004}
  [\href{https://arxiv.org/abs/1604.01404}{{\ttfamily 1604.01404}}].

\bibitem{Almelid:2017qju}
O.~Almelid, C.~Duhr, E.~Gardi, A.~McLeod and C.D.~White, \emph{{Bootstrapping
  the QCD soft anomalous dimension}},
  \href{https://doi.org/10.1007/JHEP09(2017)073}{\emph{JHEP} {\bfseries 09}
  (2017) 073} [\href{https://arxiv.org/abs/1706.10162}{{\ttfamily
  1706.10162}}].

\bibitem{Dixon:2020bbt}
L.J.~Dixon, A.J.~McLeod and M.~Wilhelm, \emph{{A Three-Point Form Factor
  Through Five Loops}},
  \href{https://doi.org/10.1007/JHEP04(2021)147}{\emph{JHEP} {\bfseries 04}
  (2021) 147} [\href{https://arxiv.org/abs/2012.12286}{{\ttfamily
  2012.12286}}].

\bibitem{Dixon:2022rse}
L.J.~Dixon, O.~Gurdogan, A.J.~McLeod and M.~Wilhelm, \emph{{Bootstrapping a
  Stress-Tensor Form Factor through Eight Loops}},
  \href{https://arxiv.org/abs/2204.11901}{{\ttfamily 2204.11901}}.

\bibitem{Abreu:2020jxa}
S.~Abreu, H.~Ita, F.~Moriello, B.~Page, W.~Tschernow and M.~Zeng,
  \emph{{Two-Loop Integrals for Planar Five-Point One-Mass Processes}},
  \href{https://doi.org/10.1007/JHEP11(2020)117}{\emph{JHEP} {\bfseries 11}
  (2020) 117} [\href{https://arxiv.org/abs/2005.04195}{{\ttfamily
  2005.04195}}].

\bibitem{Chicherin:2020umh}
D.~Chicherin, J.M.~Henn and G.~Papathanasiou, \emph{{Cluster algebras for
  Feynman integrals}},
  \href{https://doi.org/10.1103/PhysRevLett.126.091603}{\emph{Phys. Rev. Lett.}
  {\bfseries 126} (2021) 091603}
  [\href{https://arxiv.org/abs/2012.12285}{{\ttfamily 2012.12285}}].

\bibitem{Dixon:2012yy}
L.J.~Dixon, C.~Duhr and J.~Pennington, \emph{{Single-valued harmonic
  polylogarithms and the multi-Regge limit}},
  \href{https://doi.org/10.1007/JHEP10(2012)074}{\emph{JHEP} {\bfseries 10}
  (2012) 074} [\href{https://arxiv.org/abs/1207.0186}{{\ttfamily 1207.0186}}].

\bibitem{Chestnov:2020ifg}
V.~Chestnov and G.~Papathanasiou, \emph{{Hexagon bootstrap in the double
  scaling limit}}, \href{https://doi.org/10.1007/JHEP09(2021)007}{\emph{JHEP}
  {\bfseries 09} (2021) 007}
  [\href{https://arxiv.org/abs/2012.15855}{{\ttfamily 2012.15855}}].

\bibitem{He:2021fwf}
S.~He, Z.~Li, Y.~Tang and Q.~Yang, \emph{{Bootstrapping octagons in reduced
  kinematics from A$_{2}$ cluster algebras}},
  \href{https://doi.org/10.1007/JHEP10(2021)084}{\emph{JHEP} {\bfseries 10}
  (2021) 084} [\href{https://arxiv.org/abs/2106.03709}{{\ttfamily
  2106.03709}}].

\bibitem{Drummond:2017ssj}
J.~Drummond, J.~Foster and {\"{O}}.~G{\"{u}}rdo{\u{g}}an, \emph{{Cluster
  Adjacency Properties of Scattering Amplitudes in $\mathcal{N}=4$
  Supersymmetric Yang-Mills Theory}},
  \href{https://doi.org/10.1103/PhysRevLett.120.161601}{\emph{Phys. Rev. Lett.}
  {\bfseries 120} (2018) 161601}
  [\href{https://arxiv.org/abs/1710.10953}{{\ttfamily 1710.10953}}].

\bibitem{Chicherin:2017dob}
D.~Chicherin, J.~Henn and V.~Mitev, \emph{{Bootstrapping pentagon functions}},
  \href{https://doi.org/10.1007/JHEP05(2018)164}{\emph{JHEP} {\bfseries 05}
  (2018) 164} [\href{https://arxiv.org/abs/1712.09610}{{\ttfamily
  1712.09610}}].

\bibitem{Caron-Huot:2018dsv}
S.~Caron-Huot, L.J.~Dixon, M.~von Hippel, A.J.~McLeod and G.~Papathanasiou,
  \emph{{The Double Pentaladder Integral to All Orders}},
  \href{https://doi.org/10.1007/JHEP07(2018)170}{\emph{JHEP} {\bfseries 07}
  (2018) 170} [\href{https://arxiv.org/abs/1806.01361}{{\ttfamily
  1806.01361}}].

\bibitem{Henn:2018cdp}
J.~Henn, E.~Herrmann and J.~Parra-Martinez, \emph{{Bootstrapping two-loop
  Feynman integrals for planar $ \mathcal{N}=4 $ sYM}},
  \href{https://doi.org/10.1007/JHEP10(2018)059}{\emph{JHEP} {\bfseries 10}
  (2018) 059} [\href{https://arxiv.org/abs/1806.06072}{{\ttfamily
  1806.06072}}].

\bibitem{He:2021non}
S.~He, Z.~Li and Q.~Yang, \emph{{Truncated cluster algebras and Feynman
  integrals with algebraic letters}},
  \href{https://arxiv.org/abs/2106.09314}{{\ttfamily 2106.09314}}.

\bibitem{He:2021eec}
S.~He, Z.~Li and Q.~Yang, \emph{{Kinematics, cluster algebras and Feynman
  integrals}},  \href{https://arxiv.org/abs/2112.11842}{{\ttfamily
  2112.11842}}.

\bibitem{Heller:2019gkq}
M.~Heller, A.~von Manteuffel and R.M.~Schabinger, \emph{{Multiple
  polylogarithms with algebraic arguments and the two-loop EW-QCD Drell-Yan
  master integrals}},
  \href{https://doi.org/10.1103/PhysRevD.102.016025}{\emph{Phys. Rev. D}
  {\bfseries 102} (2020) 016025}
  [\href{https://arxiv.org/abs/1907.00491}{{\ttfamily 1907.00491}}].

\bibitem{Heller:2021gun}
M.~Heller, \emph{{Planar two-loop integrals for $\mathbf{\mu e}$ scattering in
  QED with finite lepton masses}},
  \href{https://arxiv.org/abs/2105.08046}{{\ttfamily 2105.08046}}.

\bibitem{Duhr:2021fhk}
C.~Duhr, V.A.~Smirnov and L.~Tancredi, \emph{{Analytic results for two-loop
  planar master integrals for Bhabha scattering}},
  \href{https://doi.org/10.1007/JHEP09(2021)120}{\emph{JHEP} {\bfseries 09}
  (2021) 120} [\href{https://arxiv.org/abs/2108.03828}{{\ttfamily
  2108.03828}}].

\bibitem{BrownNotes}
F.~{Brown}, \emph{{Notes on Motivic Periods}},
  \href{https://arxiv.org/abs/1512.06410}{{\ttfamily 1512.06410}}.

\bibitem{abel1841}
N.H.~Abel, \emph{Mémoire sur une propriété générale d'une classe très
  étendue de fonctions transcendantes},  in \emph{Oeuvres complètes de Niels
  Henrik Abel: Nouvelle édition}, L.~Sylow and S.~Lie, eds., vol.~1 of
  \emph{Cambridge Library Collection - Mathematics}, p.~145–211, Cambridge
  University Press (2012),
  \href{https://doi.org/10.1017/CBO9781139245807.013}{DOI}.

\bibitem{Zagier2000}
D.~Zagier and H.~Gangl, \emph{Classical and elliptic polylogarithms and special
  values of l-series},  in \emph{The Arithmetic and Geometry of Algebraic
  Cycles}, B.B.~Gordon, J.D.~Lewis, S.~M{\"u}ller-Stach, S.~Saito and N.~Yui,
  eds., (Dordrecht), pp.~561--615, Springer Netherlands (2000),
  \href{https://doi.org/10.1007/978-94-011-4098-0_21}{DOI}.

\bibitem{bloch2011higher}
S.J.~Bloch, \emph{Higher regulators, algebraic $ K $-theory, and zeta functions
  of elliptic curves}, vol.~11, American Mathematical Soc. (2000),
  \href{https://doi.org/10.1090/crmm/011}{10.1090/crmm/011}.

\bibitem{Broedel:2019tlz}
J.~Broedel and A.~Kaderli, \emph{{Functional relations for elliptic
  polylogarithms}}, \href{https://doi.org/10.1088/1751-8121/ab81d7}{\emph{J.
  Phys. A} {\bfseries 53} (2020) 245201}
  [\href{https://arxiv.org/abs/1906.11857}{{\ttfamily 1906.11857}}].

\bibitem{Bolbachan:2019dsu}
V.~Bolbachan, \emph{{On functional equations for the elliptic dilogarithm}},
  \href{https://arxiv.org/abs/1906.05068}{{\ttfamily 1906.05068}}.

\bibitem{Broedel:2014vla}
J.~Broedel, C.R.~Mafra, N.~Matthes and O.~Schlotterer, \emph{{Elliptic multiple
  zeta values and one-loop superstring amplitudes}},
  \href{https://doi.org/10.1007/JHEP07(2015)112}{\emph{JHEP} {\bfseries 07}
  (2015) 112} [\href{https://arxiv.org/abs/1412.5535}{{\ttfamily 1412.5535}}].

\bibitem{GriffithsHarris}
P.A.~Griffiths and J.~Harris, \emph{{Principles of algebraic geometry}}, Wiley
  classics library, Wiley, New York, NY (1994),
  \href{https://doi.org/10.1002/9781118032527}{10.1002/9781118032527}.

\bibitem{Tarasov:2017yyd}
O.V.~Tarasov, \emph{{Methods for deriving functional equations for Feynman
  integrals}}, \href{https://doi.org/10.1088/1742-6596/920/1/012004}{\emph{J.
  Phys. Conf. Ser.} {\bfseries 920} (2017) 012004}
  [\href{https://arxiv.org/abs/1709.07058}{{\ttfamily 1709.07058}}].

\bibitem{Tarasov:slides}
O.V.~Tarasov, ``Functional equations for feynman integrals and abel’s
  addition theorem.'' \url{http://theor.jinr.ru/sqs17/Talks/Tarasov.pdf}.

\bibitem{CaronHuot:2012ab}
S.~Caron-Huot and K.J.~Larsen, \emph{{Uniqueness of two-loop master contours}},
  \href{https://doi.org/10.1007/JHEP10(2012)026}{\emph{JHEP} {\bfseries 10}
  (2012) 026} [\href{https://arxiv.org/abs/1205.0801}{{\ttfamily 1205.0801}}].

\bibitem{Hodges:2009hk}
A.~Hodges, \emph{{Eliminating spurious poles from gauge-theoretic amplitudes}},
  \href{https://doi.org/10.1007/JHEP05(2013)135}{\emph{JHEP} {\bfseries 05}
  (2013) 135} [\href{https://arxiv.org/abs/0905.1473}{{\ttfamily 0905.1473}}].

\bibitem{ArkaniHamed:2010kv}
N.~Arkani-Hamed, J.L.~Bourjaily, F.~Cachazo, S.~Caron-Huot and J.~Trnka,
  \emph{{The All-Loop Integrand For Scattering Amplitudes in Planar
  $\mathcal{N}=4$ SYM}},
  \href{https://doi.org/10.1007/JHEP01(2011)041}{\emph{JHEP} {\bfseries 01}
  (2011) 041} [\href{https://arxiv.org/abs/1008.2958}{{\ttfamily 1008.2958}}].

\bibitem{integrability_in_progress}
R.~Morales, A.~Spiering, M.~Wilhelm and C.~Zhang. In progress.

\bibitem{Drummond:2010mb}
J.M.~Drummond and J.M.~Henn, \emph{{Simple loop integrals and amplitudes in
  $\mathcal{N}=4$ SYM}},
  \href{https://doi.org/10.1007/JHEP05(2011)105}{\emph{JHEP} {\bfseries 05}
  (2011) 105} [\href{https://arxiv.org/abs/1008.2965}{{\ttfamily 1008.2965}}].

\bibitem{Panzer:2014caa}
E.~Panzer, \emph{{Algorithms for the symbolic integration of hyperlogarithms
  with applications to Feynman integrals}},
  \href{https://doi.org/10.1016/j.cpc.2014.10.019}{\emph{Comput. Phys. Commun.}
  {\bfseries 188} (2015) 148}
  [\href{https://arxiv.org/abs/1403.3385}{{\ttfamily 1403.3385}}].

\bibitem{Duhr:2019tlz}
C.~Duhr and F.~Dulat, \emph{{PolyLogTools — polylogs for the masses}},
  \href{https://doi.org/10.1007/JHEP08(2019)135}{\emph{JHEP} {\bfseries 08}
  (2019) 135} [\href{https://arxiv.org/abs/1904.07279}{{\ttfamily
  1904.07279}}].

\bibitem{Arkani-Hamed:2013jha}
N.~Arkani-Hamed and J.~Trnka, \emph{{The Amplituhedron}},
  \href{https://doi.org/10.1007/JHEP10(2014)030}{\emph{JHEP} {\bfseries 10}
  (2014) 030} [\href{https://arxiv.org/abs/1312.2007}{{\ttfamily 1312.2007}}].

\bibitem{Yang:2022gko}
Q.~Yang, \emph{{Schubert Problems, Positivity and Symbol Letters}},
  \href{https://arxiv.org/abs/2203.16112}{{\ttfamily 2203.16112}}.

\bibitem{Duhr:2020gdd}
C.~Duhr and F.~Brown, \emph{{A double integral of dlog forms which is not
  polylogarithmic}}, \href{https://doi.org/10.22323/1.383.0005}{\emph{PoS}
  {\bfseries MA2019} (2022) 005}
  [\href{https://arxiv.org/abs/2006.09413}{{\ttfamily 2006.09413}}].

\bibitem{OKUN1960261}
L.~Okun and A.~Rudik, \emph{On a method of finding singularities of feynman
  graphs},
  \href{https://doi.org/https://doi.org/10.1016/0029-5582(60)90307-2}{\emph{Nuclear
  Physics} {\bfseries 15} (1960) 261}.

\bibitem{Golden:2013xva}
J.~Golden, A.B.~Goncharov, M.~Spradlin, C.~Vergu and A.~Volovich,
  \emph{{Motivic Amplitudes and Cluster Coordinates}},
  \href{https://doi.org/10.1007/JHEP01(2014)091}{\emph{JHEP} {\bfseries 01}
  (2014) 091} [\href{https://arxiv.org/abs/1305.1617}{{\ttfamily 1305.1617}}].

\bibitem{Golden:2014pua}
J.~Golden and M.~Spradlin, \emph{{A Cluster Bootstrap for Two-Loop MHV
  Amplitudes}}, \href{https://doi.org/10.1007/JHEP02(2015)002}{\emph{JHEP}
  {\bfseries 02} (2015) 002} [\href{https://arxiv.org/abs/1411.3289}{{\ttfamily
  1411.3289}}].

\bibitem{Drummond:2018dfd}
J.~Drummond, J.~Foster and {\"{O}}.~G{\"{u}}rdo{\u{g}}an, \emph{{Cluster
  adjacency beyond MHV}},
  \href{https://doi.org/10.1007/JHEP03(2019)086}{\emph{JHEP} {\bfseries 03}
  (2019) 086} [\href{https://arxiv.org/abs/1810.08149}{{\ttfamily
  1810.08149}}].

\bibitem{Zhang:2019vnm}
S.~He, Z.~Li and C.~Zhang, \emph{{Two-loop Octagons, Algebraic Letters and
  $\bar{Q}$ Equations}},
  \href{https://doi.org/10.1103/PhysRevD.101.061701}{\emph{Phys. Rev. D}
  {\bfseries 101} (2020) 061701(R)}
  [\href{https://arxiv.org/abs/1911.01290}{{\ttfamily 1911.01290}}].

\bibitem{He:2020vob}
S.~He, Z.~Li and C.~Zhang, \emph{{The symbol and alphabet of two-loop NMHV
  amplitudes from $\bar{Q}$ equations}},
  \href{https://doi.org/10.1007/JHEP03(2021)278}{\emph{JHEP} {\bfseries 03}
  (2021) 278} [\href{https://arxiv.org/abs/2009.11471}{{\ttfamily
  2009.11471}}].

\bibitem{He:2020lcu}
S.~He, Z.~Li, Q.~Yang and C.~Zhang, \emph{{Feynman Integrals and Scattering
  Amplitudes from Wilson Loops}},
  \href{https://doi.org/10.1103/PhysRevLett.126.231601}{\emph{Phys. Rev. Lett.}
  {\bfseries 126} (2021) 231601}
  [\href{https://arxiv.org/abs/2012.15042}{{\ttfamily 2012.15042}}].

\bibitem{Li:2021bwg}
Z.~Li and C.~Zhang, \emph{{The three-loop MHV octagon from $ \overline{Q} $
  equations}}, \href{https://doi.org/10.1007/JHEP12(2021)113}{\emph{JHEP}
  {\bfseries 12} (2021) 113}
  [\href{https://arxiv.org/abs/2110.00350}{{\ttfamily 2110.00350}}].

\bibitem{He:2021esx}
S.~He, Z.~Li and Q.~Yang, \emph{{Notes on cluster algebras and some all-loop
  Feynman integrals}},
  \href{https://doi.org/10.1007/JHEP06(2021)119}{\emph{JHEP} {\bfseries 06}
  (2021) 119} [\href{https://arxiv.org/abs/2103.02796}{{\ttfamily
  2103.02796}}].

\bibitem{Drummond:2019cxm}
J.~Drummond, J.~Foster, O.~G\"urdogan and C.~Kalousios, \emph{{Algebraic
  singularities of scattering amplitudes from tropical geometry}},
  \href{https://doi.org/10.1007/JHEP04(2021)002}{\emph{JHEP} {\bfseries 04}
  (2021) 002} [\href{https://arxiv.org/abs/1912.08217}{{\ttfamily
  1912.08217}}].

\bibitem{Arkani-Hamed:2019rds}
N.~Arkani-Hamed, T.~Lam and M.~Spradlin, \emph{{Non-perturbative geometries for
  planar $ \mathcal{N} $ = 4 SYM amplitudes}},
  \href{https://doi.org/10.1007/JHEP03(2021)065}{\emph{JHEP} {\bfseries 03}
  (2021) 065} [\href{https://arxiv.org/abs/1912.08222}{{\ttfamily
  1912.08222}}].

\bibitem{Henke:2019hve}
N.~Henke and G.~Papathanasiou, \emph{{How tropical are seven- and
  eight-particle amplitudes?}},
  \href{https://doi.org/10.1007/JHEP08(2020)005}{\emph{JHEP} {\bfseries 08}
  (2020) 005} [\href{https://arxiv.org/abs/1912.08254}{{\ttfamily
  1912.08254}}].

\bibitem{Herderschee:2021dez}
A.~Herderschee, \emph{{Algebraic branch points at all loop orders from positive
  kinematics and wall crossing}},
  \href{https://doi.org/10.1007/JHEP07(2021)049}{\emph{JHEP} {\bfseries 07}
  (2021) 049} [\href{https://arxiv.org/abs/2102.03611}{{\ttfamily
  2102.03611}}].

\bibitem{Henke:2021ity}
N.~Henke and G.~Papathanasiou, \emph{{Singularities of eight- and nine-particle
  amplitudes from cluster algebras and tropical geometry}},
  \href{https://doi.org/10.1007/JHEP10(2021)007}{\emph{JHEP} {\bfseries 10}
  (2021) 007} [\href{https://arxiv.org/abs/2106.01392}{{\ttfamily
  2106.01392}}].

\bibitem{Ren:2021ztg}
L.~Ren, M.~Spradlin and A.~Volovich, \emph{{Symbol alphabets from tensor
  diagrams}}, \href{https://doi.org/10.1007/JHEP12(2021)079}{\emph{JHEP}
  {\bfseries 12} (2021) 079}
  [\href{https://arxiv.org/abs/2106.01405}{{\ttfamily 2106.01405}}].

\end{thebibliography}\endgroup
\bibliographystyle{JHEP}

\end{document}